\documentclass[a4paper,11pt]{article}
\usepackage{graphicx} 
\usepackage{subcaption}

\pdfoutput=1 

\usepackage{jcappub} 

\usepackage{hanging}
\usepackage{orcidlink}
\usepackage[T1]{fontenc} 
\usepackage{multirow} 

\usepackage{pifont}
\newcommand{\cmark}{\ding{51}}%
\newcommand{\xmark}{\ding{55}}%

\newcommand{\bq}{\boldsymbol q}
\newcommand{\bx}{\boldsymbol x}
\newcommand{\bk}{\boldsymbol k}
\newcommand{\bPsi}{\boldsymbol{\Psi}}

\newcommand{\ihmpc}{\,h{\rm Mpc}^{-1}}

\newcommand{\Ckg}{C_{\ell}^{\kappa g}}
\newcommand{\Cgg}{C_{\ell}^{g g}}

\title{A joint analysis of 3D clustering and galaxy × CMB-lensing cross-correlations with DESI DR1 galaxies}

\emailAdd{mark.maus@berkeley.edu}
\affiliation{Affiliations are in Appendix \ref{sec:affiliations}}

\author[1,2]{{M.~Maus}\orcidlink{0000-0002-9020-911X},}
\author[3,1]{{M.~White}\orcidlink{0000-0001-9912-5070},}
\author[1,2]{{N.~Sailer}\orcidlink{0000-0002-5333-8983},}
\author[3,2]{{A.~Baleato Lizancos}\orcidlink{0000-0002-0232-6480},}
\author[2,1]{{S.~Ferraro}\orcidlink{0000-0003-4992-7854},}
\author[4]{{S.~Chen}\orcidlink{0000-0002-5762-6405},}
\author[5]{{J.~DeRose}\orcidlink{0000-0002-0728-0960},}
\author[2]{{J.~Aguilar},}
\author[6]{{S.~Ahlen}\orcidlink{0000-0001-6098-7247},}
\author[7]{{S.~BenZvi}\orcidlink{0000-0001-5537-4710},}
\author[8,9]{{D.~Bianchi}\orcidlink{0000-0001-9712-0006},}
\author[10]{{D.~Brooks},}
\author[11]{{E.~Burtin},}
\author[12,13]{{F.~J.~Castander}\orcidlink{0000-0001-7316-4573},}
\author[2]{{E.~Chaussidon}\orcidlink{0000-0001-8996-4874},}
\author[2]{{T.~Claybaugh},}
\author[2]{{A.~Cuceu}\orcidlink{0000-0002-2169-0595},}
\author[14]{{A.~de la Macorra}\orcidlink{0000-0002-1769-1640},}
\author[11]{{A.~de~Mattia}\orcidlink{0000-0003-0920-2947},}
\author[10]{{P.~Doel},}
\author[15]{{A.~Font-Ribera}\orcidlink{0000-0002-3033-7312},}
\author[16,17]{{J.~E.~Forero-Romero}\orcidlink{0000-0002-2890-3725},}
\author[12,18,13]{{E.~Gaztañaga},}
\author[2]{{S.~Gontcho A Gontcho}\orcidlink{0000-0003-3142-233X},}
\author[19]{{G.~Gutierrez},}
\author[2]{{J.~Guy}\orcidlink{0000-0001-9822-6793},}
\author[20,21,22]{{K.~Honscheid}\orcidlink{0000-0002-6550-2023},}
\author[23]{{C.~Howlett}\orcidlink{0000-0002-1081-9410},}
\author[24]{{M.~Ishak}\orcidlink{0000-0002-6024-466X},}
\author[25]{{R.~Kehoe},}
\author[26]{{D.~Kirkby}\orcidlink{0000-0002-8828-5463},}
\author[2]{{T.~Kisner}\orcidlink{0000-0003-3510-7134},}
\author[2]{{A.~Kremin}\orcidlink{0000-0001-6356-7424},}
\author[10]{{O.~Lahav},}
\author[27]{{C.~Lamman}\orcidlink{0000-0002-6731-9329},}
\author[2]{{M.~Landriau}\orcidlink{0000-0003-1838-8528},}
\author[28]{{L.~Le~Guillou}\orcidlink{0000-0001-7178-8868},}
\author[2]{{M.~E.~Levi}\orcidlink{0000-0003-1887-1018},}
\author[29,15]{{M.~Manera}\orcidlink{0000-0003-4962-8934},}
\author[30]{{A.~Meisner}\orcidlink{0000-0002-1125-7384},}
\author[31,15]{{R.~Miquel},}
\author[18]{{S.~Nadathur}\orcidlink{0000-0001-9070-3102},}
\author[32]{{J.~ A.~Newman}\orcidlink{0000-0001-8684-2222},}
\author[11,2]{{N.~Palanque-Delabrouille}\orcidlink{0000-0003-3188-784X},}
\author[33,34,35]{{W.~J.~Percival}\orcidlink{0000-0002-0644-5727},}
\author[36]{{F.~Prada}\orcidlink{0000-0001-7145-8674},}
\author[37]{{I.~P\'erez-R\`afols}\orcidlink{0000-0001-6979-0125},}
\author[20,38,22]{{A.~J.~Ross}\orcidlink{0000-0002-7522-9083},}
\author[39]{{G.~Rossi},}
\author[40,41,42]{{L.~Samushia}\orcidlink{0000-0002-1609-5687},}
\author[43]{{E.~Sanchez}\orcidlink{0000-0002-9646-8198},}
\author[2]{{D.~Schlegel},}
\author[44,45]{{M.~Schubnell},}
\author[46]{{H.~Seo}\orcidlink{0000-0002-6588-3508},}
\author[2]{{J.~Silber}\orcidlink{0000-0002-3461-0320},}
\author[30]{{D.~Sprayberry},}
\author[45]{{G.~Tarl\'{e}}\orcidlink{0000-0003-1704-0781},}
\author[30]{{B.~A.~Weaver},}
\author[28]{{P.~Zarrouk}\orcidlink{0000-0002-7305-9578},}
\author[2]{{R.~Zhou}\orcidlink{0000-0001-5381-4372},}
\author[47]{{H.~Zou}\orcidlink{0000-0002-6684-3997},}

\abstract{The spectroscopic data from DESI Data Release 1 (DR1) galaxies enables the analysis of 3D clustering by fitting galaxy power spectra and reconstructed correlation functions in redshift space. Given low measurements of the amplitude of structure from cosmic shear at $z\sim1$, redshift space distortions (RSD) + Baryon Acoustic Oscillation (BAO) signals from DESI galaxies combined with weak lensing can break degeneracies and provide a tight alternative constraint on the $z\sim1$ amplitude of structure. In this paper we perform joint analyses that combine full-shape + post-reconstruction information from the DESI DR1 BGS and LRG samples along with angular cross-correlations with Planck PR4 and ACT DR6 CMB lensing maps. We show that adding galaxy-lensing cross-correlations tightens clustering amplitude constraints, improving $\sigma_8$ uncertainties by $30\%$ over RSD+BAO alone. We also include angular galaxy-galaxy and galaxy-lensing spectra using photometric samples from the DESI Legacy Survey to further improve constraints. Our headline results are $\sigma_8 = 0.803\pm 0.017$, $\Omega_{\rm m} = 0.3037\pm 0.0069$, and $S_8 = 0.808\pm 0.017$. Given DESI's preference for higher $\sigma_8$ compared to lower values from BOSS, we perform a catalog-level comparison of LRG samples from both surveys. We test sensitivity to dark energy assumptions by relaxing our $\Lambda$CDM prior and allowing for evolving dark energy via the $w_0-w_a$ parameterization. We find our $S_8$ constraints to be relatively unchanged despite a $~3.5\sigma$ tension with the cosmological constant model when combining with the Union3 supernova likelihood. Finally we test general relativity (GR) by allowing the gravitational slip parameter ($\gamma$) to vary, and find $\gamma = 1.17\pm0.11$ in mild ($\sim1.5\sigma$) tension with the GR value of $1.0$.}

\begin{document}
\maketitle
\flushbottom

\section{Introduction}
\label{sec: intro}
Understanding the large-scale structure (LSS) of the universe is essential for constraining fundamental cosmological parameters and testing the standard 
$\Lambda$CDM paradigm \cite{P5_2024,Snowmass22_a,Snowmass22_b,Snowmass2013.Levi}. The statistical distribution of galaxies provides a wealth of information about cosmic expansion, the growth of structure, and the nature of dark matter and dark energy. Many telescope missions are planned or already underway that aim to extract as much cosmological information from the LSS as current observational technology permits, including the The Legacy Survey of Space and Time (LSST, \cite{LSST}) and its Dark Energy Science Collaboration (DESC, \cite{LSSTDesc,LSSTDesc2}), the Euclid Space Telescope \cite{Euclid,Amendola18}, SPHEREx \cite{Spherex}, the Dark Energy Spectroscopic Instrument (DESI, \cite{DESI}) and more. LSS observations from these missions take the form of photometric (or spectroscopic) surveys that catalog the angular (or 3D) positions of as many galaxies as possible across the Universe's expansion in order to trace the evolution of structure. DESI, the furthest advanced of these missions, is a spectroscopic survey that aims to measure spectra of over 50 million galaxies and quasars over the course of 5 years of observation \cite{DESI,Spectro.Pipeline.Guy.2023,SurveyOps.Schlafly.2023}. The state-of-the-art instrument, located at Kitt Peak National Observatory and mounted on the Mayall 4-meter telescope \cite{DESI2022.KP1.Instr}, uses 5000 robotic fibers that allow it to measure 5000 astronomical targets at once \cite{DESI2016b.Instr,FocalPlane.Silber.2023,Corrector.Miller.2023}. This will provide us with the largest 3D map of the universe to date, enabling competitive constraints on cosmological phenomena, such as the nature of dark energy, that are sensitive to the expansion history of the universe \cite{Snowmass2013.Levi}.



Following the completion of the first year of observation, the DESI collaboration produced two primary measurements of galaxy clustering: the multipoles of the power spectrum, $P_{\ell}(k)$, and the post-reconstruction correlation function multipoles, $\xi_{\ell}(s)$ \cite{DESI2024.II.KP3}. These statistics provide complementary insights into the distribution of large-scale structure. After undoing some of the nonlinear damping/smearing of the Baryon Acoustic Oscillation (BAO) signal caused by gravitational evolution, the post-reconstruction correlation function offers sharpened geometric sensitivity to the background evolution of the Universe \cite{Eisenstein_recon2007}. Fits to the `full-shape' of the redshift-space power spectrum multipoles, on the other hand, are in principle sensitive to the full range of physical processes encapsulated in the two-point statistics of the galaxy field. This includes, in particular, the redshift-space distortions of galaxy positions induced by their peculiar velocities \cite{Kaiser87,Hamilton92}. Since these velocities are sourced by gravity and can be inhibited by dark energy, they are valuable probes of both phenomena. Following careful testing and validation of the analysis methods and sources of systematics \cite{KP4s2-Chen,KP4s3-Chen,KP4s4-Paillas,KP4s5-Valcin,KP4s6-Forero-Sanchez,KP4s7-Rashkovetskyi,KP4s8-Alves,KP4s9-Perez-Fernandez,KP4s10-Mena-Fernandez,KP4s11-Garcia-Quintero,KP5s1-Maus,KP5s2-Maus,KP5s3-Noriega,KP5s4-Lai,KP5s5-Ramirez,KP5s6-Zhao} the Y1 full-shape and BAO analyses were unblinded with results presented in Refs.~\cite{DESI2024.III.KP4,DESI2024.V.KP5,DESI2024.VI.KP7A,DESI2024.VII.KP7B}.

Gravitational lensing of cosmic microwave background (CMB) photons offers a complementary window into the cosmic structures hosting these DESI galaxies, one that helps better characterize galaxy bias (when cross-correlating with galaxies) and is sensitive to a different set of systematic effects, which makes cross-correlations robust against additive biases. Such combinations of lensing and clustering data also enable the tomographic decomposition of lensing information --- which otherwise would be smeared out in projection --- into the redshift bins defined by the galaxy survey. Cross-correlations between CMB lensing and the galaxy overdensity field involves a different scaling between galaxy bias and the clustering amplitude $\sigma_8$ than galaxy-galaxy auto correlations ($\propto b\sigma_8^2$ vs $b^2\sigma_8^2$) so the joint analysis of galaxy-galaxy and galaxy-lensing spectra helps break degeneracies and tightens constraints of $\sigma_8$.
For these reasons, references \cite{Sailer24,Kim2024} analyzed cross-correlations between the photometric sample DESI luminous red galaxies (LRGs) and CMB lensing maps from both \textit{Planck} and the Atacama Cosmology Telescope (ACT) in combination with the angular galaxy auto-spectrum. Ref.~\cite{Sailer2025} repeated this analysis but included galaxies from the photometric Bright Galaxy Survey (BGS) sample from the DESI Legacy survey. Ref.~\cite{Karim2025} similarly combined photometric DESI Emission Line Galaxies (ELG) data with Planck CMB lensing. However, the galaxy samples used in those studies lacked spectroscopic information, which is crucial for tightening constraints through a more precise understanding of the 3D galaxy distribution. Although they incorporated post-reconstruction BAO measurements from the earlier Baryon Oscillation Spectroscopic Survey (BOSS) as an external dataset, integrating the spectroscopic data from DESI galaxies would allow for a more powerful combination of full-shape and BAO information with the cross-correlations—ultimately leading to significantly stronger constraints. 

In this work, we set out to combine measurements of the redshift-space clustering of DESI BGS and LRGs with the gravitational lensing effect they imprint on the background CMB. These being complementary probes of the same structures, their joint analysis will lead to degeneracy breaking and an improvement of constraints going beyond the sheer increase in number of modes. Previous works have combined 3D clustering with weak lensing measurements in pursuit of these gains, often with the goal of testing gravity through the different sensitivity of RSD and lensing to the Newtonian and Weyl potentials, respectively~\cite{pullenConstrainingGravityLargest2016a,singhProbingGravityJoint2019, wenzlAtacamaCosmologyTelescope2024, wenzlConstrainingGravityNew2024,Chen24, Chen22_2}.

In this paper we present our methodology combining the aforementioned data sets into a single joint analysis. Namely we combine the 3D clustering from the spectroscopic galaxy samples, cross-correlations of the spectroscopic LRGs with CMB lensing, and the angular galaxy-lensing cross- and galaxy-galaxy auto- spectra from the photometric LRG samples. The purpose of this work is both to present some of the tightest constraints (we find $\sigma_8 = 0.803\pm 0.017$ which is a $2.1\%$ constraint) on the clustering amplitude using currently available galaxy data, and also to develop the methodology and pipeline\footnote{\url{https://github.com/mmaus96/C3PO}} for this setup in anticipation for the Y3 DESI data release. While this work only incorporates cross-correlations using spectroscopic BGS and LRG samples, a companion paper is in development presenting cross-correlations using the Y1 quasar (QSO) sample~\cite{deBelsunce2025}. 

This paper is organized as follows. We begin by describing the galaxy and CMB lensing data in \S\ref{sec: data}, followed by a description of the joint likelihood and estimators in \S\ref{sec: likelihood}. \S\ref{sec: thy} presents the theoretical models employed in our analysis including a description of the free parameters and their priors. We present results in \S\ref{sec: results} and conclude in \S\ref{sec: conc}.

\section{Data}
\label{sec: data}

\begin{figure}[htb]
    \centering
    \includegraphics[width=0.98\linewidth]{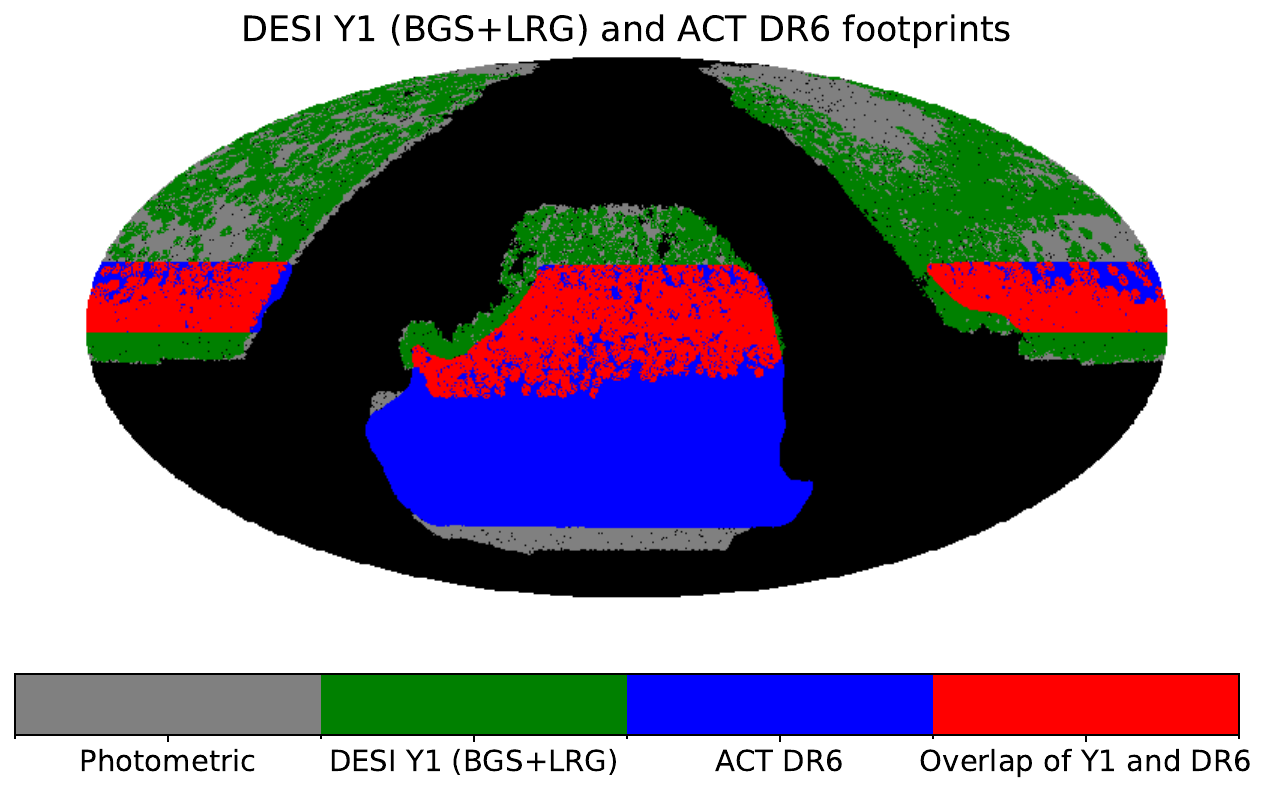}
    \caption{Overlapping maps of the data samples used in this paper. DESI Y1 (BGS+LRG) and ACT DR6 footprints are shown in green and blue, respectively, while the red regions show their overlap. The footprint of the full photometric LRG sample is shown in grey. The \textit{Planck} PR4 map overlaps almost entirely with each of these footprints and is therefore not shown here. 
    }
    \label{fig:y1_dr6_overlap}
\end{figure}

Our analysis involves both spectroscopic and imaging galaxy catalogs from the Dark Energy Spectroscopic Instrument (DESI) along with lensing convergence maps, masks, and noise curves obtained from \textit{Planck} and the Atacama Cosmology Telescope (ACT). We briefly describe each of these data sets in the following two subsections. Figure \ref{fig:y1_dr6_overlap} shows the overlap of the various survey footprints. We do not explicitly plot the \textit{Planck} PR4 footprint as it almost entirely overlaps with the other surveys except in regions with bright galactic or extragalactic sources that are masked out.

\subsection{DESI}
\label{sec: data.desi}

\begin{figure}[htb]
    \centering
    \includegraphics[width=0.98\linewidth]{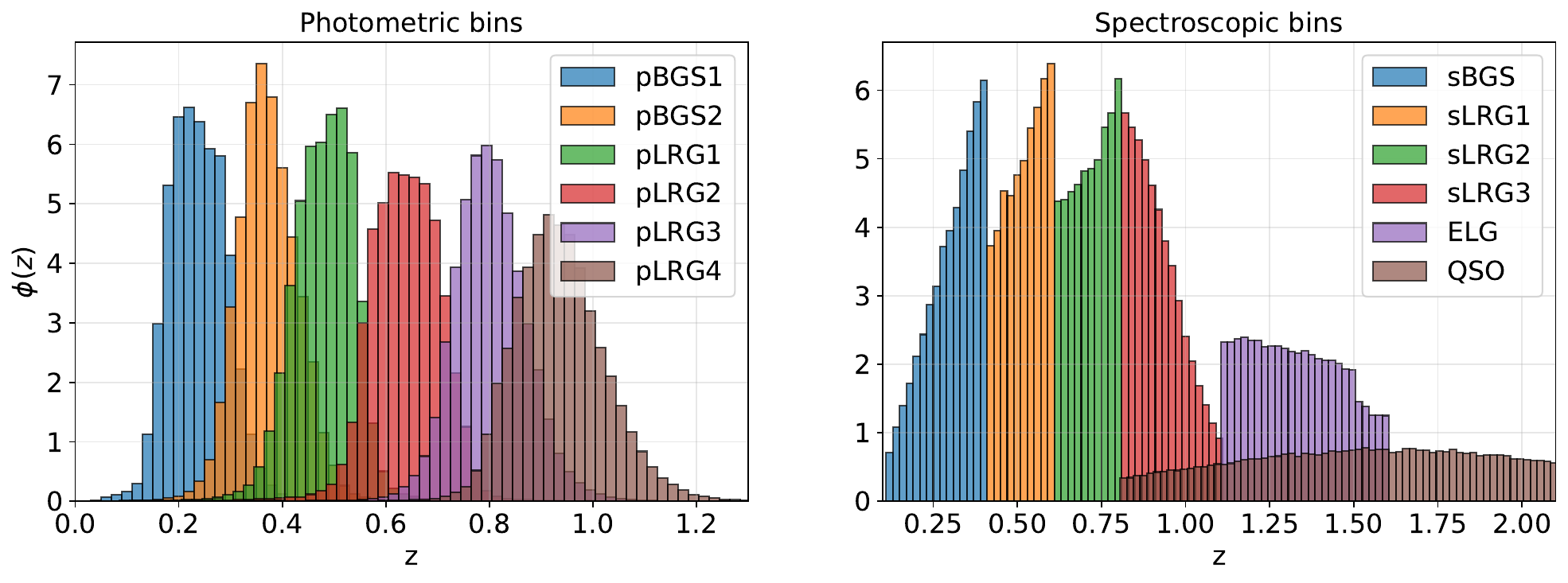}
    \caption{Normalized redshift distributions for the Photometric and Spectroscopic samples. Note that the ELG and QSO samples are only used for 3D clustering.}
    \label{fig:desi_dndz}
\end{figure}

The Dark Energy Spectroscopic Instrument (DESI) is a Stage-IV galaxy redshift survey that aims to measure spectra from approximately 50 million targets over the course of 5 years to create the largest 3D map of the universe to date. The DESI instrument is mounted on the Mayall 4-meter telescope at Kitt Peak National Observatory in Arizona. The sample of objects targeted in the survey consists of four different classes: (1) bright galaxies spanning the redshift range $0.1\leq z\leq 0.4$ which are observed in the Bright Galaxy Survey (BGS) at times when moonlight inhibits observations of fainter objects; (2) luminous red galaxies (LRG) in the redshift range $0.4\leq z\leq 1.1$; (3) emission line galaxies (ELG) spanning $0.8\leq z\leq 1.6$; and (4) quasi-stellar objects/quasars (QSO) of redshifts $0.8\leq z\leq 2.1$. All of these targets are selected from the DESI Legacy Imaging Survey DR9 \cite{DESI_legacy}. Our analysis in this paper will include a hybrid of samples from the DESI DR1 data release \cite{DESI2024.I.DR1}, for which we have spectroscopic redshifts, and the non-overlapping subset of the photometric targeting sample, from which we can only measure and analyze angular summary statistics in broad $z$-bins based on photometric redshifts \cite{LRG.TS.Zhou.2023}. We show in Table~\ref{tab:samp_vals} relevant estimated properties and in  Fig.~\ref{fig:desi_dndz} the redshift distributions for the photometric and spectroscopic samples. The spectroscopic LRG sample is split into three redshift bins with cuts $0.4\leq z \leq 0.6$, $0.6< z \leq 0.8$, and $0.8< z \leq 1.1$ while the BGS\footnote{For the DR1 BGS sample (that we use) an absolute magnitude cut of $M_r < -21.5$ was applied to ensure a more constant number density (see \S3 of \cite{DESI2024.II.KP3}). For DR2 the magnitude cut is changed to $M_r<-21.35$\cite{DESIDR2_bao}.} sample covers a single redshift bin $0.1\leq z\leq 0.4$. The photometric LRG sample, whose selection is described in \cite{LRG.TS.Zhou.2023}, is grouped into four photometrically defined redshift bins, as presented in \cite{Zhou23}. We use the ``main'' LRG sample as opposed to the ``extended'' sample observed before the main survey\cite{LRG.TS.Zhou.2023}. The photometric BGS sample we use was separated into two bins by \cite{ChenDeRose24}, with redshift distributions calibrated from a spectroscopic subsample in a similar procedure as the binning of the LRGs.

\begin{table}
    \centering
    \begin{tabular}{|c|c|c|c|c|c|c|c|c|}
        Tracer ($z_{\rm eff}$) & $10^3$SN$_0$ & $10^4$SN$_2$ & 10$^6$SN$_{\rm 2D}$ & $s_0$ & $s_2$ &  $s_{\mu}$& $\bar{n}_{\theta}$ (deg$^{-2}$)  \\
        \hline\hline
        sBGS (0.295) & $5.72$ & $5.49$  &  29.1 & 759.46 & 30.34 &  1.113 & 40.2 \\
        sLRG1 (0.510) & $5.08$ & $7.31$ & 21.8 & 325.43 & 20.83 &  1.016 & 88.3 \\
        sLRG2 (0.706) & $5.23$ & $7.52$ & 13.9 & 850.23 & 47.14 &  0.996 & 134.5 \\
        sLRG3 (0.910) & 9.57 & 13.77 & 8.79 & 1185.87 & 60.89 &  1.032 & 149.8 \\
        pBGS1 (0.211) & -- & -- & 0.463 & -- & -- &  0.81 & 627 \\
        pBGS2 (0.352) & -- & -- & 0.918 & -- & -- &  0.80 & 317\\
        pLRG1 (0.470) & -- & -- & 4.07 & -- & -- &  0.972 & 81.9\\
        pLRG2 (0.625) & -- & -- & 2.25 & -- & -- &  1.044 & 148.1\\
        pLRG3 (0.785) & -- & -- & 2.05 & -- & -- &  0.974 & 162.4\\
        pLRG4 (0.914) & -- & -- & 2.25 & -- & -- &  0.988 & 148.3\\
        ELG (1.317) & $10.7$ & $2.58$ &  -- & 447.49 & 8.32 &  -- & 239.0 \\
        QSO (1.491) & $47.38$ & $11.46$ &  -- & 814.55 & 63.27 & -- & 118.2 \\
    \end{tabular}
    \caption{Sample specific values measured/estimated for each tracer.  
    There are missing entries for values that are not relevant/used from that sample, e.g.\ we only use ELG and QSO samples for 3D clustering so magnification bias and SN$_{\rm 2D}$ have not been measured for those samples. SN$_0$ and SN$_{\rm 2D}$ are Poisson shot-noise values in 3D and 2D, respectively. SN$_2$ is a stochastic noise component from the Finger of God effect that multiplies $\mu^2$ in our power spectrum model. Its estimated value is therefore SN$_0$ multiplied by factors of the satellite fraction and virial velocity dispersion (squared) using values typical for these tracers (see section 4.2 and Appendix C of \cite{KP5s2-Maus} for a more detailed discussion). $s_0$ and $s_2$ are linear coefficients multiplying the rotation templates in the power spectrum associated with the transformation that diagonalizes the window matrix (see \S\ref{sec: weights}). $s_{\mu}$ is the number count slope for each sample used for estimating magnification bias contributions to angular spectra, and $\bar{n}_{\theta}$ is the angular number density of each sample.}
    \label{tab:samp_vals}
\end{table}

\subsection{\textit{Planck} PR4 and ACT DR6}
\label{sec: data.cmb}

We use the lensing convergence $\kappa_{\ell m}$'s from the PR4 and DR6 public data releases from \textit{Planck} \cite{Carron_2022} and the Atacama Cosmology Telescope (ACT; \cite{ACT:2023dou,ACT:2023ubw,ACT:2023kun}) respectively. Reconstruction of the \textit{Planck} PR4 convergence was performed with a global minimum variance estimator \cite{Maniyar:2021msb} and a more optimal (and anisotropic) filtering scheme over the scale range $100\leq \ell \leq 2048$, resulting in a detection significance of 42$\sigma$ for the CMB lensing autocorrelation and an increase in signal-to-noise (SNR) of $\sim20\%$ over PR3.
In our analysis, we make use of the public lensing convergence multipoles, noise curves, and reconstruction mask, which are publicly available\footnote{\url{https://github.com/carronj/planck_PR4_lensing}}. The ACT DR6 lensing convergence\footnote{\url{https://portal.nersc.gov/project/act/dr6_lensing_v1/maps/baseline/}} was reconstructed from a nearly-optimal linear combination of temperature- and polarization-based quadratic estimators (QE) over the scale range $600 < \ell < 3000$, yielding a 43$\sigma$ detection of the CMB lensing autocorrelation. To mitigate noise biases in the lensing reconstruction disjoint data splits were used in each leg of the QE \cite{Madhavacheril:2020ido}, while a profile hardened estimator was used in the temperature-only QE to reduce biases arising from extragalactic foregrounds \cite{2013MNRAS.431..609N,Osborne:2013nna,Sailer:2020lal,Sailer:2022jwt}.
For both the \textit{Planck} and ACT $\kappa_{\ell m}$ data we apply a low-pass filter to suppress the convergence for modes above $\ell = 3000$.

\subsection{Blinding}

Each of the aforementioned data sets that we use in this paper were unblinded at the time that this analysis was started. On the modeling side, each piece of the pipeline presented here was tested on mocks and blinded data to determine optimal fiducial settings and scale cuts, from which we do not deviate in this work. In particular the 3D clustering part of our analysis pipeline was extensively tested and validated on \texttt{AbacusSummit} \cite{Maksimova21,Garrison21} mocks in Refs.~\cite{KP5s1-Maus,KP5s2-Maus} and then tested again on Y1 DESI data that was blinded at the catalog level. The final fiducial settings and scale cuts for the 3D clustering analysis, as presented in Ref.~\cite{DESI2024.V.KP5}, are the exact settings we use in this paper with the exception that \cite{DESI2024.V.KP5} combined full-shape with BAO $\alpha_{\parallel,\perp}$ rather than combining with the post-reconstruction correlation function as we do here. Likewise, Refs.~\cite{Sailer24,Kim2024} blinded the cosmological parameters and cross-correlations with ACT DR6 when testing the pipeline for fitting angular spectra and then froze their analysis choices before testing again on mocks. In this paper, the analysis of 2D spectra is based on the pipeline from Refs.~\cite{Sailer2025} and \cite{Sailer24} and we use the same settings and priors as those chosen from their blinded tests. 

\section{Likelihood and Estimators}
\label{sec: likelihood}

In this analysis, we construct a joint likelihood with independent contributions from (1) 3D clustering in the DESI Y1 spectroscopic BGS, LRG, ELG, and QSO samples (pre-reconstruction power spectra and post-reconstruction correlation functions); (2) 2D angular spectra from cross-correlations of the DESI Y1 BGS and LRG samples with the CMB lensing convergence ($\kappa$) maps; and (3) 2D angular spectra from galaxy auto correlations using the DESI photometric BGS and LRG samples as well as their cross-correlations with CMB lensing. 
On the 3D clustering side we combine the fullshape power spectrum with the post-reconstruction correlation function into a single data vector for each bin and take into account their cross-correlation in the covariance matrix. The reconstruction procedure aims to undo nonlinear damping of the BAO feature and sharpens the peak in order to obtain tighter constraints on $\Omega_{\rm m}$ and $H_0$ which enter the distance-redshift relation \cite{Eisenstein_recon2007,Padmanabhan_recon2009,Noh_recon2009,White_recon2015,Chen_recon2019,KP4s2-Chen}. When combining with the angular spectra the better measurement of $\Omega_{\rm m}$ from BAO will also improve $\sigma_8$ constraints by helping break the degeneracy between the two parameters \cite{Chen_BOSSrecon2022,Sailer24}.
The joint (log) likelihood can be written as:
\begin{align}
    \log \mathcal{L}_{\rm tot} &= \log \mathcal{L}^{3D}_{\rm spec-z} + \log \mathcal{L}^{2D}_{\rm spec-z}+ \log \mathcal{L}^{2D}_{\rm photo-z} \\
    &\log \mathcal{L}^{3D}_{\rm spec-z} = \sum_{i=1}^6\log \mathcal{L}\left[ P^{\rm pre}_{\ell}(k),\xi^{\rm post}_{\ell}(s) \right]_{i} \label{eqn:logL_3dspec} \\
    &\log \mathcal{L}^{2D}_{\rm spec-z} = \log \mathcal{L}\left[\left( C^{\kappa g}_{\ell}\right)_{\mathrm{sBGS}},\left( C^{\kappa g}_{\ell}\right)_{\mathrm{sLRG}1-3}\right] \label{eqn:logL_2Dspec} \\
    &\log \mathcal{L}^{2D}_{\rm photo-z} = \log \mathcal{L}\left[ \left( C^{g g}_{\ell},C^{\kappa g}_{\ell}\right)_{\mathrm{pBGS}1,2}, \left( C^{g g}_{\ell},C^{\kappa g}_{\ell}\right)_{\mathrm{pLRG}1-4}\right] \label{eqn:logL_2D_phot}
\end{align}
where the sum in Eq.~\ref{eqn:logL_3dspec} includes the three spectroscopic LRG bins as well as the BGS, ELG, and QSO tracers. The likelihood in Eq.~\ref{eqn:logL_2Dspec} involves spectroscopic BGS (``sBGS'') and three LRG (``sLRG'') bins, while the likelihood Eq.~\ref{eqn:logL_2D_phot} encompasses the contributions from the two BGS (``pBGS'') and four LRG (``pLRG'') samples from the imaging catalogs. We treat the $\log \mathcal{L}_{\rm spec-z}^{3D}$ and $\log \mathcal{L}_{\rm spec-z}^{2D}$ likelihoods as uncorrelated because the 3D and 2D spectra are mostly sensitive to different line-of-sight Fourier modes (i.e.\ $k_{\parallel}$), except on large angular scales with low information content (see e.g.\ ref.~\cite{Taylor22} for an analysis of the covariance between 3D and 2D spectra). We can neglect the covariance between the power spectra in the spectroscopic versus photometric bins by applying a mask to the photometric catalog/maps to remove the overlapping regions of the Y1 and imaging footprints. 

The measurements of these 3D ($P_{\ell}$ and $\xi_{\ell}$) and 2D ($C_{\ell}$) spectra (and their window matrices and covariances) are described in the following subsections, along with a discussion of relevant weights and systematic corrections and the handling of fiber collisions.

\subsection{Spectra, windows, and covariance}

\begin{figure}
\centering
\resizebox{0.99\columnwidth}{!}{\includegraphics{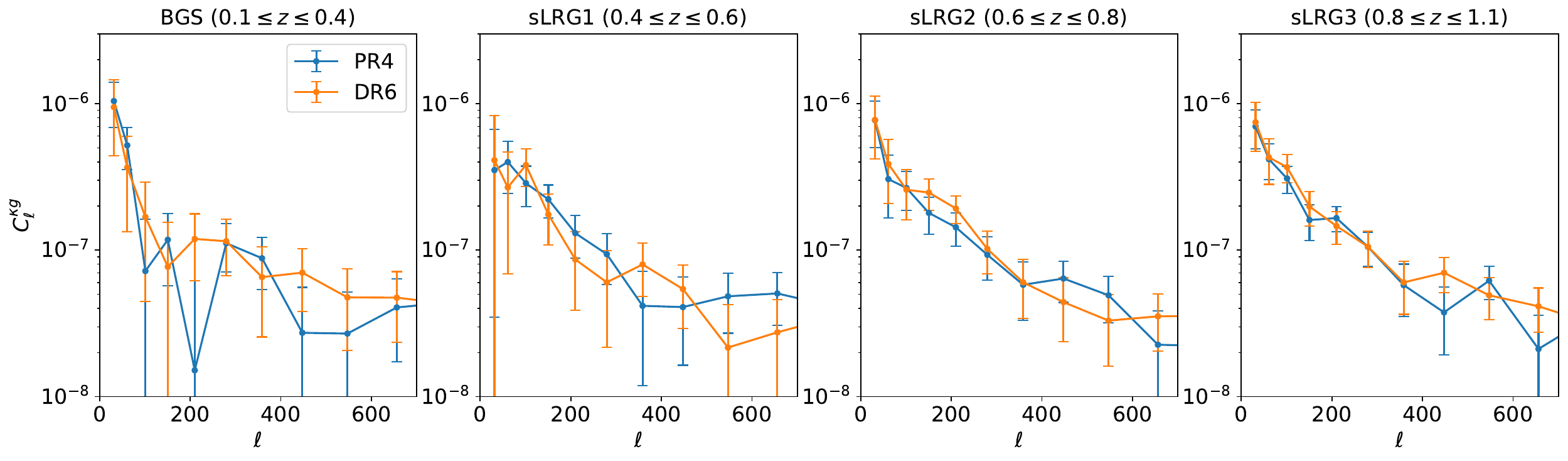}}
\resizebox{0.99\columnwidth}{!}{\includegraphics{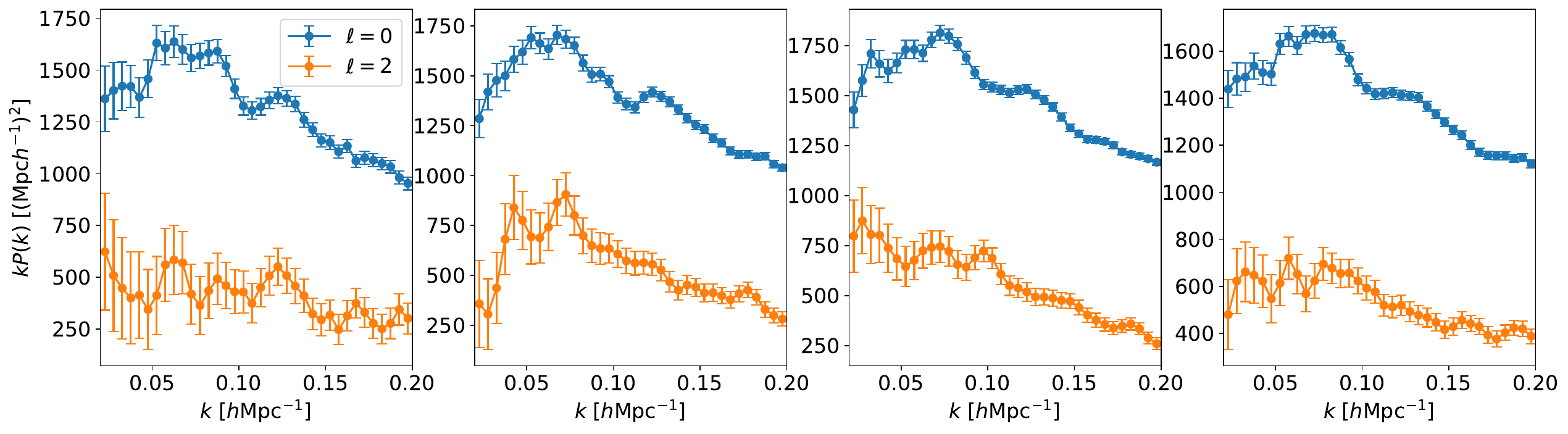}}
\resizebox{0.99\columnwidth}{!}{\includegraphics{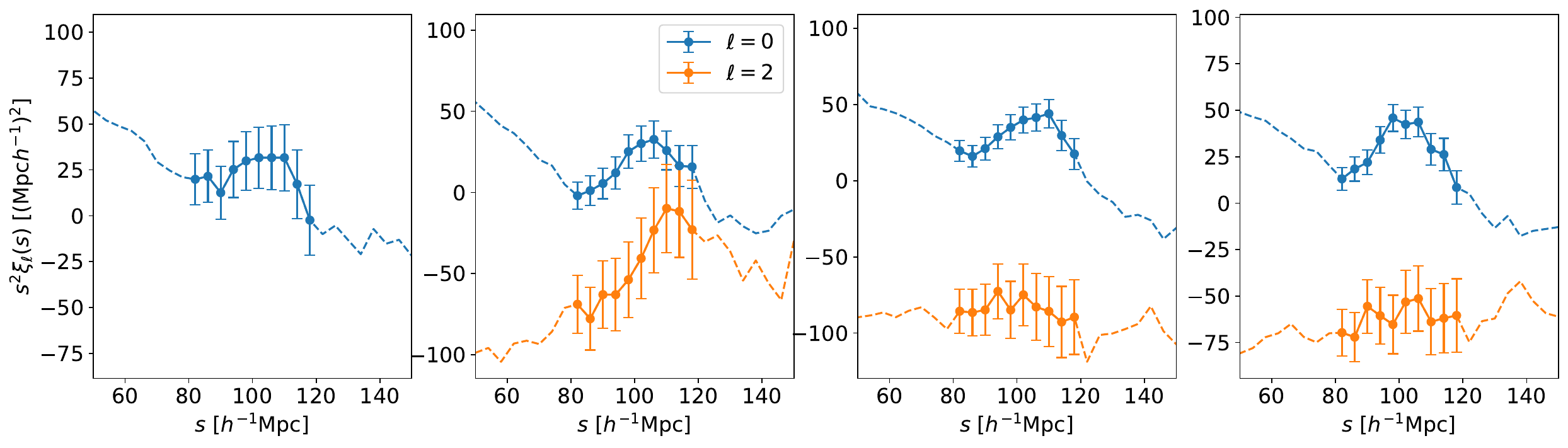}}
\caption{We show in the top row the pseudo-$\Ckg$ spectra computed from cross correlations between the DESI Y1 spectroscopic BGS and LRG galaxies and \textit{Planck} PR4 and ACT DR6 CMB lensing maps. The middle and bottom rows show the 3D power spectrum multipoles and post-reconstruction correlation function multipoles, respectively, from the same BGS and LRG samples. For the post-reconstruction correlation function multipoles we show the data used for fitting as points with errorbars in the fit range $80\leq s\leq 120$ $h$Mpc$^{-1}$ while the dashed lines show the continuation of these curves outside of the fitting range. We use only the monopole and quadrupole moments, consistent with the fiducial choice of the DESI Y1 analysis\cite{DESI2024.V.KP5} that was based on findings in Refs.~\cite{KP5s2-Maus,KP5s3-Noriega,KP5s4-Lai} that the hexadecapole contributed negligibly to $\Lambda$CDM constraints. 
\label{fig: LRG_spectra}}
\end{figure}

The 3D power spectra and correlation functions are computed after converting observed angles and redshifts into physical coordinates by assuming a fiducial cosmology based on a best-fitting \textit{Planck} 2018 flat-$\Lambda$CDM model \cite{PlanckParams18,DESI2024.II.KP3}. We use the 3D clustering estimators produced by DESI for the year-1 (Y1) full-shape \cite{DESI2024.V.KP5,DESI2024.VII.KP7B} and BAO \cite{DESI2024.III.KP4,DESI2024.VI.KP7A} key projects that are now publicly available as part of the DR1 data release \cite{DESI2024.I.DR1}. The power spectrum is based on the Feldman-Kaiser-Peacock (FKP) field \cite{FKP1994} and estimated using the method described in \cite{Hand2017}.  The measurement of this (along with the associated window matrices) is implemented in the \texttt{pypower}\footnote{\url{https://github.com/cosmodesi/pypower}} code. The (reconstructed) correlation function is based on the Landy-Szalay estimator \cite{Landy1993} and is measured from the catalogs using the \texttt{PYCORR}\footnote{\url{https://github.com/cosmodesi/pycorr}} package. BAO reconstruction is performed by shifting data and randoms in order to partially undo the effect of bulk-flows that shift and damp the BAO signal \cite{Padmanabhan_2012}. Further details on the spectroscopic catalogs and two-point clustering estimators are given in \cite{DESI2024.II.KP3}. 

The $P_{\ell}(k)$-auto, $\xi_{\ell}^{\rm post}(s)$-auto, and cross covariance contributions are computed using a suite of 1000 `EZmocks' \cite{Chuang_2014}. These computationally cheap simulations use the Zel'dovich approximation to evolve the dark matter density field before populating halos with galaxies based on an assumed bias model. For the BGS galaxies we use mocks created in boxes with side length of $2\,h^{-1}$Gpc, while the LRG, ELG, and QSO mocks were created in boxes with side lengths of $6\,h^{-1}$Gpc. For each tracer the 1000 EZmock simulations are used to compute the pre-reconstruction power spectrum and post-reconstruction correlation function multipoles and compute the joint covariance matrix numerically. We assume no covariance between the different tracers (and between redshift bins within the LRG sample), which allows us to simply add the $\log \mathcal{L}^{3D}$ contributions from the spectroscopic bins. 

The accuracy of these mock-based covariances was tested against analytical counterparts (\textsc{RascalC} \cite{KP4s7-Rashkovetskyi} for configuration space and \textsc{TheCov} \cite{KP4s8-Alves} in Fourier space) with fits to Abacus mocks with and without fiber-assignment \cite{KP4s6-Forero-Sanchez}. In those papers it was found that the mock-based covariance underestimated the errors by $\approx10\%$.  To correct for this a rescaling factor, described in \S4 of ref.~\cite{KP4s6-Forero-Sanchez}, was introduced. Following the DR1 official fullshape analysis we use this corrected covariance in this paper but note that purely analytical or more accurate mock based covariances are desirable in the future. While analytical covariances exist for the pre-reconstruction power spectrum and post-reconstruction correlation function separately, there currently does not exist an analytic joint covariance and thus we are limited to numerical covariances based on mocks for the time being. Finally, we include rescaling factors \cite{Hartlap07,Percival2014} to account for the finite number ($N=1000$) EZmocks, though these are smaller corrections than the rescaling factors described above \cite{DESI2024.V.KP5}.

For computing the angular galaxy-galaxy auto and galaxy-lensing cross spectra we use the ``Direct-SHT'' method developed by ref.~\cite{Baleato24}\footnote{\url{https://github.com/martinjameswhite/directsht}} and implemented in $\texttt{NaMaster}$\footnote{\url{https://github.com/LSSTDESC/NaMaster}}~\cite{Alonso18} via the \texttt{NmtFieldCatalogClustering} routine \cite{Wolz_2025}. This approach is close in spirit to the FKP method discussed above, and is better suited to cross-correlation with galaxy samples than earlier methods (e.g.\  \cite{Kitanidis21,White22,Sailer24}) that created a pixelized map from the galaxy catalogs. This catalog-based method avoids aliasing, sub-pixel effects, etc.\ that can arise from working with pixelized maps of discrete objects in Fourier space (see e.g.\  refs.~\cite{Hall_2025,Euclid2025_LIX} for a recent exploration of these issues).  We also keep the small variations in the mask directly from the random catalogs, rather than imposing a binary mask.  

While this pixel-free estimator is expected to become standard for future analyses involving cross-correlations with fields from discrete objects such as galaxies, routines for computing analytic covariances from catalog-based fields are still in development. We therefore still create pixelized galaxy masks (and overdensity maps-though these are technically not needed for analytic covariances) in order to use the standard $\texttt{gaussian\_covariance}$ method in NaMaster to compute analytic gaussian covariances \cite{Garc_a_Garc_a_2019}. We cannot ignore correlations between $\Ckg$'s in different LRG redshift bins so the $i\neq j$ terms in cov$[C_{\ell}^{\kappa gi}C_{\ell}^{\kappa gj}]$ are computed as well.
Following the method described at the end of \S3 of \cite{Sailer24}, we obtain the fiducial spectra for $C_{\ell}^{\kappa gi}$ and $C_{\ell}^{gi gj}$ needed to compute cov$[C_{\ell}^{\kappa gi}C_{\ell}^{\kappa gj}]$ by simply fitting theoretical models to $C_{\ell}^{\kappa gi}$ and $C_{\ell}^{gi gi}$. This is done by minimizing $\sum_{\ell}(C_\ell - C_\ell^{\rm model})^2/C_\ell^2$, where $C_\ell^{\rm model}$ is either a theoretical $\Ckg$ or $\Cgg$ vector convolved with a window function. For the fiducial $C_{\ell}^{gi gj}(i\neq j)$ spectra we linearly interpolate nuisance parameters obtained from the $(i=j)$ auto-spectrum fits in order to approximate their redshift dependence. Lastly, when including both the Planck PR4 and and ACT DR6 lensing data, we take into account their cross-correlations in the covariance.

Following Refs.~\cite{Sailer24,Sailer2025} we bin the $C_{\ell}$ measurements into bandpowers equally spaced in $\sqrt{\ell}$ between $10\leq \ell \leq 6143$ with edges:

[10, 20, 44, 79, 124, 178, 243, 317, 401, 495, 600, 713, 837, 971, 1132, 1305, 1491, 1689, 1899, 2122, 2357, 2605, 2865, 3137, 3422, 3719, 4028, 4350, 4684, 5030, 5389, 5760, 6143].

\noindent For each measurement of $C_{\ell}$ we also compute a window matrix using the 

\noindent$\texttt{get\_bandpower\_windows}$ routine in $\texttt{NaMaster}$. Theory predictions are then convolved with these window functions to produce models that can be compared to the observed data bandpowers. We discuss scale cuts used for fitting theory to data in section \S\ref{sec: priors}.

To summarize, the data vectors used for analysis in this paper include (1) the 3D (pre-reconstruction) galaxy power spectrum and reconstructed correlation funtion multipoles ($P_{\ell}(k)$, $\xi_{\ell}(s)$; $\ell=0,2$) computed from the spectroscopic BGS, LRG (three bins), ELG, and QSO samples with covariances computed numerically from EZmocks; (2) angular galaxy-lensing cross spectra ($\Ckg$) between the spectroscopic BGS and LRG samples and both the \textit{Planck} PR4 and ACT DR6 lensing maps with analytic gaussian covariances; and (3) angular galaxy-galaxy and galaxy-lensing spectra between the photometric BGS and LRG samples and the PR4 and DR6 lensing maps with analytic gaussian covariances. We show the 3D $P_{\ell}(k)$, $\xi_{\ell}(s)$, and angular $\Ckg$ spectra computed from the spectroscopic BGS and LRG samples in Fig.~\ref{fig: LRG_spectra} and refer readers to Appendix~\ref{app:alt_spectra} for figures showing the other Y1 tracers and spectra computed from the photometric LRG samples. 

We do not include the quadrupole moments of the correlation functions ($\xi_2(s)$) for the BGS and QSO tracers as it was determined in the unblinding tests for the Y1 BAO analysis \cite{DESI2024.III.KP4} that the quadrupole moments for these tracers were too noisy to contribute meaningfully to the constraints and therefore only 1D fits to the monopole were performed. Following this decisions made in the Y1 DESI analyses, we exclude the BGS and QSO quadrupole moments from $\xi_{\ell}^{\rm post}(s)$ in our $P_{\ell}(k)+\xi_{\ell}^{\rm post}(s)$ data vectors.



\begin{figure}[htb]
    \centering
    \includegraphics[width=\linewidth]{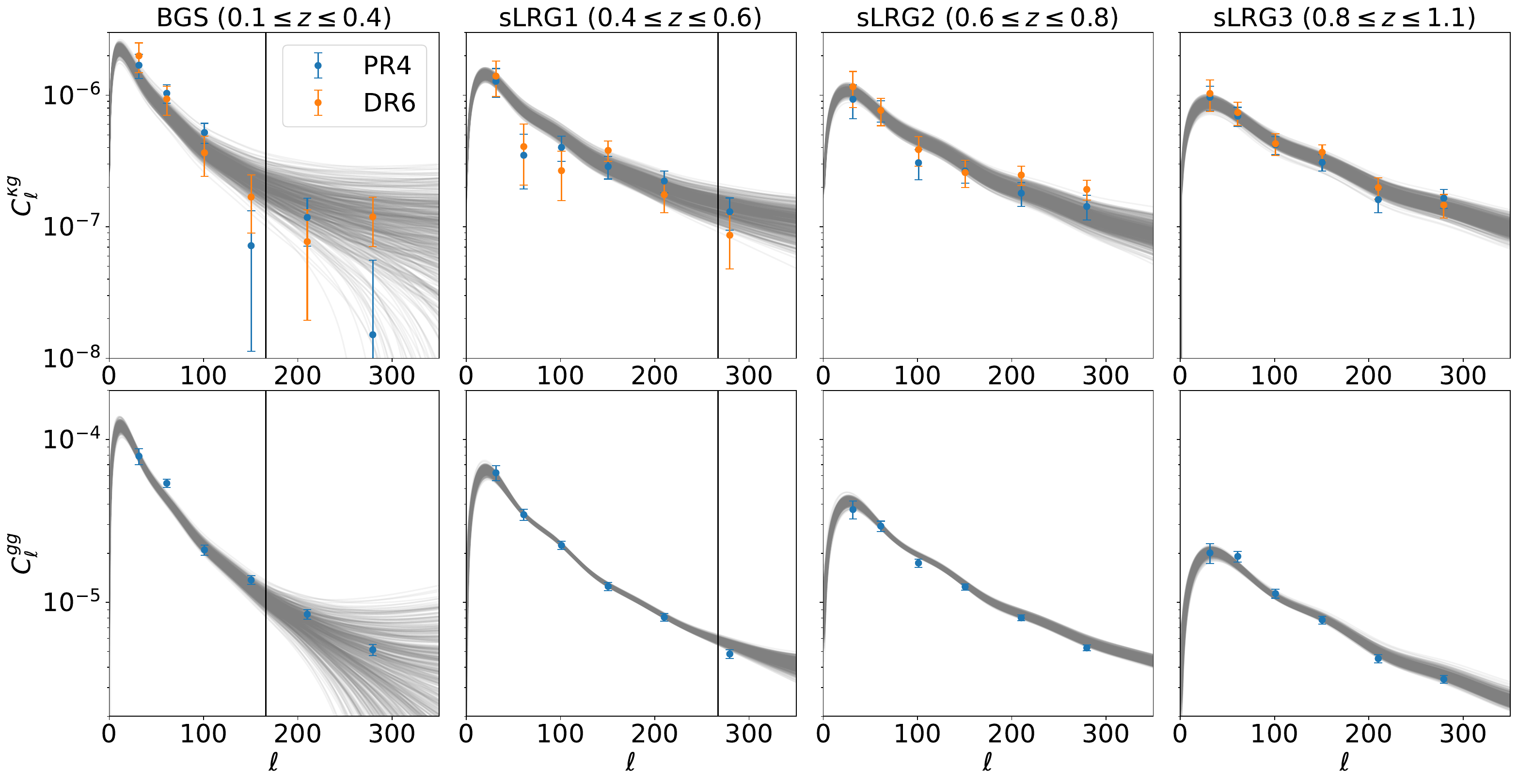}
    \caption{Posterior Predictive Distributions for $\Ckg$ and $\Cgg$ in the spectroscopic bins, using samples from the 3D full-shape + post-recon BAO fit. 500 samples are randomly drawn from the 1$\sigma$ region of the 3D analysis chains and used to predict corresponding $\Ckg$ and $\Cgg$ curves. The data points in the top row show the cross-correlations of DESI galaxies with Planck PR4 (blue) and ACT DR6 (orange) lensing, while the data in the bottom row are the angular auto-spectra measured from the DESI galaxies. In the first two columns we also show a vertical line to denote the $\ell = k\times\chi(z)-1/2$ corresponding to the highest wavelength, $k_{\rm max}=0.2$ $\ihmpc$, used in the full-shape fits. We do not expect the 3D clustering to provide information for $C_{\ell}$ at higher bandpowers. 
    }
    \label{fig:desi_ppd}
\end{figure}

In Fig.~\ref{fig:desi_ppd} we show the posterior predictive distribution (PPD) of $\Ckg$ conditioned on samples from a 3D full-shape + post-reconstruction BAO fit. There are two parameters in $\Ckg$, ie. $\alpha_x$ and $s_{\mu}$ (defined in \S4.4), that do not enter the power spectrum directly. For $\alpha_x$ we relate its value to the $\alpha_0$ parameter of $P_{\ell}(k)$ with a correction sampled from a Gaussian centered at 0 with a width of 2. For the magnification bias we sample from its prior given in Tab.~\ref{tab: priors}. A PPD is useful both for seeing what information a new dataset will contribute as well as a consistency check of our models and data. We find that in the low-$\ell$ (large scale) regime several of the sLRG1 and sLRG2 data points lie below the predicted distribution, which suggests that the $\Ckg$ data may have a preference for lower values of parameters relating to the amplitude of clustering. The opposite is true in the BGS bin whose large-scale cross-correlations appear to prefer higher clustering amplitude than the 3D analysis predicts. This is in line with the results in the tomographic analysis using photometric BGS and LRG galaxies in ref.~\cite{Sailer2025} where the authors found a preference for higher $S_8$ from the BGS sample and lower $S_8$ from LRG's. We expect cosmological constraints to be driven more by the large scale data while deviations in clustering at smaller-scales can often be absorbed when marginalizing over nuisance parameters. In addition to the cross-correlations, we also show in the bottom row of Fig.~\ref{fig:desi_ppd} the PPD for the galaxy autocorrelation $\Cgg$. We find that the information content from 3D clustering predicts a narrow distribution of $\Cgg$ models that are consistent with the data.

\subsection{Weights, fiber assignment and systematic corrections}
\label{sec: weights}

In this section we discuss a few of the more technical details regarding weights, fiber collisions, imaging systematics, integral constraints and the Monte-Carlo norm correction. A summary can be found in Tab.~\ref{tab: summ wts}. To begin with, the power spectrum and correlation function estimators make use of both the weighted galaxy catalog as well as a catalog of weighted randoms that define the footprint. The weights used for these estimators are a product of the Feldman‑Kaiser‑Peacock (FKP; \cite{FKP1994}) weights and those from systematic corrections. The weights are applied to galaxies (and randoms) to account for spatial variations in the survey selection function and number density that would bias the power spectrum estimator and to approximate an optimal weighting for reduced variance. The computation of these weights in DESI DR1 is described in \S8 of ref.~\cite{DESI2024.II.KP3}. 

One of the contributions to the systematic weights is due to fiber assignment incompleteness, which is the effect of missing observations in a spectroscopic survey due to instrumental limitations such as the minimum separation of optical fibers in the focal plane. One possible method for mitigating fiber assignment incompleteness is via inverse probability weighting through methods described extensively in refs.~\cite{Bianchi_2017,Bianchi20}. For DESI DR1 the fiducial spectra were produced with each galaxy and random getting assigned a completeness weight of $1/f_{\rm TLID}$, where $f_{\rm TLID}$ is related to the number of competing DESI targets for a given fiber (see \S 5 of ref.~\cite{DESI2024.II.KP3} for more details).\footnote{A similar but more robust version of this is the individual inverse probability weighting (IIP) scheme, in which a fiber assignment code is run many times with changing random seed and then each galaxy is weighted by the inverse of the resulting probability of assignment ($N_{\rm assign}/N_{\rm realizations}$). However, the computational cost of running the fiber assign code many times prevented this method from being used for DESI Y1. That being said, both methods involve weighting each galaxy to give a modified density field for which the large-scale impact of fiber assignment on $\delta_g$ is removed. \label{fn: IIP}} For cross-correlations with lensing ($\Ckg$) these completeness weights are adequate for correcting for fiber-assigment, but not for 3D galaxy-galaxy spectra. Since the inverse weights do not account for all fiber assignment incompleteness on small scales, unbiased two-point statistics are obtained by modifying our summary statistic (i.e.\ the multipoles of the power spectrum) to be explicitly insensitive to small scales. For Y1 this was done by applying a ``$\theta$-cut'' \cite{KP3s5-Pinon} to the clustering by removing galaxy pairs with angular separations of $\leq 0.05^{\circ}$ (see also \S5.3 of ref.~\cite{DESI2024.II.KP3}). This modified statistic sufficiently suppresses the effect of missing galaxy pairs in the power spectrum to within 1/5 of DR1 precision \cite{KP3s5-Pinon}. As discussed in \S3 of ref.~\cite{KP3s5-Pinon}, the modification of the power spectrum to remove close pairs must also be included in the theoretical model, which is done by modifying the survey window function to remove the geometrical effect of close pairs. 

In this paper we follow the fiducial DESI Y1 method for the 3D $P_{\ell}(k)$ measurements, namely by individually weighting galaxies by $1/f_{\rm TLID}$ and applying the $\theta$-cut as described above. However, we note that in the future there may be a preference towards applying pairwise inverse probability weights (PIP) in which pairs of galaxies are weighted based on their probability of being observed \cite{Bianchi_2017}. This approach does not actually modify the galaxy density itself but just corrects pairs entering the 2-point function estimator, and thus doesn't really have any meaning for e.g. $\Ckg$. These PIP weights can be obtained using the IIP(see Footnote \ref{fn: IIP}) weights and alternate fiber assign realizations. So long as there are few/no pairs with 0 probability, PIP weighting results in an unbiased estimator for two-point functions; however, this was not the case for Y1 as much of the footprint area was only covered by one or two tiles (this is discussed at the end of \S5.3 in \cite{DESI2024.II.KP3}). The use of PIP weights will be revisited for the Y3 or Y5 3D clustering analyses, provided that a significantly faster fiber assignment code is available, as it will also allow for fitting to smaller scales. However, we also stress that one of the primary advantages in jointly fitting galaxy-galaxy and galaxy-lensing measurements is the breaking of degeneracies between large-scale bias and the clustering amplitude. If inconsistent weights are applied between $P_\ell$ and $C_\ell^{\kappa g}$ then the two galaxy fields are inherently different and don't share the same set of biases.

Returning to the $\theta$-cut technique, accounting for the cut in the window function results in a matrix that is highly non-diagonal by mixing large- and small- scale modes such that predictions for observed $P_{\ell}(k\leq 0.2)$ require computing the theory up to very high $k$. While biases in $P_{\ell}(k\leq 0.2)$ incurred from truncating the input theory are absorbed by the nuisance parameters in the model without any effect on cosmological constraints, ref.~\cite{KP3s5-Pinon} (see \S5) developed a method that transforms the power spectrum, window, and covariance in such a way that recovers a more diagonal window function. In order to account for the distortion in the power spectrum due to this rotation we marginalize over a set of templates (one for each multipole) with linear coefficients $s_{\ell}$. 


The overdensity of galaxies is inferred via the difference in the observed distribution of galaxies and the expected distribution in the absence of clustering, characterized by a survey selection function. If this selection function were perfect then the measured density contrast and summary statistics computed from it would directly describe the clustering of galaxies. However, the selection function itself is usually derived from the data itself, and given the finite volume of a survey, this can lead to a mis-estimation of the mean density field and introduces a bias in the measured clustering along the line of sight (the ``radial integral constraint'', or RIC \cite{deMattia2019}). While this is a small effect ($\lesssim0.15\%$ of DR1 measurement uncertainties for $k\leq 0.05$ $\ihmpc$ ) we account for it by subtracting a template correction from the data power spectrum. This template is produced (see \S10.1.2 of ref.~\cite{DESI2024.II.KP3}) by measuring power spectra from 50 EZmock realizations with and without the RIC effect (by changing how redshifts of randoms are chosen). The difference in mean $P(k)$ with and without RIC is fit with a polynomial to get the template that approximates the RIC effect in our data vector. 

In addition, in order to account for the effect of observation/imaging conditions on the target selection, which then cause variations in number density and measured clustering, we apply imaging systematic weights to the data. These weights are derived from a linear regression method as was done for the BOSS and eBOSS LSS catalogs \cite{Reid16,Ross_2020}. As discussed in \S10.1.2 of ref.~\cite{DESI2024.II.KP3}, it was found that for the ELG and QSO samples different choices of regression methods (linear vs.\ random forest \cite{Chaussidon_2021} vs.\ neural net \cite{Rezaie_2020}) resulted in significant variations in large scale power. We mitigate this dependence on imaging weights by introducing a template correction in the power spectrum model with a free parameter, $s_{\rm ph}$ that we marginalize over. In addition, we correct for the removal of large-scale angular modes by regression methods with the inclusion of a second polynomial template correction. Eq.~5.1 in \S~5.5  of ref.~\cite{DESI2024.V.KP5} shows explicitly how these two template corrections are included in the model, but in practice the constant ``mode-removal'' correction is just subtracted from the data vector rather than additively included in the model in our pipeline. \S10.1.2 of ref.~\cite{DESI2024.II.KP3} describes how the template corrections are measured from mocks. After implementing these RIC and imaging systematic corrections, the residual systematic errors at the parameter level were found to be $\lesssim 0.2\times$ DR1 statistical errors \cite{DESI2024.V.KP5}. Fig.~13 of ref.~\cite{DESI2024.II.KP3} demonstrates how well these polynomial templates fit the RIC and imaging weight effects in the Power Spectrum but also show that these effects are subdominant ($\lesssim0.2\%$) compared to the DR1 measurement uncertainties for $k\leq 0.05$ $\ihmpc$). 

Regarding the cross-correlations with CMB lensing, there are only a few subtleties. As stated previously, we employ the new ``DirectSHT'' \cite{Baleato24} method for computing the angular power spectrum directly from catalogs in a way that resembles the FKP field used for measuring the 3D power spectrum. In addition, we follow refs.~\cite{Sailer24, ACT:2023oei, Farren:2024rla} by including a Monte Carlo normalization correction to $C_{\ell}^{\kappa g}$ to correct for mode-couplings from masking/filtering in the CMB maps that result in a mis-normalization in the lensing $\kappa_{\ell m}$ data. This normalization correction is computed from a large set of simulated CMB-lensing reconstructions (see eq.~C.1 of ref.~\cite{Sailer24}) constructed for each redshift bin and then multiplied mode-by-mode to each bandpower of the measured $\Ckg$ data. These norm corrections are $\lesssim 50\%$ of the $\Ckg$ errors $\ell$'s within the fit ranges in the spectroscopic bins and $\lesssim30\%$ in the photometric samples (after the DR1 footprint has been removed).

\begin{table}
    \centering
    \begin{tabular}{|c|c|c|}
        Systematic & Mitigation method & Reference(s)  \\
        \hline\hline
        Spatial variations in $\bar{n}$ & FKP weights & \cite{FKP1994},\cite{DESI2024.II.KP3}(\S8)\\
        Fiber Assignment & Comp. weights and $\theta$-cut & \cite{Bianchi_2017,Bianchi20,KP3s5-Pinon}\\
        Radial Integral Constraint & Template correction & \cite{deMattia2019},\cite{DESI2024.II.KP3}(\S10.1.2)\\
        Imaging systematics & Im. weights, mode-removal temp. & \cite{Reid16,Ross_2020,Chaussidon_2021,Rezaie_2020},\cite{DESI2024.II.KP3}(\S10.1.2)\\
        \hline
        Pixelization effects & DirectSHT & \cite{Baleato24}\\
        Mis-normalization of $\kappa$-maps & MC norm correction & \cite{Sailer24, ACT:2023oei, Farren:2024rla}
    \end{tabular}
    \caption{Summary of systematic effects in the data and how they are mitigated, with references to the methods papers. The first four lines are relevant in the 3D power spectrum while the last two are for the angular spectra (and the norm correction is only relevant in $\Ckg$). }
    \label{tab: summ wts}
\end{table}

\section{Theory and models}
\label{sec: thy}

In the following subsections we briefly describe the theoretical models that we use in our analysis for the 3D clustering and angular spectra. These models have undergone significant validation in previous works and we refer readers to refs.\ \cite{KP5s1-Maus,KP5s2-Maus,Sailer24} for more detailed descriptions of the theories and validation tests on mocks. Both the 3D full-shape and the angular clustering models are based on the Lagrangian framework for describing the evolution of galaxy overdensities and their relationship with the underlying matter field, as we describe in the first subsection. For all spectra we employ emulators that mimic the true theoretical predictions but significantly reduce computational costs. On the 3D clustering side we train emulators based on Taylor series expansions of the model with respect to cosmological parameters, as has been described in refs.\ \cite{Chen22,KP5s2-Maus}. Predictions of real space spectra that enter the angular $\Ckg$ and $\Cgg$ models are emulated with neural networks. 

\subsection{Lagrangian Perturbation Theory}

The Lagrangian framework for cosmological perturbation theory within the context of large scale structure is a theoretical approach to model the evolution of cosmic matter density fields under gravitational interactions, emphasizing the trajectories of mass elements from initial conditions to their positions at later times. In this framework, we track particles from their initial (Lagrangian) positions, $\bq$, to their advected (Eulerian) positions, $\bx$, via the displacement field $\bx = \bq + \bPsi(\bq, t)$.
Mass conservation allows us to relate the galaxy overdensity at late times to the initial proto-halo distribution by \cite{Mat08a,Mat08b}
\begin{align}
    1 + \delta_g(\bx) = \int d^3\bq\ F[\delta_0(\bq)] \, \delta_D(\bx - \bq - \bPsi(\bq))
\end{align}
where $\delta_D(\cdots)$ is the Dirac delta function and the functional $F[\delta_0(\bq)]$ connects the galaxy fluctuations to the  initial linear matter distribution, $\delta_0(\bq)$, with a perturbative expansion of operators allowed by symmetry \cite{Cat00,Ang15,Des18,ChenCasWhi19,FujitaVlah20}:
\begin{align}
    F[\delta_0(\bq)] = 1 +  b_1\delta_0 + \frac{1}{2}b_2(\delta_0(\bq)^2 - \left\langle\delta_0^2\right\rangle)+b_s(s_0^2(\bq) - \left\langle s_0^2\right\rangle) + b_3 \mathcal{O}_3(\bq) + \cdots
    \label{eq: Fq}
\end{align}
where $b_1$, $b_2$, and $b_s$ are the linear, quadratic, and shear bias parameters, respectively. The $b_3$ term encompasses all third order contributions in the bias expansion, most of which are degenerate with each other and therefore can be sufficiently modeled as a single term \cite{Chen20}. In Fourier space the galaxy overdensity becomes
\begin{align}
    (2\pi)^3 \delta_D(\bk)+\delta_g(\bk) = \int d^3\bq\ F[\delta_0(\bq)]\, e^{-i\bk \cdot(\bq + \bPsi(\bq))}
\end{align}

\subsection{RSD power spectrum}

An advantage of having spectroscopic data is the ability to use redshifts of tracers in addition to their angular coordinates to compute the 3D power spectrum, once a fiducial cosmology has been assumed to convert the angles and redshifts into Cartesian coordinates. There are two caveats that must be taken into account: (1) the conversion of galaxy redshifts and angles into physical distances when estimating $P(\bk)$ requires the assumption of a reference cosmology. Deviations between this fiducial and the true cosmology introduces geometrical distortions that must be accounted for when comparing the theoretical power spectra to those measured from the data (the ``Alcock-Paczyński'' (AP) effect \cite{Alcock79}); (2) Distortions/anisotropy are introduced due to line-of-sight peculiar velocities affecting the radial distances inferred from redshifts (``Redshift-space Distortions (RSD)''). We handle the first of these by rescaling the theoretical power spectrum and coordinates by scaling factors\footnote{Note that in the literature surrounding 3D clustering analyses one often encounters similarly defined parameters, $\alpha_{\parallel}$ and $\alpha_{\perp}$, that differ from $q_{\parallel,\perp}$ by factors of $r_d^{\rm fid}/r_d$, where $r_d$ refers to the sound horizon scale at the drag epoch. The ``standard BAO analyses'' assume a fiducial template linear power spectrum and then vary $\alpha_{\parallel,\perp}$ to simulate how changes in sound horizon and $q_{\parallel,\perp}$ affect the BAO signal. Cosmological constraints are then obtained by interpreting the $\alpha_{\parallel}$ and $\alpha_{\perp}$ constraints under the assumption of a specific cosmological model. In our analysis we use the ``direct fitting'' technique in which the cosmology and linear power spectrum are directly and consistently varied and thus the effect of the changing sound horizon scale is already accounted for. We refer readers to \S4 of \cite{KP5s2-Maus} and Appendix~C of \cite{Chen22} for further discussion on this matter and the advantages of our method.}
\begin{align}
    q_\parallel = \frac{H^{\rm ref}(z)}{H(z)}
    \quad , \quad
    q_\perp = \frac{D_A(z)}{D^{\rm ref}_A(z)} \quad.
    \label{eq: AP_geo}
\end{align}
RSD is handled by a boost of the displacements in the LOS direction:
\begin{align}
    \bPsi_s = \bPsi + \dot{\bPsi} = \bPsi + \frac{\hat{n}(\textbf{v}\cdot\hat{n})}{\mathcal{H}}.
\end{align}
The redshift-space galaxy-galaxy power spectrum is then given by 
\begin{align}
    P_{s,g}(\bk) = \int d^3\bq \left\langle e^{i\bk \cdot (\bq + \Delta_s)}F(\bq_1)F(\bq_2) \right\rangle_{\bq = \bq_1-\bq_2}.
    \label{eq: VPint} 
\end{align}
where $\Delta_s = \Psi_s(\bq_1) - \Psi_s(\bq_2)$. The full expression for the \texttt{velocileptors} LPT model, including counterterms and stochastic terms that control the sensitivity to small-scale (galaxy/halo formation) physics, is presented in ref.~\cite{KP5s2-Maus}. The combination of full-shape and post-reconstruction BAO modeling is also described in \S5.4 of ref.~\cite{KP5s2-Maus}.

\subsection{HEFT and Aemulus $\nu$}
\label{sec: HEFT}

For modeling the angular power spectra we make use of hybrid effective field theory (HEFT; \cite{Modi20,Hadzhiyska:2021xbv, anzu21,Zennaro2022,DeRose_2023,Ibanez2023,Nicola2024}) which aims to capture nonlinear gravitational dynamics and baryonic effects more accurately than standard perturbative techniques (see Appendix~\ref{app: EFT_HEFT} for a comparison of constraints using HEFT vs EFT for the real-space power spectra). This method combines the symmetry-based bias expansion from LPT with nonlinear displacements of particles advected from their initial coordinates using N-body simulations. We make use of a neural network emulator trained on the \texttt{Aemulus $\nu$} \cite{DeRose_2023}\footnote{\url{https://github.com/AemulusProject/aemulus_heft}} suite of simulations to produce the basis spectra for the nonlinear real-space power spectrum model. Eqs 4.4 and 4.5 of ref.~\cite{Sailer24} (and references therein) detail how the galaxy-galaxy and galaxy-matter power spectra are built up out of the emulated basis spectra ($P_{XY}$ where $(X,Y)$ refer to the CDM+baryon field, total matter field, or contributions to the galaxy density contrast from the $(1,2,s)$ terms in $F[(\delta_0(\bq)]$ in Eq.~\ref{eq: Fq}) produced from the \texttt{Aemulus $\nu$} project. Additionally \S4.5 of ref.~\cite{Sailer24} describes how the 2D angular auto and cross spectra are obtained through integration over the $P_{gg},P_{gm},$ and $P_{mm}$ spectra and projection kernels under the Limber approximation \cite{Limber1954,Kaiser1992}. 

\subsection{Parameters, Priors and Scale cuts}
\label{sec: priors}

For our baseline analysis we vary four cosmological parameters ($n_s,H_0,\omega_b,\omega_{cdm},\log(10^{10}A_s)$) with uniform priors on $H_0$, $\omega_{cdm}$ and $\log(10^{10}A_s)$. For the baryon abundance, $\omega_b$ we apply a Gaussian prior informed by Big Bang Nucleosynthesis \cite{Schoeneberg2024BBN}. We apply a Gaussian prior on $n_s$ matching the constraints from \textit{Planck} 2018 \cite{PlanckParams18} and keep the sum of neutrino masses fixed to $\Sigma m_{\nu} = 0.06\,$eV. In Table~\ref{tab: priors} we list the priors on these cosmological parameters as well as all nuisance parameters entering our models that we describe below.

In combining the pre- and post-reconstruction signals in the 3D full-shape +BAO part of our analysis we follow the procedure described in \S~5.4 of \cite{KP5s2-Maus} and repeat some clarifying details below. The post-reconstruction BAO signal is modeled with uniform priors on the linear bias $B_1$ and growth rate $F$, while Gaussian priors are applied to the six broadband parameters, ($M_{0,1}$, $Q_{0,1}$, and $Q^{\rm sp}_{0,1}$). The form of the broadband model is described in detail in \S~5.2 of \cite{KP4s2-Chen} (also summarized in \S~4.3.2 of \cite{DESI2024.III.KP4}). Note that we have used a different naming convention as Refs. \cite{KP4s2-Chen,DESI2024.III.KP4} in order to avoid confusion with other parameters in our setup. Namely we use (``$B_1$'',``$F$'') for parameters named (``$b$'',``$f$'') in \cite{KP4s2-Chen} and (``$b_1$'',``$f$'') in \cite{DESI2024.III.KP4} to highlight that we allow them to vary independently of the linear bias and growth rate appearing in our power spectrum model. Furthermore, \cite{KP4s2-Chen} uses $\tilde{a}_{\ell,(0,1)}$ for the two sets of polynomial broadband terms in the monopole and quadrupole and \cite{DESI2024.III.KP4} names these $b_{\ell,(0,2)}$(the first index refers to the multipole and second to the power on $k$), whereas we call them $M_{0,1}$ and $Q_{0,1}$). We call the spline terms in the quadrupole ($Q^{\rm sp}_0$,$Q^{\rm sp}_1$) whereas \cite{KP4s2-Chen} and \cite{DESI2024.III.KP4} name them ($a_{2,0}$,$a_{2,1}$). Our avoidance of ``$b$'' and ``$a$'' parameter names for the broadband model is to prevent confusion with the galaxy bias $b_i$ and counterterm $\alpha_i$ parameters. Regarding priors the widths of the broadband parameter priors are large, an order of magnitude larger than the typical large scale power of the power spectrum. 

The full-shape power spectrum model includes counterterms with coefficients $\alpha_i$ that are parameterized such that these terms scale with the linear theory multipoles and the priors are chosen so that each counterterm is approximately kept within 50$\%$ of linear theory at $k=0.2\ihmpc$ (this reparameterization is shown in eq 3.6 \cite{KP5s2-Maus} and explained in the following text). For the stochastic SN parameters we choose prior widths that scale with the Poisson shot-noise (the inverse of galaxy number density), satellite fraction, velocity dispersion $\sigma_v$, and typical halo radius. These quantities are dependent on the properties of the galaxy sample, and therefore we list the final prior values in Table \ref{tab:samp_vals}. The $s_{0,2,\rm ph}$ parameters that are involved with rotations of the window function and imaging systematics also have Gaussian priors with widths dependent on the sample and provided in Table \ref{tab:samp_vals}. The marginalization over photometric templates is only necessary for the ELG and QSO samples, so we do not include $s_{\rm ph}$ in the model for the BGS or any LRG samples.

In each spectroscopic redshift bin we rescale the galaxy‑bias parameters by a factor of $\sigma_{8}^{\,n}(z)$. This parametrization follows the way the monopole of the power spectrum depends on the combination of bias and amplitude, making it closer to the actual observable and thereby reducing parameter‑projection effects (see Appendix~B.2 of Ref~\cite{KP5s2-Maus} wherein the choice of $b_n\sigma_8^n(z)$ parameterization is motivated). In these bins we use the same bias parameters for the cross correlations with lensing as for modeling $P_{\ell}(k)$.  Then $\Ckg$ has two additional parameters, a counterterm $\alpha_x$ for which we apply a broad Gaussian prior, and the magnification bias $s_\mu$ whose Gaussian priors are centered on values measured in ref.~\cite{Heydenreich2025} with widths of 0.1.

For fitting the photometric BGS and LRG sample we directly follow the parameterization and prior choices of refs.~\cite{Sailer2025}(BGS) and \cite{Sailer24}(LRG) with uniform priors on $b_1$ and $b_2$ and Gaussian priors on all the rest. As discussed at the end of \S~4.2 of \cite{Sailer24}(see eq. 4.7), the counterterm $\alpha_x$ appearing in $P_{gm}$ can be related to the auto $\alpha_a$ parameter rather than varying independently. Small discrepancies in this relationship are controlled by $\epsilon$ on which we place a tight prior with width of 2. The magnification bias and projected shot noise values on which their priors are centered are given for each sample in Table \ref{tab:samp_vals} and were measured for the LRG sample in ref.~\cite{Zhou23} and \cite{ChenDeRose24} for the BGS sample. 

The ``Analytic Marg.'' column in Table \ref{tab: priors} specifies parameters that enter linearly in the models and that we analytically marginalize over rather than sampling them. Refs.~\cite{KP5s2-Maus,Sailer24} describe the method of analytical marginalization employed and validate it on mocks for the 3D and 2D analyses respectively.  

When fitting our models to data we apply the following scale cuts: For $P_{\ell}(k)$ we follow the fiducial choice of DESI Y1 \cite{DESI2024.V.KP5} and apply a fit range of $0.02\leq k \leq 0.2$ $\ihmpc$. For $\xi_{\ell}^{\rm post}(s)$ we use the range $80\leq s \leq 130$ Mpc$h^{-1}$, which encloses the BAO peak with minimal broadband signal that would be highly correlated with the power spectrum. For the photometric bins we follow refs.~\cite{Sailer24,Sailer2025} by choosing $79\leq\ell\leq243$ and $79\leq\ell\leq401$ for $\Cgg$ in the two BGS bins and $79\leq\ell\leq600$ for all the LRG bins. For photometric $\Ckg$ we use $\ell_{\rm min} = 20$ and $\ell_{\rm min} = 44$ for Planck PR4 and ACT DR6 respectively, while using the same upper bounds as for $\Cgg$. For the spectroscopic $\Ckg$ fits we match the photometric scale cuts for the LRG samples but use $\ell_{\rm max}=400$ for BGS.

\begin{table}
\centering
\begin{tabular}{c|c|c|c}
Parameters & Likelihood & Analytic Marg. &Prior \\
\hline\hline
$n_s$ &  \multirow{4}{*}{All} & \xmark &$\mathcal{N}[0.9649,0.0042]$ \\
$H_0$ & &\xmark  & $\mathcal{U}[55,80]$\\
$\Omega_{b}h^2$ & &\xmark & $\mathcal{N}[0.02218,0.00055]$\\
$\Omega_{cdm}h^2$ & &\xmark & $\mathcal{U}[0.08,0.16]$\\
$\log(10^{10} A_\mathrm{s})$& &\xmark &$\mathcal{U}[2.0,4.0]$\\
\hline\hline
$B_1 (z_i^{s})$ & \multirow{5}{*}{BAO ($\xi_{\ell}^{\rm post}(s)$)} & \xmark &  $\mathcal{U}[0,5.0]$\\
$F (z_i^{s})$ &  & \xmark &  $\mathcal{U}[0,5.0]$\\
$M_{0,1}(z_i^{s})$ & & \cmark & $\mathcal{N}[0,5e5]$ \\
$Q_{0,1}$ & & \cmark & $\mathcal{N}[0,5e5]$ \\
$Q^{\rm sp}_{0,1}(z_i^{s})$ & & \cmark & $\mathcal{N}[0,5e5]$ \\
\hline
$\alpha_0(z_i^{s})$ & \multirow{9}{*}{FS ($P_{\ell}(k)$)} & \cmark & $\mathcal{N}[0,12.5]$\\ 
$\alpha_2(z_i^{s})$ &  & \cmark & $\mathcal{N}[0,12.5]$\\
SN$_0(z_i^{s})$ & & \cmark & $\mathcal{N}[0,\rm Tab.~\ref{tab:samp_vals}]$ \\
SN$_2(z_i^{s})$ & & \cmark & $\mathcal{N}[0,\rm Tab.~\ref{tab:samp_vals}]$ \\
$s_0(z_i^{s})$ & & \cmark & $\mathcal{N}[0,\rm Tab.~\ref{tab:samp_vals}]$ \\
$s_2(z_i^{s})$ & & \cmark & $\mathcal{N}[0,\rm Tab.~\ref{tab:samp_vals}]$ \\
$s_{\rm ph}(z_i^{s})$ & & \cmark & $\mathcal{N}[0,0.2]$ \\
\hline
$(1+b_1)\sigma_8(z_i^{s})$ & \multirow{3}{*}{FS, $\Ckg$(spec-z)}  & \xmark & $\mathcal{U}[0.5,3.0]$\\
$b_2\sigma_8(z_i^{s})^2$ &  & \xmark & $\mathcal{N}[0,5]$\\
$b_s\sigma_8(z_i^{s})^2$ &  & \xmark &$\mathcal{N}[0,5]$\\
\hline
$\alpha_x(z_i^{s})$ & \multirow{2}{*}{$\Ckg$(spec-z)} & \cmark & $\mathcal{N}[0,50]$ \\
$s_{\mu}(z_i^{s})$ &  & \xmark & $\mathcal{N}[\rm Tab.~\ref{tab:samp_vals},0.1]$ \\
\hline
$b_1(z_i^{p})$ & \multirow{6}{*}{$\Cgg$, $\Ckg$(photo-z)}  & \xmark & $\mathcal{U}[-1,2]_{\rm (BGS)};\mathcal{U}[0,3]_{\rm (LRG)}$\\
$b_2(z_i^{p})$ &  & \xmark & $\mathcal{U}[-5,5]$\\
$b_s(z_i^{p})$ &  & \xmark &$\mathcal{N}[0,1]$\\
$\alpha_a(z_i^{p})$ &  & \cmark & $\mathcal{N}[0,50]$ \\
$\epsilon(z_i^{p})$ &  & \cmark & $\mathcal{N}[0,2]$ \\
$s_{\mu}(z_i^{p})$ &  & \xmark & $\mathcal{N}[\rm Tab.~\ref{tab:samp_vals},0.1]$ \\
SN$_{\rm 2D}(z_i^{p})$ & $\Cgg$ (photo-z) & \cmark & $\mathcal{N}[\rm Tab.~\ref{tab:samp_vals},0.3]$ \\
\end{tabular}
\caption{Parameters and priors used for each likelihood component of the analysis. We use the notation $\mathcal{U}$[min,max] for uniform priors and $\mathcal{N}[\mu,\sigma]$ for gaussian priors centered at $\mu$ with standard deviation $\sigma$. 
$z_i^s$ and $z_i^p$ refer to the $i$th spectroscopic or photometric redshift bin. The parameters with \cmark in the Analytic Marg. coloumn are linear parameters are not sampled as they appear linearly in the model and can be analytically marginalized over. For the parameters that reference Tab.~\ref{tab:samp_vals} the typical values are very specific to the sample and therefore the priors are different for each tracer/redshift bin.
}
\label{tab: priors}
\end{table}

\section{Results}
\label{sec: results}

\begin{figure}[htb]
    \centering
    \includegraphics[width=\linewidth]{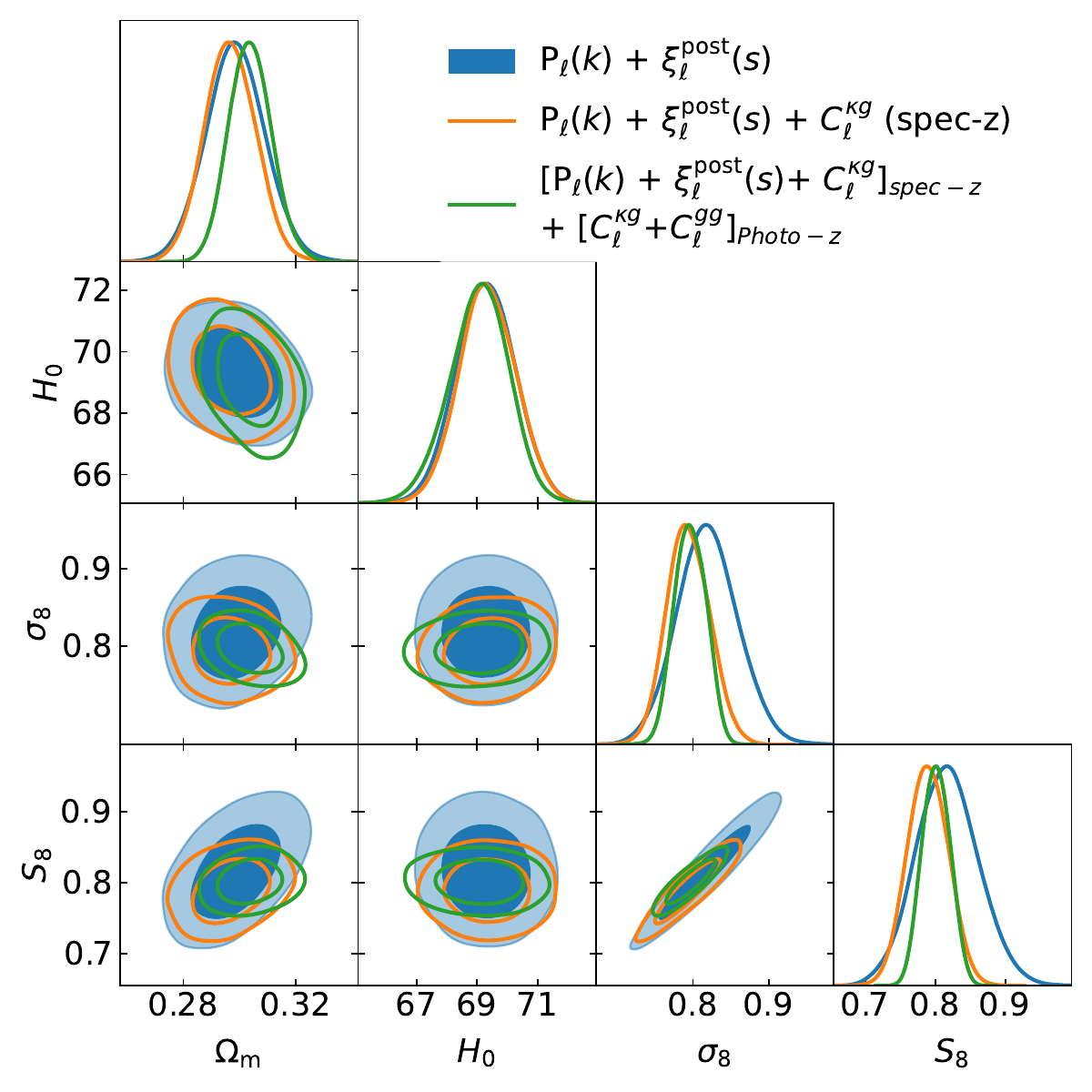}
     \caption{Comparison of FS+BAO only, FS+BAO combined with cross-correlations with lensing (\textit{Planck} PR4 and ACT DR6), 
     and the whole combination of 3D clustering and angular spectra from both spectroscopic and photometric BGS and LRG galaxies from DESI along with PR4 and DR6 lensing data sets.} 
    \label{fig:joint_contours_BGSLRG}
\end{figure}

\begin{figure}[htb]
    \centering
    \includegraphics[width=\linewidth]{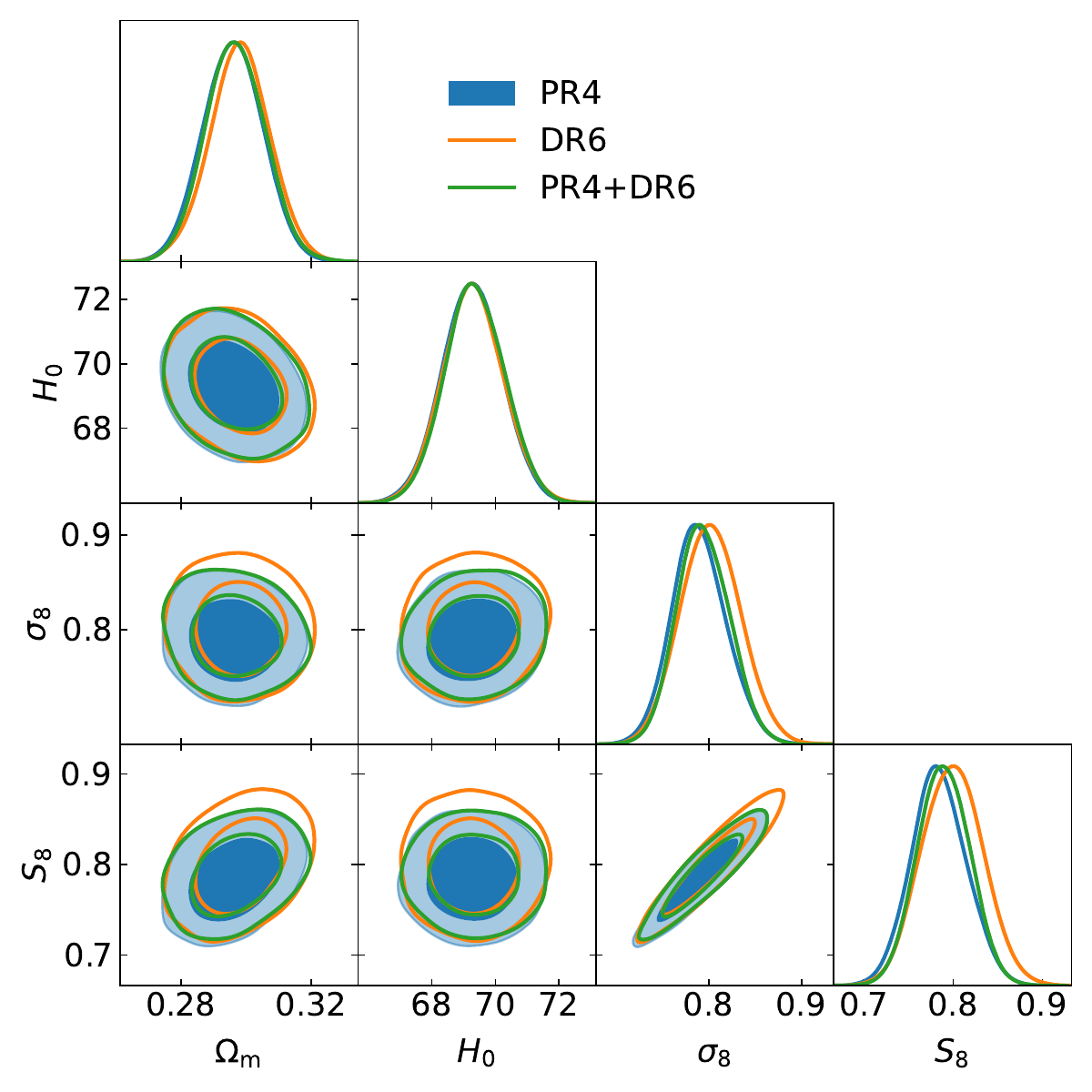}
    \caption{Comparison of constraints using 3D+$\Ckg$ within the spectroscopic BGS and LRG bins for different combinations with CMB lensing maps.} 
    \label{fig:3d_Ckg_PR4vsDR6}
\end{figure}

\begin{figure}[htb]
    \centering
    \includegraphics[width=\linewidth]{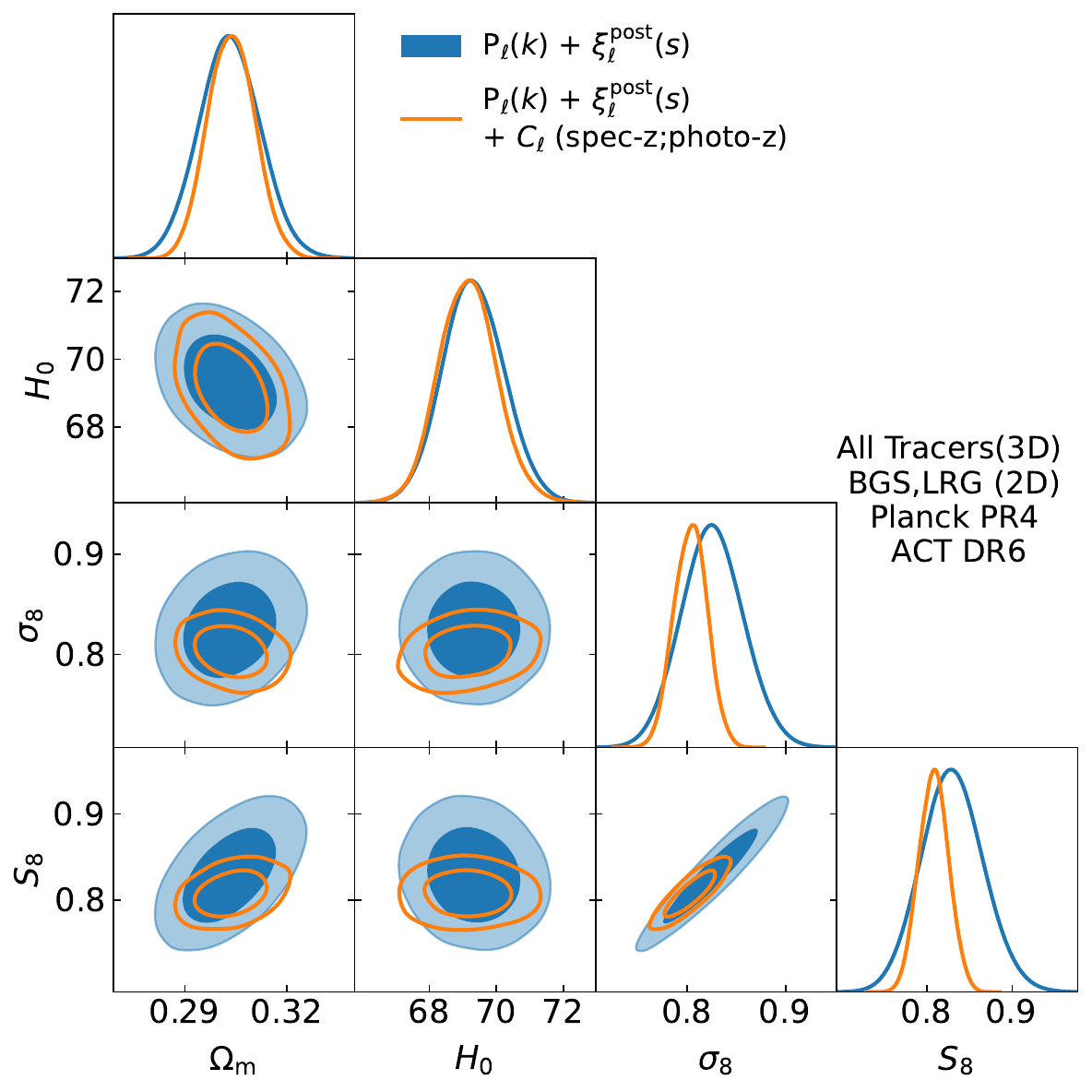}
    \caption{Constraints using all six DESI Y1 tracers in the 3D clustering. The blue contours show $P_{\ell}(k)+\xi_{\ell}(s)$ alone and the orange contour combines with angular spectra using BGS and LRG galaxies within the spectroscopic bins as well as from the photometric sample.} 
    \label{fig:joint_contours_pr4_dr6_allTracers}
\end{figure}

\begin{table}
\caption{Summary of constraints with different data/likelihood combinations, starting with full-shape (FS) in the spectroscopic LRG bins and then incrementally adding $\Ckg$ (spectroscopic), $\Ckg$ + $\Cgg$ (photometric), the other spectroscopic tracers, and finally adding the reconstructed BAO correlation function. For all of these fits we applied a gaussian \textit{Planck} prior on $n_s$ of $\mathcal{N}[\mu=0.9649,\sigma=0.0042]$. All cross correlations with CMB lensing involve both the \textit{Planck} PR4 and ACT DR6 maps.
}             
\label{table: constraints}      
\centering                          
\begin{tabular}{c||c|c|c|c}        
\hline\hline                 
Data Combination &  $\Omega_{m}$ & $\sigma_8$ & $S_8$ \\
\hline \hline
$P_{\ell}+\xi_{\ell}$ (sBGS+sLRG) &  $0.299\pm 0.011$ & $0.819\pm 0.040$ & $0.817\pm 0.045$ \\
\hline
$P_{\ell}+\xi_{\ell}+\Ckg$ (sBGS+sLRG;PR4) &  $0.2960\pm 0.0091$ & $0.790^{+0.027}_{-0.031}$ & $0.784\pm 0.031$ \\
\hline
$P_{\ell}+\xi_{\ell}+\Ckg$ (sBGS+sLRG;DR6) &  $0.2981\pm 0.0093$ & $0.802\pm 0.032$ & $0.799\pm 0.034$ \\
\hline
$P_{\ell}+\xi_{\ell}+\Ckg$ (sBGS+sLRG;PR4+DR6) &  $0.2967\pm 0.0091$ & $0.794\pm 0.028$ & $0.789\pm 0.029$ \\
\hline
$+\Ckg,\Cgg$ (pBGS+pLRG)&  $0.3042\pm 0.0077$ & $0.797\pm 0.019$ & $0.803\pm 0.019$ \\
\hline
$+P_{\ell}+\xi_{\ell}$(ELG,QSO) &  $0.3037\pm 0.0069$ & $0.803\pm 0.017$ & $0.808\pm 0.017$ \\
\hline \hline
\end{tabular}
\end{table}

We begin by showing in Fig.~\ref{fig:joint_contours_BGSLRG} a comparison of constraints using BGS and LRG galaxies and both Planck PR4 and ACT DR6 lensing maps. We compare the 3D clustering analysis only ($P_{\ell}(k) + \xi_{\ell}(s)$) to the combination of 3D + $\Ckg$ (using spectroscopic BGS and LRG galaxies only) and the combination of 3D + angular spectra using both spectroscopic and photometric BGS and LRG galaxy samples. In 3D clustering, the primary source of information on $\Omega_{\rm m}$ and $H_0$ (assuming a BBN prior on $\omega_b$) comes from the BAO signal that constrains angular diameter and Hubble distances. The post-reconstruction correlation function has a sharpened BAO peak that provides better measurement of distances than from the same feature in the power spectrum alone. However, the full-shape galaxy power spectrum also mildly constrains the combination $\Omega_{\rm m}h^2$ due to the shape from the $k_{\rm eq}$ scale and thus has a rotated degeneracy direction on the $\Omega_{\rm m}-H_0$ plane relative to $\xi_{\ell}^{\rm post}$. The combination of $P_{\ell}$ and $\xi_{\ell}^{\rm post}$ therefore provide tighter constraints on the two parameters than either data set alone. The amplitude $\sigma_8$ constraints come entirely from the full-shape power spectrum. The angular spectra alone do not provide new information on $H_0$ but are sensitive to the combination $\sigma_8\sqrt{\Omega_{\rm m}}$ and thus constrain the parameter $S_8 \equiv \sigma_8\sqrt{\Omega_{\rm m}/0.3}$. In addition, galaxy-galaxy clustering and galaxy-lensing cross correlation spectra scale differently with galaxy bias ($\propto b_1^2P(k)$ vs $\propto b_1P(k)$) such that their combination helps break the $b_1-\sigma_8$ degeneracy. We therefore find that by simply adding spectroscopic $\Ckg$ to the 3D clustering analysis our $\sigma_8$ and $S_8$ constraints improve by $30\%$ and $35\%$, respectively. In contrast, various analyses employing the bispectrum in addition to the power spectrum when fitting BOSS-like mocks \cite{Philcox_2022}, BOSS data \cite{Chen24} or even DESI Y1 \cite{Masot2025} have found only $10-20\%$ improvements over power spectrum (+BAO) analyses alone. When adding the angular spectra from the photometric BGS and LRG samples we find an additional $\sim~30\%$ reduction in $\sigma_8$ and $S_8$ errors. Due to the breaking of the $\Omega_{\rm m}-\sigma_8$ degeneracy present in angular spectra when combining them with 3D data, we also find improvements in $\Omega_{\rm m}$ compared to 3D data alone. Just adding $\Ckg$ in the spectroscopic bins to the 3D analysis tightens $\Omega_{\rm m}$ by $\sim 20\%$ while the additional photometric data set results in a $30\%$ improvement over $P_{\ell}(k)+\xi_{\ell}(s)$. For $\Omega_{\rm m}$ and $H_0$ we do not find significant projection effects in the 3D or 3D+$C_{\ell}$ analysis, as best-fit values obtained using the \texttt{BOBYQA} \cite{Powell2009,Cartis2018,Cartis2021} minimizer are are within $0.1\sigma$ of the means of the posteriors. However, there is a $\sim0.5\sigma$ projection in $\sigma_8$ in the 3D+$\Ckg$ analysis that is slightly larger than the $\sim0.4\sigma$ projection in the 3D only case. This may be due to us not relating $\alpha_{x}$ to the 3D counterterms in the spectroscopic bins as we do for the auto and cross counterterms in the photometric sample. We can modify this aspect of the pipeline in the future if necessary but also expect a significant reduction in projection effects in Y3 simply due to the substantial increase in constraining power of the data.

To check the dependence of our constraints on the two different lensing data sets, we test the $P_{\ell}(k)+\xi_{\ell}^{\rm post}(s)+ \Ckg$ analysis within the spectroscopic samples using just Planck PR4, just ACT DR6, and their combination (for the cross-correlation with galaxies). We show this comparison in Fig.~\ref{fig:3d_Ckg_PR4vsDR6}. We find a slight preference for higher $\sigma_8$ (and $S_8$) when combining with the ACT DR6 dataset compared to Planck PR4. Using both lensing data sets improves the amplitude constraints by $\sim 5\%$. 

\begin{figure}[htb]
    \centering
    \includegraphics[width=0.99\linewidth]{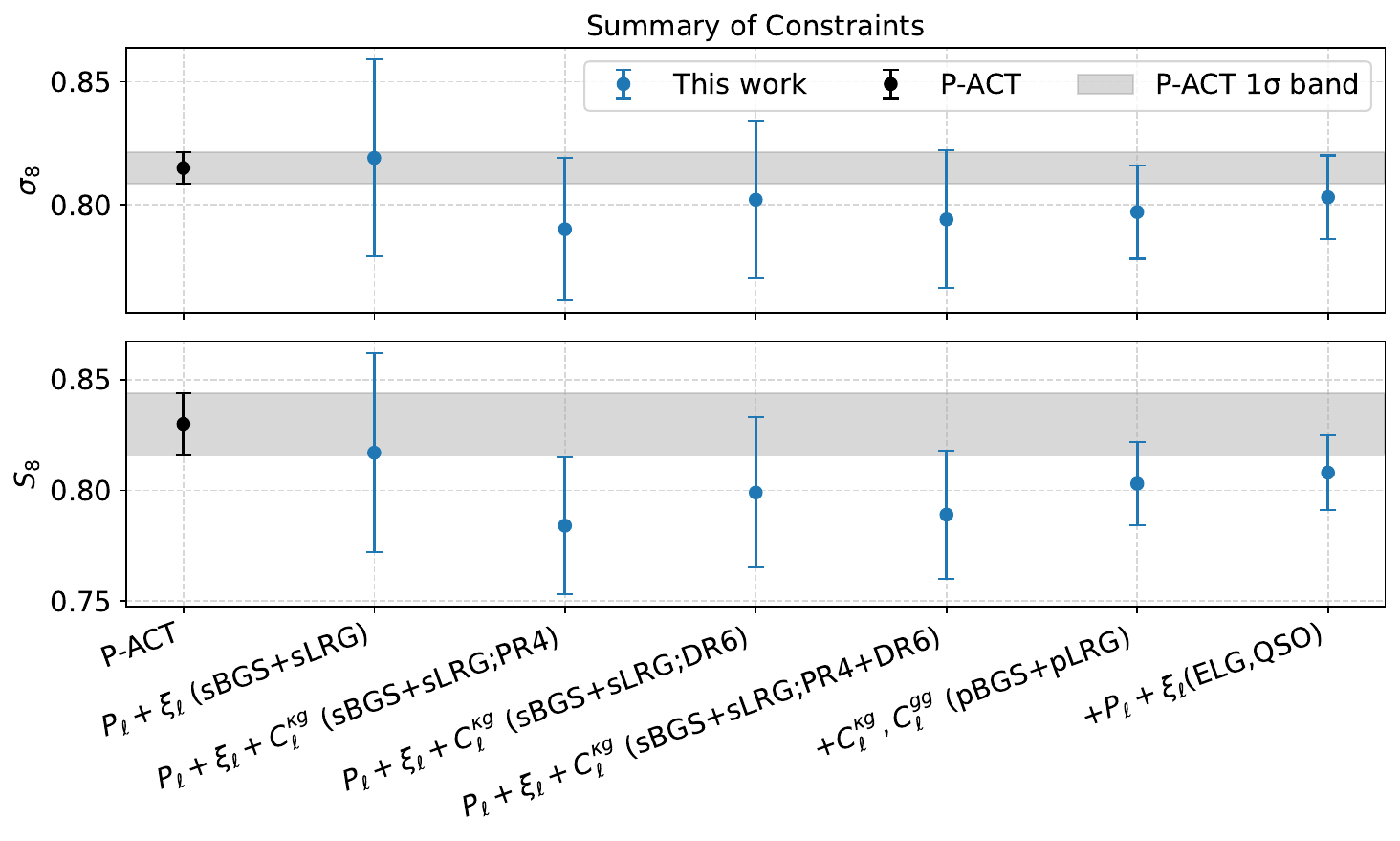}
    \caption{Summary of $\sigma_8$ and $S_8$ from this work, with values corresponding to the results of \\ Table~\ref{table: constraints}. We also show for reference the constraints from the recent ``P-ACT'' \cite{PACT2025} result using Planck and ACT CMB data. }
    \label{fig:s8 summary}
\end{figure}

Thus far we have only focused on comparisons of constraints using BGS and LRG galaxies from the Y1 spectroscopic catalogs as well as the legacy survey imaging samples. The Y1 DESI spectroscopic data also includes ELG and QSO samples, and while we do not currently consider their cross-correlations with lensing in this paper\footnote{Cross-correlations between DESI QSO and CMB lensing are currently being analyzed in parallel to this work, in ref.~\cite{deBelsunce2025}}, we are at liberty to include them in the $P_{\ell}(k)+\xi_{\ell}^{\rm post}(s)$ part of our analysis. Including these additional data, we find $\sigma_8 = 0.803\pm 0.017$ and $S_8 = 0.808\pm 0.017$ for the joint 3D and 2D analysis, improving on the 3D-only constraints by $57\%$ and $62\%$ respectively. The inclusion of the ELG and QSO tracers also improves on our $\sigma_8$ and $S_8$ by $\sim11\%$ compared to our constraints in Fig.~\ref{fig:joint_contours_BGSLRG} that analyzed the 3D and 2D clustering using only the BGS and LRG galaxy samples. These results and those discussed earlier in this section are listed in Table~\ref{table: constraints} and are shown in Fig.~\ref{fig:s8 summary}.


\section{Discussion and Conclusion}
\label{sec: conc}

In this paper we present constraints from a joint analysis combining the 3D clustering information from the DESI Y1 spectroscopic galaxies with angular cross correlations with CMB lensing reconstruction maps from \textit{Planck} PR4 and ACT DR6 data.  Our final fits include the 3D power spectrum and post-reconstruction correlation function multipoles from the six spectroscopic galaxy samples (three of which are LRG's split into three redshift bins). The spectroscopic BGS and LRG samples are cross-correlated with CMB lensing to produce four angular power spectra. We additionally include angular galaxy auto and galaxy-lensing cross spectra using the imaging BGS and LRG samples from the DESI Legacy Survey DR9. With our full dataset we obtain 2.1$\%$ constraints on the amplitude of structure, $\sigma_8 = 0.803\pm 0.017$.  This is a $\sim21\%$ improvement over the constraints ($\sigma_8 = 0.772^{+0.020}_{-0.023}$) from ref.~\cite{Sailer24} that fit $\Ckg$ and $\Cgg$ using the photometric LRGs and $\sim19\%$ improvement over the $\sigma_8=0.791\pm0.021$ constraint using photometric BGS+LRGs presented in \cite{Sailer2025}. Our constraints are statistically consistent with those from the primary CMB from Planck 2018 ($\sigma_8=0.811\pm0.006$, \cite{PlanckParams18}), the recent ``P-ACT'' result that combined ACT DR6 with large-scale Planck data to get $\sigma_8=0.813\pm0.005$ \cite{PACT2025}, and the constraint of $\sigma_8 = 0.815\pm0.012$ by ref.~\cite{Farren:2024rla} combining unWise galaxies with ACT DR6 CMB lensing. Other results using CMB include $0.8067 \pm 0.0067$ from \cite{Rosenberg22} that used the Planck PR4 temperature and polarization maps and the $0.8136\pm0.0063$ constraint from \cite{deBelsunce21} that combined Planck 2018 with a low-$\ell$ \texttt{SRoll2} likelihood. On the weak galaxy lensing side there has been a trend of low $S_8$ constraints (see e.g refs.~\cite{kids1000,Amon21,Secco21,Dalal23,Li23,DESKids23} and others) that was broken only recently with the cosmic shear analysis of the final data release of the Kilo-Degree Survey (KiDS), whose constraints of $S_8=0.815^{+0.016}_{-0.021}$ \cite{KiDS-Legacy25} are consistent both with Planck and our analysis. Another recent joint cosmic shear analysis combining Dark Energy Survey (DES), KiDS, and Hyper-Suprime Cam (HSC) datasets and fitting to small scales found $S_8=0.795^{+0.015}_{-0.017}$ which is less than $1\sigma$ below our constraint\cite{GarciaGarcia24}.

Conversely, ref.~\cite{DESI2024.V.KP5} obtained $\sigma_8 = 0.841 \pm 0.034$ using only the 3D power spectrum multipoles combined with post-reconstruction BAO, which is double the error of our constraint. Including the cross-correlation with lensing appears to not shift the mean of the $\sigma_8$ constraints significantly when compared to 3D clustering alone but does tighten the errorbars by nearly a factor of two. We therefore do not find a `low' $\sigma_8$, in stark contrast with similar analyses using BOSS data \cite{Chen22,Chen22_2,Chen24}. Ref.~\cite{ChenDeRose24} used BGS and LRG galaxies from the DESI targeting sample to combine the galaxy auto-correlation and weak galaxy-galaxy lensing with DES to obtain constraints of $\sigma_8 = 0.840\pm 0.044$, consistent with our constraints. This suggests that previously low constraints on the clustering amplitudes may be a feature/systematic in the BOSS galaxy sample itself. In Appendix~\ref{app: boss_vs_desi} we perform a catalog-level comparison between BOSS and DESI galaxies in search of signs for a potential systematic in the BOSS dataset. We cut the BOSS galaxies to the same $0.4\leq z\leq0.6$ redshift slice as the lowest redshift LRG bin from DESI. Then we use a mask to filter both DESI and BOSS samples in that redshift range to just their overlapping footprints. Finally we reweight the BOSS galaxies to match the redshift distribution of the DESI sample. Using these subsamples we compute angular galaxy auto spectra from BOSS and DESI as well as their cross correlations and analyze their residual. We find a small $\ell^{-1.5}$ feature in the residual that may indicate a contamination in the BOSS data set. 

We also test two extensions to the standard $\Lambda$CDM model of cosmology in the appendices of this paper. In Appendix~\ref{app: w0wa} we test the robustness of our amplitude constraints in the presence of evolving dark energy by varying the $w_0$ and $w_a$ dark energy equation of state parameters and combining with the Union3 \cite{Rubin2023} Type 1a supernova dataset. We find that despite a $\sim3.5\sigma$ discrepancy with the $\Lambda$CDM model ($w_0=-0.755^{+0.083}_{-0.055}$, $w_a = -1.03^{+0.25}_{-0.36}$) our constraints on $S_8$ are consistent with those obtained in the body of this paper while assuming a cosmological constant for dark energy. In Appendix~\ref{app: slip} we also test modified gravity through the gravitational slip parameter and find $\gamma=1.17\pm0.11$, in mild tension with General Relativity but consistent with being a projection effect (Appendix \ref{app: slip}). 

In this paper we focus our analysis on the Y1 DESI spectroscopic galaxy sample, which was publicly released in March 2025. The Y1 key projects focused primarily on full-shape and BAO modeling without any cross-correlations with lensing. However, observations for Y3 have already concluded, with the first BAO analysis on the Y3 DESI data presented recently \cite{DESIDR2_bao}. The DESI Y3 data analyses will include a much larger emphasis on the joint constraints from spectroscopic galaxy samples and cross correlations with lensing (weak galaxy lensing and CMB). This work therefore plays a dual purpose by developing a methodology and pipeline in anticipation of the Y3 data release in addition to the presentation of constraints discussed above. The effective volume of year 3 data is roughly 2.2$\times$ that of Y1, so we expect a reduction in statistical errors by $1/\sqrt{2.2}$ in the 3D clustering data. We also expect gains in the constraining power from the cross-correlation with ACT lensing due to a larger overlap expected between the DESI Y3 and ACT DR6 footprints compared to Y1. 

\section{Data availability}
Data/chains from the plots in this paper will be made available on Zenodo as part of DESI’s Data Management Plan. A link will be provided upon acceptance of this paper by the journal.

\section*{Acknowledgments}

We would like to thank Alex Krolewski and Gerrit Farren for helpful discussions during the preparation of this manuscript. 

This material is based upon work supported by the U.S. Department of Energy (DOE), Office of Science, Office of High-Energy Physics, under Contract No. DE–AC02–05CH11231, and by the National Energy Research Scientific Computing Center, a DOE Office of Science User Facility under the same contract. Additional support for DESI was provided by the U.S. National Science Foundation (NSF), Division of Astronomical Sciences under Contract No. AST-0950945 to the NSF’s National Optical-Infrared Astronomy Research Laboratory; the Science and Technology Facilities Council of the United Kingdom; the Gordon and Betty Moore Foundation; the Heising-Simons Foundation; the French Alternative Energies and Atomic Energy Commission (CEA); the National Council of Humanities, Science and Technology of Mexico (CONAHCYT); the Ministry of Science, Innovation and Universities of Spain (MICIU/AEI/10.13039/501100011033), and by the DESI Member Institutions: \url{https://www.desi.lbl.gov/collaborating-institutions}.

The DESI Legacy Imaging Surveys consist of three individual and complementary projects: the Dark Energy Camera Legacy Survey (DECaLS), the Beijing-Arizona Sky Survey (BASS), and the Mayall $z$-band Legacy Survey (MzLS). DECaLS, BASS and MzLS together include data obtained, respectively, at the Blanco telescope, Cerro Tololo Inter-American Observatory, NSF's NOIRLab; the Bok telescope, Steward Observatory, University of Arizona; and the Mayall telescope, Kitt Peak National Observatory, NOIRLab. NOIRLab is operated by the Association of Universities for Research in Astronomy (AURA) under a cooperative agreement with the National Science Foundation. Pipeline processing and analyses of the data were supported by NOIRLab and the Lawrence Berkeley National Laboratory. Legacy Surveys also uses data products from the Near-Earth Object Wide-field Infrared Survey Explorer (NEOWISE), a project of the Jet Propulsion Laboratory/California Institute of Technology, funded by the National Aeronautics and Space Administration. Legacy Surveys was supported by: the Director, Office of Science, Office of High Energy Physics of the U.S. Department of Energy; the National Energy Research Scientific Computing Center, a DOE Office of Science User Facility; the U.S. National Science Foundation, Division of Astronomical Sciences; the National Astronomical Observatories of China, the Chinese Academy of Sciences and the Chinese National Natural Science Foundation. LBNL is managed by the Regents of the University of California under contract to the U.S. Department of Energy. The complete acknowledgments can be found at \url{https://www.legacysurvey.org/}.

Any opinions, findings, and conclusions or recommendations expressed in this material are those of the author(s) and do not necessarily reflect the views of the U. S. National Science Foundation, the U. S. Department of Energy, or any of the listed funding agencies.

The authors are honored to be permitted to conduct scientific research on Iolkam Du'ag (Kitt Peak), a mountain with particular significance to the Tohono O'odham Nation.

\appendix

\section{Other spectra}
\label{app:alt_spectra}
\begin{figure}
\centering
\resizebox{0.99\columnwidth}{!}{\includegraphics{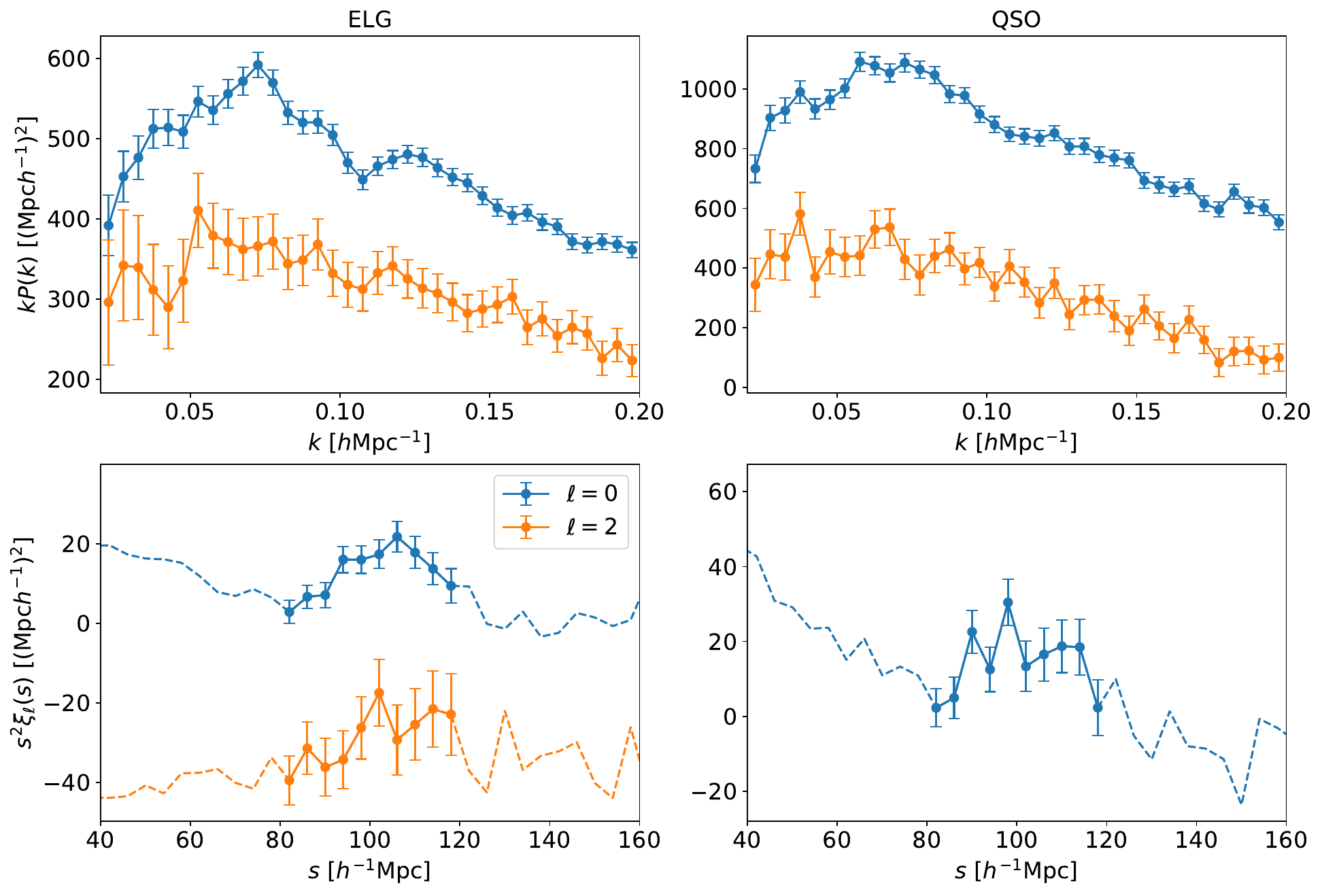}}
\caption{3D Power spectrum multipoles(top) and Post-recon correlation function multipoles (bottom) computed from the DESI Y1  ELG($1.1\leq z\leq 1.6$) and QSO($0.8\leq z\leq 2.1$) samples. For the correlation function the data points with errorbars span the range that we fit to, whereas the dashed lines show the continuation of the data to higher and lower scales.
\label{fig: other_Y1_spectra}}
\end{figure}

\begin{figure}[htb]
    \centering
    \includegraphics[width=\linewidth]{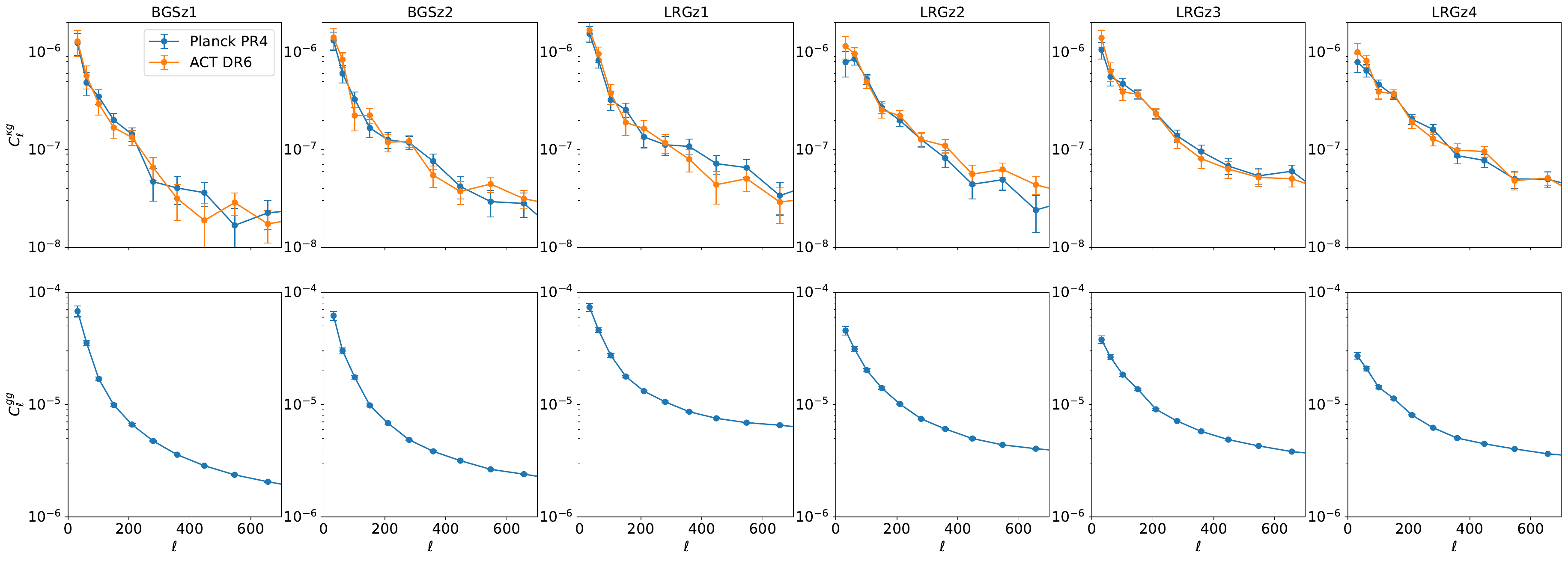}
    \caption{Pseudo-$\Ckg$ and Pseudo-$\Cgg$ spectra computed from cross and auto correlations between the DESI photometric sample (separated into four tomographic redshift bins) and the \textit{Planck} PR4 and ACT DR6 lensing maps.
    } 
    \label{fig:cell_photoz}
\end{figure}

We show in Fig.~\ref{fig: other_Y1_spectra} the 3D clustering spectra for the other DESI Y1 tracers:BGS, ELG, and QSO. In Fig.~\ref{fig:cell_photoz} we show the angular spectra from the photometric BGS and LRG bins.

\section{Comparison of BOSS and DESI}
\label{app: boss_vs_desi}

\begin{figure}[htb]
    \centering
    \includegraphics[width=\linewidth]{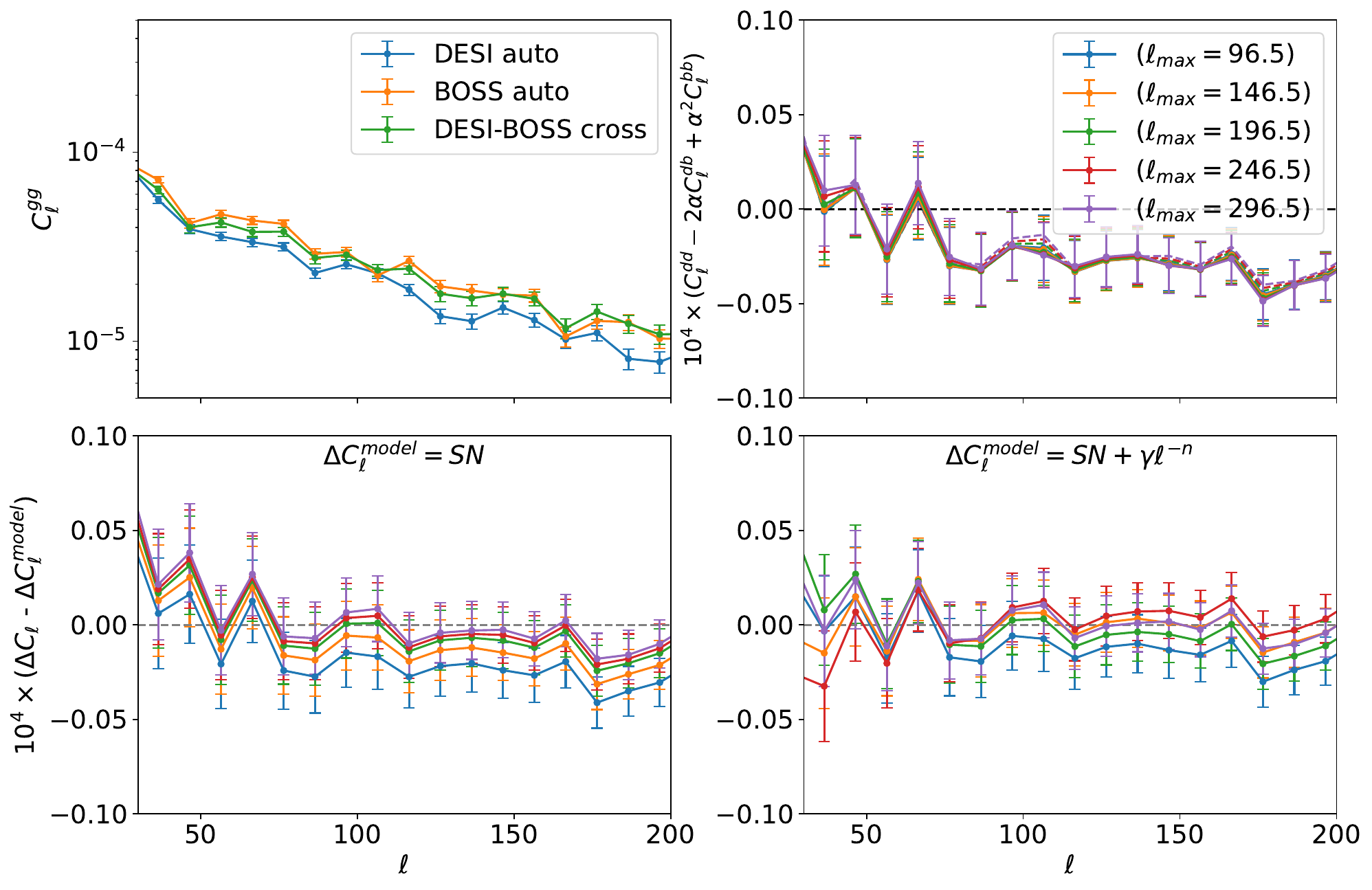}
    \caption{Comparison of auto and cross spectra between BOSS and DESI galaxies. The upper left panel shows the DESI auto-, BOSS auto- and DESI-BOSS $\Cgg$ cross  spectra. In the upper right panel we compute the residuals (see text) for values of $\alpha$ obtained with a minimizer choosing different scale cuts. In the bottom two panels we attempt to fit the residuals using purely a shot-noise term (left) and a shot-noise plus power-law model (right). The different curves in the bottom panels again correspond to different scale cuts used in the experiment.
    } 
    \label{fig:boss_desi_cl}
\end{figure}

As noted in the discussion, the results presented in this paper as well as others using DESI Y1 data (e.g.\ ref.\ \cite{DESI2024.V.KP5}) show $\sigma_8$ constraints that are no longer in tension with the primary CMB. In this section we briefly investigate the possibility that a systematic is present in the BOSS catalogs that may be the reason for such discrepancies. In the following we use the \textsc{CMASSLOWZTOT} catalog from BOSS DR12 but find similar behavior when using the \textsc{CMASSLOWZ} sample instead. The simplest approach towards comparing the two samples is through the computed angular galaxy-galaxy spectra. However, an apples-to-apples comparison requires us to select subsamples from these catalogs that have overlapping footprints, redshift ranges, and redshift distributions. We first use the randoms from both catalogs to create binary masks (with \texttt{HEALPix} and {\sc Nside}=128) defining their respective footprints and by extension their overlapping footprint. Using this we filter both catalogs (galaxies and randoms). We then choose a redshift bin of $0.4\leq z\leq 0.6$ in order to retain enough galaxies from both samples. Finally, we use the redshift distributions of the two samples within this redshift range in order to reweight the BOSS galaxies to match the $dN/dz$ of the DESI galaxies when computing the angular spectra using the DirectSHT method. We do this by interpolating the redshift distributions of DESI and BOSS within the $0.4\leq z\leq 0.6$ range and computing their ratio at the redshift of each BOSS galaxy and applying this as another weight to the BOSS galaxies. These spectra\footnote{In the DESI catalogs, the `WEIGHTS' column is the total weight that takes into account completeness, imaging systematics, and redshift failures. We therefore weight the DESI galaxies and randoms by \textsc{catalog[`WEIGHT']}$\times$\textsc{catalog[`WEIGHT\_FKP']}. In the BOSS data catalog, we create a new `WEIGHT' column using the respective columns in the original catalogs for systematics (`WEIGHT\_SYSTOT'), completeness (`WEIGHT\_CP'), and redshift failure (`WEIGHT\_NOZ') weights and compute: \newline $\textsc{catalog[`WEIGHT']} = \textsc{catalog[`WEIGHT\_SYSTOT']}\times(\textsc{catalog[`WEIGHT\_NOZ']}+\textsc{catalog[`WEIGHT\_CP']} - 1)$. Since these weights are not relevant in the random catalog we set the values in the `WEIGHT' column to 1.The final weights we use in the BOSS data and randoms is then \textsc{catalog[`WEIGHT']}$\times$\textsc{catalog[`WEIGHT\_FKP']}.
} are shown in Fig.~\ref{fig:boss_desi_cl}. We attempt to compare the BOSS and DESI galaxies in the following way. 

We define the residual power spectrum
\begin{align}
    \Delta P =  \langle(\delta_{D}(\bk)-\alpha\delta_{B}(\bk))^2\rangle,
\end{align}
where $\delta_{D}$ is the overdensity of DESI galaxies, $\delta_{B}$ is the overdensity of BOSS galaxies, and $\alpha$ is a parameter related to the difference in large-scale bias of the samples. When $\Delta P$ is minimized with respect to $\alpha$, then the remaining residual should be related to stochastic noise, non-linear bias and evolution, and any systematics that may be present in one of the samples (which we are searching for). If we neglect maginification bias in this experiment (using the same $dN/dz$ for the two samples we estimate the difference in their magnification bias contributions to be $\mathcal{O}(10^{-8})$.) then under the Limber approximation we can write the projected galaxy auto spectrum as 
\begin{align}
    C_{\ell} \simeq \int\frac{d\chi}{\chi^2}\ [W_g(\chi)]^2P_{gg}(k)
    \quad .
\end{align}
The projection kernel $W_g(z) = H(z)\ dN/dz$ is the same for the two samples if we reweight the BOSS galaxies to match the redshift distribution of the DESI sample. We can then write the 2D residual as
\begin{align}
\left\langle\left(\delta_{\ell m}^D - \alpha\,\delta_{\ell m}^B\right)^2\right\rangle
    &\simeq \int\frac{d\chi}{\chi^2}[W_g(\chi)]^2 \left[P^{DD}(\bk) - 2\alpha P^{DB}(\bk) + \alpha^2P^{BB}(\bk)\right] \nonumber\\
    &= C_\ell^{DD} -2\alpha C_\ell^{DB} + \alpha^2 C_\ell^{BB} \equiv \Delta C_\ell^{\rm residual}.
    \label{eq: residual}
\end{align}
If we restrict ourselves to very large ($\ell\lesssim 200$) scales, the decorrelation from non-linear effects is small. The residual is minimized when $\alpha = C_\ell^{DB}/C_\ell^{BB}$. We show in the upper right plot of Fig.~\ref{fig:boss_desi_cl} the residual for different $\alpha$ obtained from averaging $C_\ell^{DB}/C_\ell^{BB}$ up to different scales $\ell_{\rm max}$. Note that for most modes above $\ell\simeq 25$ the residual is negative despite it being defined as a perfect square. This is because when computing $C_{\ell}$, \texttt{NaMaster} internally removes the shot-noise for auto-spectra but not for cross-spectra between two different fields. This results in a constant shift from the $-2\alpha C_{\ell}^{DB}$ term in Eq.~\ref{eq: residual}. We next fit the residual with two models, a constant shot-noise $SN$, and a shot-noise plus a power-law term, $SN+\gamma\ell^{-n}$. These are shown in the lower two panels of Fig.~\ref{fig:boss_desi_cl}. We find that there is a mild preference for an $\ell^{-n}$ power law in the residual with best fit parameters: $n\simeq1.7\pm0.2$, $SN \simeq (-2.9\pm0.8)\times10^{-6}$, $\gamma\simeq0.0023\pm0.0015$ (using the standard deviations across the different $\ell_{\rm max}$ choices). This may hint at the presence of a systematic in at least one of the galaxy catalogs, and given the disparities in depth and uniformity of the imaging we believe it is most likely the BOSS galaxy catalog. 
The catalog-level comparison shown here is not a complete analysis and suggests the need for further investigation. Here we have just used 2D spectra but it may be more informative to repeat this experiment with the 3D power spectra as well, and would also alleviate the need to make assumptions about magnification bias, which we ignore in this experiment. As stated previously, we estimate the difference in magnification bias contributions in the two samples is $\mathcal{O}(10^{-8})$, which is about two orders of magnitude smaller than the residuals we see in Fig.~\ref{fig:boss_desi_cl}. Our results also do not distinguish between the possibility of a systematic being present in the DESI data versus BOSS; however, ref.~\cite{Chen22_2} (\S5.3) previously measured the cross-correlation between galaxies in the non-overlapping $z_1$ and $z_3$ redshift bins in BOSS and found correlations inconsistent with magnification that suggest the presence of systematics in the BOSS data. In any case, our measurement suggests that a systematic in at least one of the two catalogs may be responsible for some of the discrepancy in $\sigma_8$ constraints between analyses using BOSS versus DESI data but we leave for future work more a comprehensive investigation into this issue.

\section{EFT vs HEFT}
\label{app: EFT_HEFT}

\begin{figure}[htb]
    \centering
    \includegraphics[width=\linewidth]{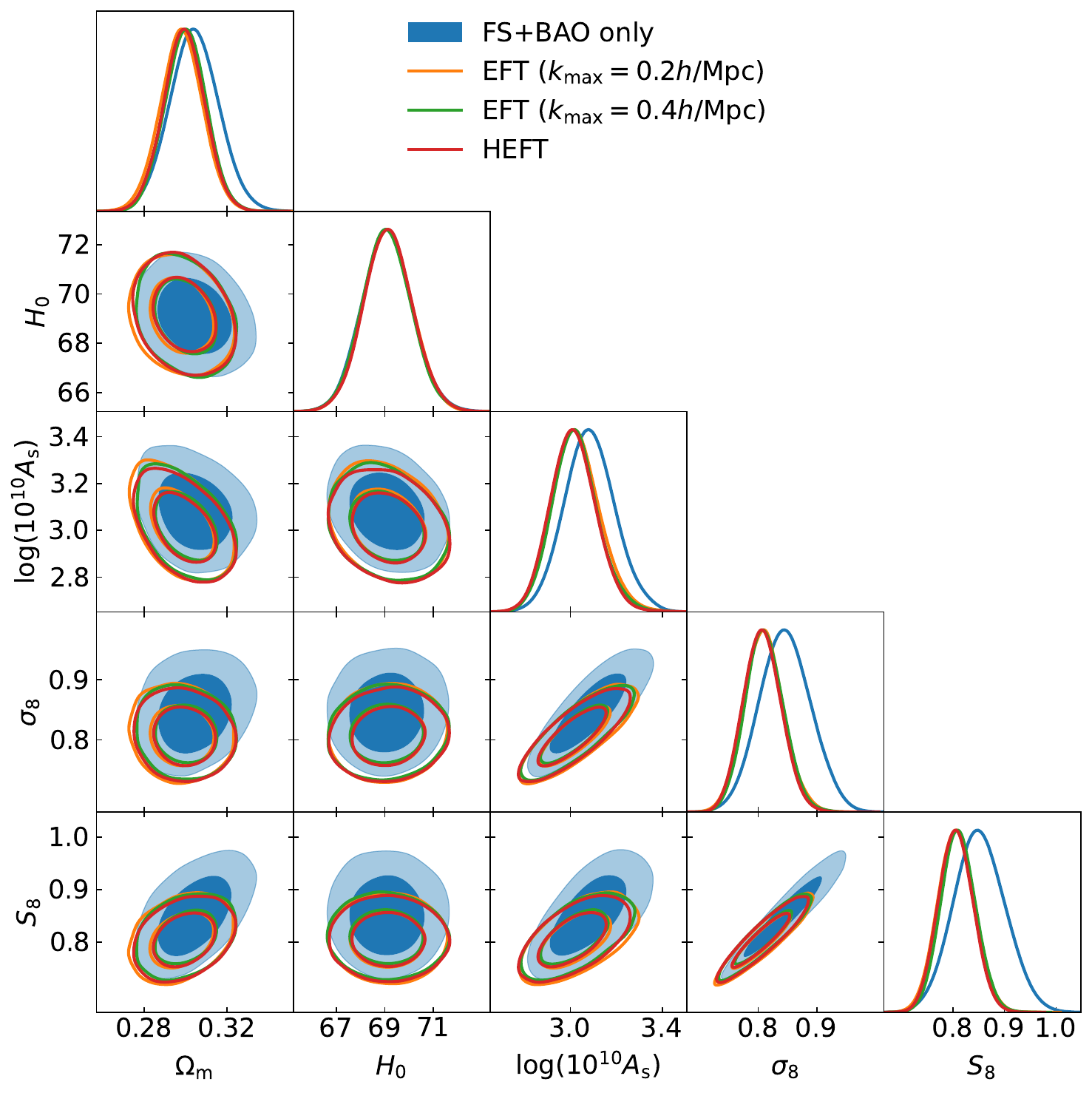}
    \caption{Comparison of constraints using EFT for the $\Ckg$ theory vs HEFT. We also show FS+BAO constraints without and angular spectra for reference. For this test we limit ourselves to the spectroscopic LRG bins only.} 
    \label{fig:cleft_test}
\end{figure}

As discussed in \S~\ref{sec: HEFT}, our baseline setup employs a Hybrid Effective Field Theory (HEFT) model using real-space power spectra computed from the \texttt{Aemulus $\nu$} simulations and emulated with a neural network, which are then passed into the integrands of the Limber integrals to produce the angular spectrum predictions. Meanwhile our theory predictions for the redshift space power spectra are computed using the effective Lagrangian perturbation theory module \texttt{LPT\_RSD} from \texttt{velocileptors}. In principle one can use this \texttt{velocileptors} EFT model for the angular spectra as well just by setting $\mu=0$ in \texttt{LPT\_RSD}. The advantage of the HEFT model is that the use of N-body simulations allow for accurate and stable predictions to smaller scales (higher $k$; up to $k_{\rm max} \lesssim 0.6 \ihmpc$ for $z\leq 1$ \cite{DeRose_2023}), and this was the motivation for ref.~\cite{Sailer24} choosing $\ell_{\rm max} = k_{\rm max} \chi(z=0.3) - 0.5 \simeq 600$ in their analysis, which we choose in the body of this work as well. In contrast, in the tests on DESI LRG-like mocks presented in ref.~\cite{KP5s2-Maus} we showed that $\mathcal{O}(1\sigma)$ biases arise in cosmological parameters when including modes above $k \simeq 0.2 \ihmpc$, leading to the fiducial choice of $k_{\rm max} = 0.2 \ihmpc$ for the DESI Y1 full-shape analysis using the EFT power spectrum. However, some of this bias comes from redshift-space effects at small scales such as Finger of God (FoG) and therefore wouldn't be a restriction for the real-space EFT power spectrum\cite{Baleato2025}.  A disadvantage of HEFT is that one needs to run new simulations in order to open up the parameter space. The \texttt{Aemulus $\nu$} simulations are compatible with $w$CDM cosmologies with varying neutrino mass, but in order to test e.g. evolving dark energy with the CPL parameterization new simulations are needed. For EFT this is not the case and the theory is compatible with any beyond-$\Lambda$CDM parameter space provided that the model doesn't significantly affect the perturbation theory kernels\footnote{Currently \texttt{velocileptors} makes use of the Einstein-deSitter (EdS) approximation for the expansion of operators in the 1-loop power spectrum. Efforts are under way to generalize the theory beyond EdS.}. 

In Fig.~\ref{fig:cleft_test} we compare constraints from 3D (FS+BAO) only, FS+BAO+$\Ckg$ using EFT for the 2D spectra with two choices of scale cuts, and the usual setup using HEFT for the $\Ckg$ spectra. For this test we limit ourselves to the spectroscopic LRG bins only for simplicity. When using EFT for $\Ckg$ we choose $\ell_{\max}(z)$ values for the three redshift bins of (250, 350, 430) and (530, 700, 850), roughly corresponding to fixed $k_{\rm max}$ of 0.2 and 0.4 $\ihmpc$ respectively. We find that the fits using EFT for $\Ckg$ agree very closely with that using HEFT and we see little change between the two choices of scale cuts. This suggests that within the noise levels of the current CMB lensing maps, the constraints are only weakly dependent on contributions to the bandpowers from beyond-linear scales. In fact, ref.~\cite{Sailer24} similarly found quite close agreement between linear theory (using $\ell_{\rm max} < 250$ in each bin) and their baseline constraints using HEFT with $\ell_{\rm max} = 600$. However, ref.~\cite{ChenDeRose24} performed a similar test of scale cuts on \texttt{Buzzard} mocks in their galaxy-galaxy analysis and found a $\sim25\%$ improvement in constraining power between $k_{\rm max} = 0.2$ and $0.4$ $\ihmpc$. This may suggest that the noise levels of the current CMB lensing measurements are limiting how much constraining power we gain from smaller scales. Note that $k_{\rm max} = 0.4$ $\ihmpc$ actually corresponds to higher $\ell_{\rm max}$ in the second and third LRG bin than the fixed $\ell_{\rm max} = 600$ we have used with HEFT throughout this paper, and we expected HEFT to be more accurate at small scales. So the fact that the EFT constraints are still consistent with HEFT suggests that the counterterms can sufficiently absorb the small-scale differences between the two models for the noise levels of our lensing data.

\section{Evolving dark energy}
\label{app: w0wa}

\begin{figure}[htb]
    \centering
    \includegraphics[width=\linewidth]{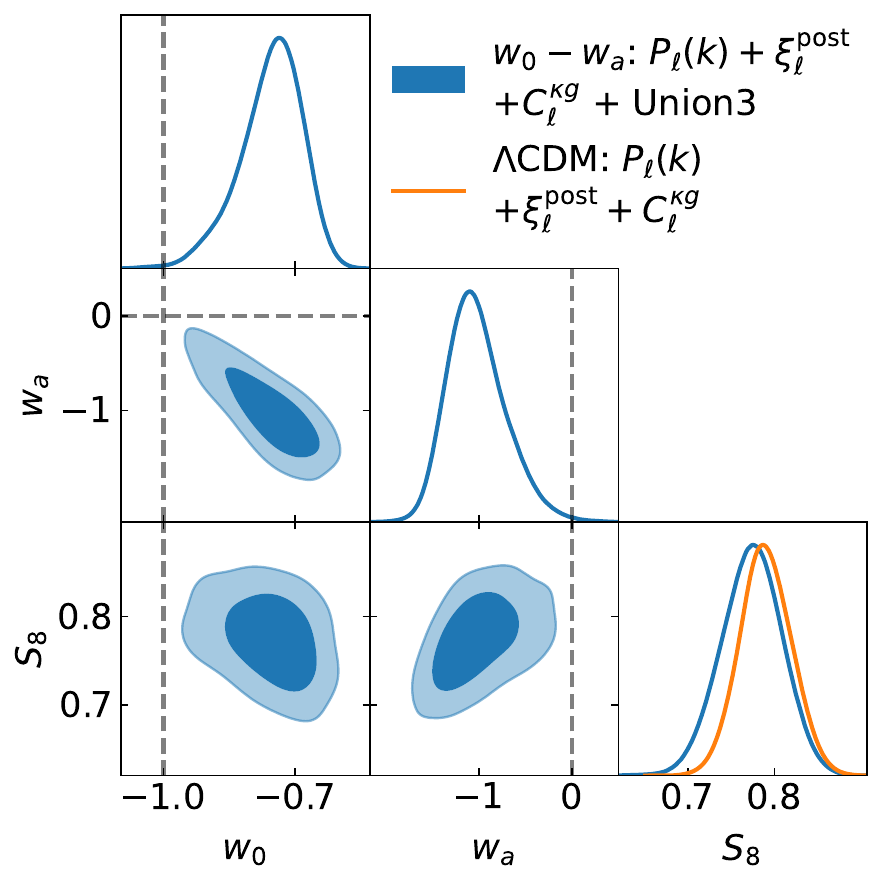}
    \caption{Constraints on evolving dark energy $w_0$ and $w_a$ using the 3D full-shape analysis along with $\Ckg$ in the spectroscopic BGS and LRG bins and including the Union3 likelihood. We also include the $\Lambda$CDM constraints on $S_8$ from the $P_{\ell}(k)+\xi_{\ell}^{\rm post}(s)+ \Ckg$ combination (without SNe) for comparison.} 
    \label{fig:w0wa_test}
\end{figure}

Recent results from the DESI collaboration based on modeling the broadband shape of the power spectrum \cite{DESI2024.VI.KP7A} or just the BAO feature \cite{DESI2024.VII.KP7B,DESIDR2_bao} (in combination with CMB and SNe) have shown preference for evolving dark energy over $\Lambda$CDM, with significances of $\sim2-4\sigma$. While dark energy is not the main focus of this paper, it will play a more prominent role in analyses of Y3 data and therefore we present a short test here to demonstrate that our constraints on the clustering amplitude are largely unaffected. The dark energy model tested in refs.~\cite{DESI2024.VI.KP7A,DESI2024.VI.KP7A,DESIDR2_bao} is the Chevallier-Polarski-Linder (CPL; \cite{Chevallier2001,Linder2003}) parameterization, commonly referred to as the $w_0w_a$CDM model, in which the dark energy equation of state scales linearly with the scale factor:
\begin{align}
    w(a) = w_0 + w_a(1-a).
\end{align}
In $\Lambda$CDM dark energy has a constant energy density, which corresponds to $w_0=-1.0$ and $w_a = 0.0$. 
In Fig.~\ref{fig:w0wa_test} we present constraints on $w_0$, $w_a$ and $S_8$ using a joint analysis that combines $P_{\ell}(k)+\xi_{\ell}^{\rm post}(s)+\Ckg$ within the spectroscopic Y1 BGS and LRG bins with the Union3 \cite{Rubin2023} supernova data set. The \texttt{Aemulus$\nu$} simulations that we use for HEFT predictions are currently not compatible with the $w_0w_a$CDM parameterization so we instead use EFT models (see previous section) for computing the real-space power spectra that enter the 2D $C_{\ell}$ models. For simplicity we choose to fix ($n_s,\omega_b,\theta_\star$) instead of including a CMB likelihood in our data combination. We choose the same scale cuts here as our baseline fits in the body of the paper, as we showed in the previous section that our EFT constraints were consistent when using both more and less conservative $\ell_{\rm max}$ than the HEFT fit. We find a $\simeq 3.5\sigma$ tension with $\Lambda$CDM, with $w_0 = -0.755^{+0.083}_{-0.055}$ and $w_a = -1.03^{+0.25}_{-0.36}$. These results are consistent with those found in ref.~\cite{DESI2024.VII.KP7B} when combining DESI Y1 3D clustering with CMB and Union3, albeit with tighter errorbars. However, this is not an apples-apples comparison as we fix ($n_s,\omega_b,\theta_\star$) instead of including a CMB likelihood in our data combination. The lower right panel of Fig.~\ref{fig:w0wa_test} compares the $S_8$ posteriors between the $w_0w_a$CDM fit and our $\Lambda$CDM fit from the main body of the paper (fourth line in Tab.~\ref{table: constraints}) using the same $P_{\ell}(k)+\xi_{\ell}^{\rm post}(s)+\Ckg$ data combination but without Union3. While this again is not a direct apples-to-apples comparison (because the $\Lambda$CDM fit does not involve fixing $n_s,\omega_b,\theta_\star$), we include it in the figure to show that $S_8$ constraints are largely unaffected by $w_0$ and $w_a$ deviations from $\Lambda$CDM.  A similar result was also found in ref.~\cite{DESI2024.VII.KP7B}. As the focus of this paper is primarily about the clustering amplitude, we are justified in assuming $\Lambda$CDM in the body of this paper.  Future analyses of the DESI Y3 data set using the methods presented here will involve a more proper treatment of evolving dark energy, including the use of a neural network emulator based on HEFT.

\section{Gravitational Slip}
\label{app: slip}

\begin{figure}[htb]
    \centering
    \includegraphics[width=\linewidth]{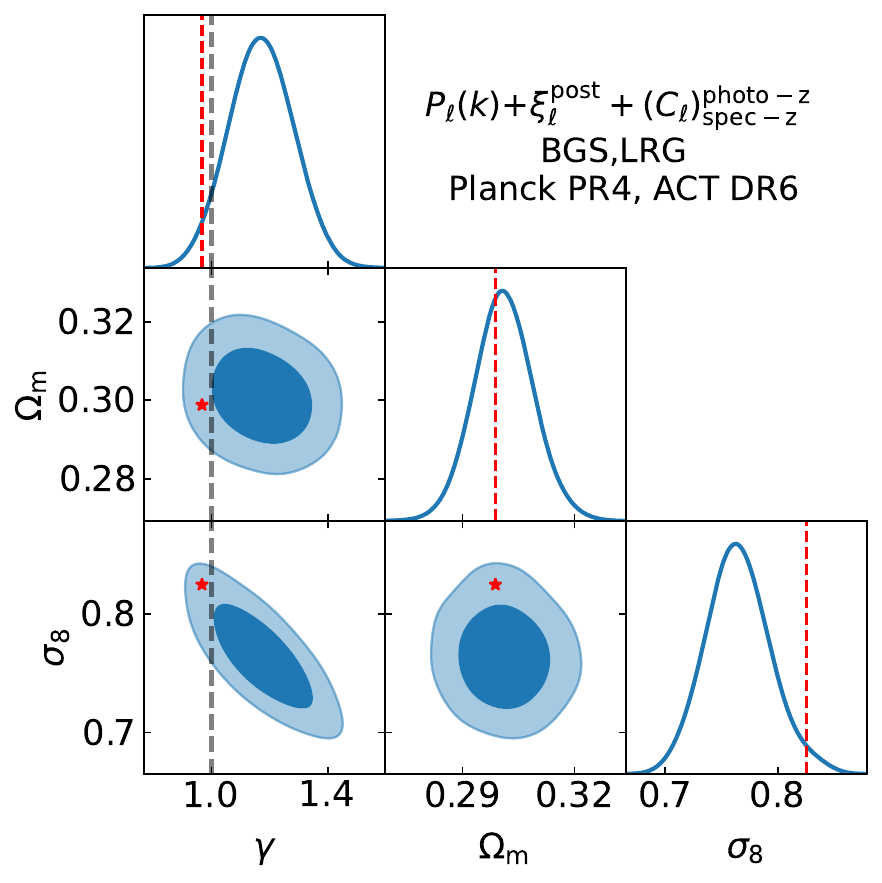}
    \caption{Constraints combining 3D full-shape and BAO analysis with $C_{\ell}$'s in the spectroscopic and photometric BGS and LRG bins while allowing the gravitaional slip parameter $\gamma$ to vary. The red dashed lines and stars show the MAP values obtained from running a minimizer, and indicate that projection effects are causing the shift to higher values of $\gamma$. 
    } 
    \label{fig:grav_slip}
\end{figure}

One simple way to test for modified gravity in our setup is through gravitational slip (see e.g.\ ref.\ \cite{Joyce2016} for a review), as was done in ref.~\cite{Chen22_2}. The gravitational interaction that affects massless photons comes from a different component of the space-time metric as that which affects massive objects. In standard general relativity these two components are equal but for some modified gravity models (e.g.\ the existence of a 5th force \cite{Grimm2025}) the Weyl potential that impacts photon trajectories can be different from the Newtonian potential which affects galaxy trajectories. Redshift space distortions are caused by the in-fall of target galaxies into gravitational wells and drive our amplitude constraints from 3D clustering. Meanwhile, the CMB lensing involves the trajectories of CMB photons being impacted by intervening mass distributions and is thus dependent on the Weyl potential. Since our joint analysis includes both RSD and the cross-correlations of galaxies with CMB-lensing, we are in a good position to test for differences in the Newtonian and Weyl potentials through the gravitational slip parameter $\gamma$ that is related to the relative strengths of the two potentials:
\begin{align}
    \Phi_{\gamma} \equiv \frac{\Phi+\Psi}{2}\equiv\Psi\frac{1+\gamma}{2} \quad , \quad \gamma = \frac{\Phi}{\Psi},
\end{align}
with $\Psi$ and $\Phi$ the scalar perturbations that appear in the time and spatial components of the metric respectively and $\Phi_{\gamma}$ is the Weyl potential that determines the trajectory of light. If $\gamma$ is allowed to vary away from 1 but taken to be spatially and temporally constant then the amplitude of $\Ckg$ is simply modulated by $c_\kappa\equiv(1+\gamma)/2$ and the magnification bias by $c_\kappa^2$. A measurement of $\gamma\neq1$ would indicate a departure from general relativity. We choose to vary $c_\kappa$ and show the implied constraints on $\gamma$ in Fig.~\ref{fig:grav_slip}, using the joint analysis setup involving $P_{\ell}(k)+\xi_{\ell}^{\rm post}(s)+C_{\ell}$ from the spectroscopic and photometric BGS and LRG samples. We find $\gamma=1.17\pm0.11$, in mild tension with the predictions from general relativity, along with low $\sigma_8 = 0.765^{+0.027}_{-0.030}$. However we obtain MAP values of $\gamma = 0.967$ and $\sigma_8 = 0.825$ when running a minimizer, suggesting that the marginal constraint on $\gamma$ suffers from projection effects heightened by the degeneracy with $\sigma_8$ as observed in the lowest left panel of Fig.~\ref{fig:grav_slip}. We also show the MAP values as red lines/stars in the figure to further illustrate this point.

\section{Author Affiliations}
\label{sec:affiliations}

\begin{hangparas}{.5cm}{1}

$^{1}${University of California, Berkeley, 110 Sproul Hall \#5800 Berkeley, CA 94720, USA}

$^{2}${Lawrence Berkeley National Laboratory, 1 Cyclotron Road, Berkeley, CA 94720, USA}

$^{3}${Department of Physics, University of California, Berkeley, 366 LeConte Hall MC 7300, Berkeley, CA 94720-7300, USA}

$^{4}${Institute for Advanced Study, 1 Einstein Drive, Princeton, NJ 08540, USA}

$^{5}${Physics Department, Brookhaven National Laboratory, Upton, NY 11973, USA}

$^{6}${Department of Physics, Boston University, 590 Commonwealth Avenue, Boston, MA 02215 USA}

$^{7}${Department of Physics \& Astronomy, University of Rochester, 206 Bausch and Lomb Hall, P.O. Box 270171, Rochester, NY 14627-0171, USA}

$^{8}${Dipartimento di Fisica ``Aldo Pontremoli'', Universit\`a degli Studi di Milano, Via Celoria 16, I-20133 Milano, Italy}

$^{9}${INAF-Osservatorio Astronomico di Brera, Via Brera 28, 20122 Milano, Italy}

$^{10}${Department of Physics \& Astronomy, University College London, Gower Street, London, WC1E 6BT, UK}

$^{11}${IRFU, CEA, Universit\'{e} Paris-Saclay, F-91191 Gif-sur-Yvette, France}

$^{12}${Institut d'Estudis Espacials de Catalunya (IEEC), c/ Esteve Terradas 1, Edifici RDIT, Campus PMT-UPC, 08860 Castelldefels, Spain}

$^{13}${Institute of Space Sciences, ICE-CSIC, Campus UAB, Carrer de Can Magrans s/n, 08913 Bellaterra, Barcelona, Spain}

$^{14}${Instituto de F\'{\i}sica, Universidad Nacional Aut\'{o}noma de M\'{e}xico,  Circuito de la Investigaci\'{o}n Cient\'{\i}fica, Ciudad Universitaria, Cd. de M\'{e}xico  C.~P.~04510,  M\'{e}xico}

$^{15}${Institut de F\'{i}sica d’Altes Energies (IFAE), The Barcelona Institute of Science and Technology, Edifici Cn, Campus UAB, 08193, Bellaterra (Barcelona), Spain}

$^{16}${Departamento de F\'isica, Universidad de los Andes, Cra. 1 No. 18A-10, Edificio Ip, CP 111711, Bogot\'a, Colombia}

$^{17}${Observatorio Astron\'omico, Universidad de los Andes, Cra. 1 No. 18A-10, Edificio H, CP 111711 Bogot\'a, Colombia}

$^{18}${Institute of Cosmology and Gravitation, University of Portsmouth, Dennis Sciama Building, Portsmouth, PO1 3FX, UK}

$^{19}${Fermi National Accelerator Laboratory, PO Box 500, Batavia, IL 60510, USA}

$^{20}${Center for Cosmology and AstroParticle Physics, The Ohio State University, 191 West Woodruff Avenue, Columbus, OH 43210, USA}

$^{21}${Department of Physics, The Ohio State University, 191 West Woodruff Avenue, Columbus, OH 43210, USA}

$^{22}${The Ohio State University, Columbus, 43210 OH, USA}

$^{23}${School of Mathematics and Physics, University of Queensland, Brisbane, QLD 4072, Australia}

$^{24}${Department of Physics, The University of Texas at Dallas, 800 W. Campbell Rd., Richardson, TX 75080, USA}

$^{25}${Department of Physics, Southern Methodist University, 3215 Daniel Avenue, Dallas, TX 75275, USA}

$^{26}${Department of Physics and Astronomy, University of California, Irvine, 92697, USA}

$^{27}${Center for Astrophysics $|$ Harvard \& Smithsonian, 60 Garden Street, Cambridge, MA 02138, USA}

$^{28}${Sorbonne Universit\'{e}, CNRS/IN2P3, Laboratoire de Physique Nucl\'{e}aire et de Hautes Energies (LPNHE), FR-75005 Paris, France}

$^{29}${Departament de F\'{i}sica, Serra H\'{u}nter, Universitat Aut\`{o}noma de Barcelona, 08193 Bellaterra (Barcelona), Spain}

$^{30}${NSF NOIRLab, 950 N. Cherry Ave., Tucson, AZ 85719, USA}

$^{31}${Instituci\'{o} Catalana de Recerca i Estudis Avan\c{c}ats, Passeig de Llu\'{\i}s Companys, 23, 08010 Barcelona, Spain}

$^{32}${Department of Physics \& Astronomy and Pittsburgh Particle Physics, Astrophysics, and Cosmology Center (PITT PACC), University of Pittsburgh, 3941 O'Hara Street, Pittsburgh, PA 15260, USA}

$^{33}${Department of Physics and Astronomy, University of Waterloo, 200 University Ave W, Waterloo, ON N2L 3G1, Canada}

$^{34}${Perimeter Institute for Theoretical Physics, 31 Caroline St. North, Waterloo, ON N2L 2Y5, Canada}

$^{35}${Waterloo Centre for Astrophysics, University of Waterloo, 200 University Ave W, Waterloo, ON N2L 3G1, Canada}

$^{36}${Instituto de Astrof\'{i}sica de Andaluc\'{i}a (CSIC), Glorieta de la Astronom\'{i}a, s/n, E-18008 Granada, Spain}

$^{37}${Departament de F\'isica, EEBE, Universitat Polit\`ecnica de Catalunya, c/Eduard Maristany 10, 08930 Barcelona, Spain}

$^{38}${Department of Astronomy, The Ohio State University, 4055 McPherson Laboratory, 140 W 18th Avenue, Columbus, OH 43210, USA}

$^{39}${Department of Physics and Astronomy, Sejong University, 209 Neungdong-ro, Gwangjin-gu, Seoul 05006, Republic of Korea}

$^{40}${Abastumani Astrophysical Observatory, Tbilisi, GE-0179, Georgia}

$^{41}${Department of Physics, Kansas State University, 116 Cardwell Hall, Manhattan, KS 66506, USA}

$^{42}${Faculty of Natural Sciences and Medicine, Ilia State University, 0194 Tbilisi, Georgia}

$^{43}${CIEMAT, Avenida Complutense 40, E-28040 Madrid, Spain}

$^{44}${Department of Physics, University of Michigan, 450 Church Street, Ann Arbor, MI 48109, USA}

$^{45}${University of Michigan, 500 S. State Street, Ann Arbor, MI 48109, USA}

$^{46}${Department of Physics \& Astronomy, Ohio University, 139 University Terrace, Athens, OH 45701, USA}

$^{47}${National Astronomical Observatories, Chinese Academy of Sciences, A20 Datun Road, Chaoyang District, Beijing, 100101, P.~R.~China}

\end{hangparas}

\bibliography{main}

\providecommand{\href}[2]{#2}\begingroup\raggedright\begin{thebibliography}{100}

\bibitem{P5_2024}
S.~{Asai}, A.~{Ballarino}, T.~{Bose}, K.~{Cranmer}, F.-Y.~{Cyr-Racine}, S.~{Demers} et~al., \emph{{Exploring the Quantum Universe: Pathways to Innovation and Discovery in Particle Physics}}, \href{https://doi.org/10.48550/arXiv.2407.19176}{\emph{arXiv e-prints} (2024) arXiv:2407.19176} [\href{https://arxiv.org/abs/2407.19176}{{\ttfamily 2407.19176}}].

\bibitem{Snowmass22_a}
R.~{Flauger}, V.~{Gorbenko}, A.~{Joyce}, L.~{McAllister}, G.~{Shiu} and E.~{Silverstein}, \emph{{Snowmass White Paper: Cosmology at the Theory Frontier}}, \href{https://doi.org/10.48550/arXiv.2203.07629}{\emph{arXiv e-prints} (2022) arXiv:2203.07629} [\href{https://arxiv.org/abs/2203.07629}{{\ttfamily 2203.07629}}].

\bibitem{Snowmass22_b}
G.~{Cabass}, M.M.~{Ivanov}, M.~{Lewandowski}, M.~{Mirbabayi} and M.~{Simonovi{\'c}}, \emph{{Snowmass white paper: Effective field theories in cosmology}}, \href{https://doi.org/10.1016/j.dark.2023.101193}{\emph{Physics of the Dark Universe} {\bfseries 40} (2023) 101193} [\href{https://arxiv.org/abs/2203.08232}{{\ttfamily 2203.08232}}].

\bibitem{Snowmass2013.Levi}
M.~{Levi}, C.~{Bebek}, T.~{Beers}, R.~{Blum}, R.~{Cahn}, D.~{Eisenstein} et~al., \emph{{The DESI Experiment, a whitepaper for Snowmass 2013}}, {\emph{arXiv e-prints} (2013) arXiv:1308.0847} [\href{https://arxiv.org/abs/1308.0847}{{\ttfamily 1308.0847}}].

\bibitem{LSST}
{LSST Science Collaboration}, P.A.~{Abell}, J.~{Allison}, S.F.~{Anderson}, J.R.~{Andrew}, J.R.P.~{Angel} et~al., \emph{{LSST Science Book, Version 2.0}}, {\emph{ArXiv e-prints} (2009) } [\href{https://arxiv.org/abs/0912.0201}{{\ttfamily 0912.0201}}].

\bibitem{LSSTDesc}
{LSST Dark Energy Science Collaboration}, \emph{{Large Synoptic Survey Telescope: Dark Energy Science Collaboration}}, \href{https://doi.org/10.48550/arXiv.1211.0310}{\emph{arXiv e-prints} (2012) arXiv:1211.0310} [\href{https://arxiv.org/abs/1211.0310}{{\ttfamily 1211.0310}}].

\bibitem{LSSTDesc2}
{\scshape LSST Dark Energy Science} collaboration, \emph{{The LSST Dark Energy Science Collaboration (DESC) Science Requirements Document}},  \href{https://arxiv.org/abs/1809.01669}{{\ttfamily 1809.01669}}.

\bibitem{Euclid}
R.~{Laureijs}, J.~{Amiaux}, S.~{Arduini}, J..~{Augu{\`e}res}, J.~{Brinchmann}, R.~{Cole} et~al., \emph{{Euclid Definition Study Report}}, {\emph{ArXiv e-prints} (2011) } [\href{https://arxiv.org/abs/1110.3193}{{\ttfamily 1110.3193}}].

\bibitem{Amendola18}
L.~{Amendola}, S.~{Appleby}, A.~{Avgoustidis}, D.~{Bacon}, T.~{Baker}, M.~{Baldi} et~al., \emph{{Cosmology and fundamental physics with the Euclid satellite}}, \href{https://doi.org/10.1007/s41114-017-0010-3}{\emph{Living Reviews in Relativity} {\bfseries 21} (2018) 2} [\href{https://arxiv.org/abs/1606.00180}{{\ttfamily 1606.00180}}].

\bibitem{Spherex}
O.~{Dor{\'e}}, J.~{Bock}, M.~{Ashby}, P.~{Capak}, A.~{Cooray}, R.~{de Putter} et~al., \emph{{Cosmology with the SPHEREX All-Sky Spectral Survey}}, \href{https://doi.org/10.48550/arXiv.1412.4872}{\emph{arXiv e-prints} (2014) arXiv:1412.4872} [\href{https://arxiv.org/abs/1412.4872}{{\ttfamily 1412.4872}}].

\bibitem{DESI}
{DESI Collaboration}, A.~{Aghamousa}, J.~{Aguilar}, S.~{Ahlen}, S.~{Alam}, L.E.~{Allen} et~al., \emph{{The DESI Experiment Part I: Science,Targeting, and Survey Design}}, {\emph{ArXiv e-prints} (2016) } [\href{https://arxiv.org/abs/1611.00036}{{\ttfamily 1611.00036}}].

\bibitem{Spectro.Pipeline.Guy.2023}
J.~{Guy}, S.~{Bailey}, A.~{Kremin}, S.~{Alam}, D.M.~{Alexander}, C.~{Allende Prieto} et~al., \emph{{The Spectroscopic Data Processing Pipeline for the Dark Energy Spectroscopic Instrument}}, \href{https://doi.org/10.3847/1538-3881/acb212}{\emph{\aj} {\bfseries 165} (2023) 144} [\href{https://arxiv.org/abs/2209.14482}{{\ttfamily 2209.14482}}].

\bibitem{SurveyOps.Schlafly.2023}
E.F.~{Schlafly}, D.~{Kirkby}, D.J.~{Schlegel}, A.D.~{Myers}, A.~{Raichoor}, K.~{Dawson} et~al., \emph{{Survey Operations for the Dark Energy Spectroscopic Instrument}}, \href{https://doi.org/10.3847/1538-3881/ad0832}{\emph{\aj} {\bfseries 166} (2023) 259} [\href{https://arxiv.org/abs/2306.06309}{{\ttfamily 2306.06309}}].

\bibitem{DESI2022.KP1.Instr}
{DESI Collaboration}, B.~{Abareshi}, J.~{Aguilar}, S.~{Ahlen}, S.~{Alam}, D.M.~{Alexander} et~al., \emph{{Overview of the Instrumentation for the Dark Energy Spectroscopic Instrument}}, \href{https://doi.org/10.3847/1538-3881/ac882b}{\emph{\aj} {\bfseries 164} (2022) 207} [\href{https://arxiv.org/abs/2205.10939}{{\ttfamily 2205.10939}}].

\bibitem{DESI2016b.Instr}
{DESI Collaboration}, A.~{Aghamousa}, J.~{Aguilar}, S.~{Ahlen}, S.~{Alam}, L.E.~{Allen} et~al., \emph{{The DESI Experiment Part II: Instrument Design}}, {\emph{arXiv e-prints} (2016) arXiv:1611.00037} [\href{https://arxiv.org/abs/1611.00037}{{\ttfamily 1611.00037}}].

\bibitem{FocalPlane.Silber.2023}
J.H.~{Silber}, P.~{Fagrelius}, K.~{Fanning}, M.~{Schubnell}, J.N.~{Aguilar}, S.~{Ahlen} et~al., \emph{{The Robotic Multiobject Focal Plane System of the Dark Energy Spectroscopic Instrument (DESI)}}, \href{https://doi.org/10.3847/1538-3881/ac9ab1}{\emph{\aj} {\bfseries 165} (2023) 9} [\href{https://arxiv.org/abs/2205.09014}{{\ttfamily 2205.09014}}].

\bibitem{Corrector.Miller.2023}
T.N.~{Miller}, P.~{Doel}, G.~{Gutierrez}, R.~{Besuner}, D.~{Brooks}, G.~{Gallo} et~al., \emph{{The Optical Corrector for the Dark Energy Spectroscopic Instrument}}, \href{https://doi.org/10.3847/1538-3881/ad45fe}{\emph{\aj} {\bfseries 168} (2024) 95} [\href{https://arxiv.org/abs/2306.06310}{{\ttfamily 2306.06310}}].

\bibitem{DESI2024.II.KP3}
{DESI Collaboration}, A.G.~{Adame}, J.~{Aguilar}, S.~{Ahlen}, S.~{Alam}, D.M.~{Alexander} et~al., \emph{{DESI 2024 II: Sample Definitions, Characteristics, and Two-point Clustering Statistics}}, \href{https://doi.org/10.48550/arXiv.2411.12020}{\emph{arXiv e-prints} (2024) arXiv:2411.12020} [\href{https://arxiv.org/abs/2411.12020}{{\ttfamily 2411.12020}}].

\bibitem{Eisenstein_recon2007}
D.J.~{Eisenstein}, H.-J.~{Seo}, E.~{Sirko} and D.N.~{Spergel}, \emph{{Improving Cosmological Distance Measurements by Reconstruction of the Baryon Acoustic Peak}}, \href{https://doi.org/10.1086/518712}{\emph{\apj} {\bfseries 664} (2007) 675} [\href{https://arxiv.org/abs/astro-ph/0604362}{{\ttfamily astro-ph/0604362}}].

\bibitem{Kaiser87}
N.~{Kaiser}, \emph{{Clustering in real space and in redshift space}}, \href{https://doi.org/10.1093/mnras/227.1.1}{\emph{\mnras} {\bfseries 227} (1987) 1}.

\bibitem{Hamilton92}
A.J.S.~{Hamilton}, \emph{{Measuring Omega and the real correlation function from the redshift correlation function}}, \href{https://doi.org/10.1086/186264}{\emph{\apjl} {\bfseries 385} (1992) L5}.

\bibitem{KP4s2-Chen}
S.F.~{Chen}, C.~{Howlett}, M.~{White}, P.~{McDonald}, A.J.~{Ross}, H.J.~{Seo} et~al., \emph{{Baryon acoustic oscillation theory and modelling systematics for the DESI 2024 results}}, \href{https://doi.org/10.1093/mnras/stae2090}{\emph{\mnras} {\bfseries 534} (2024) 544} [\href{https://arxiv.org/abs/2402.14070}{{\ttfamily 2402.14070}}].

\bibitem{KP4s3-Chen}
X.~{Chen}, Z.~{Ding}, E.~{Paillas}, S.~{Nadathur}, H.~{Seo}, S.~{Chen} et~al., \emph{{Extensive analysis of reconstruction algorithms for DESI 2024 baryon acoustic oscillations}}, \href{https://doi.org/10.48550/arXiv.2411.19738}{\emph{arXiv e-prints} (2024) arXiv:2411.19738} [\href{https://arxiv.org/abs/2411.19738}{{\ttfamily 2411.19738}}].

\bibitem{KP4s4-Paillas}
E.~{Paillas}, Z.~{Ding}, X.~{Chen}, H.~{Seo}, N.~{Padmanabhan}, A.~{de Mattia} et~al., \emph{{Optimal reconstruction of baryon acoustic oscillations for DESI 2024}}, \href{https://doi.org/10.1088/1475-7516/2025/01/142}{\emph{\jcap} {\bfseries 2025} (2025) 142} [\href{https://arxiv.org/abs/2404.03005}{{\ttfamily 2404.03005}}].

\bibitem{KP4s5-Valcin}
{D.~Valcin et al.}, \emph{{Combined tracer analysis for DESI 2024 BAO analysis}}, {\emph{in preparation} (2025) }.

\bibitem{KP4s6-Forero-Sanchez}
D.~{Forero-S{\'a}nchez}, M.~{Rashkovetskyi}, O.~{Alves}, A.~{de Mattia}, S.~{Nadathur}, P.~{Zarrouk} et~al., \emph{{Analytical and EZmock covariance validation for the DESI 2024 results}}, \href{https://doi.org/10.48550/arXiv.2411.12027}{\emph{arXiv e-prints} (2024) arXiv:2411.12027} [\href{https://arxiv.org/abs/2411.12027}{{\ttfamily 2411.12027}}].

\bibitem{KP4s7-Rashkovetskyi}
M.~{Rashkovetskyi}, D.~{Forero-S{\'a}nchez}, A.~{de Mattia}, D.J.~{Eisenstein}, N.~{Padmanabhan}, H.~{Seo} et~al., \emph{{Semi-analytical covariance matrices for two-point correlation function for DESI 2024 data}}, \href{https://doi.org/10.1088/1475-7516/2025/01/145}{\emph{\jcap} {\bfseries 2025} (2025) 145} [\href{https://arxiv.org/abs/2404.03007}{{\ttfamily 2404.03007}}].

\bibitem{KP4s8-Alves}
{O.~Alves et al.}, \emph{{Analytical covariance matrices of DESI galaxy power spectra}}, {\emph{in preparation} (2025) }.

\bibitem{KP4s9-Perez-Fernandez}
A.~{P{\'e}rez-Fern{\'a}ndez}, L.~{Medina-Varela}, R.~{Ruggeri}, M.~{Vargas-Maga{\~n}a}, H.~{Seo}, N.~{Padmanabhan} et~al., \emph{{Fiducial-cosmology-dependent systematics for the DESI 2024 BAO analysis}}, \href{https://doi.org/10.1088/1475-7516/2025/01/144}{\emph{\jcap} {\bfseries 2025} (2025) 144} [\href{https://arxiv.org/abs/2406.06085}{{\ttfamily 2406.06085}}].

\bibitem{KP4s10-Mena-Fernandez}
J.~{Mena-Fern{\'a}ndez}, C.~{Garcia-Quintero}, S.~{Yuan}, B.~{Hadzhiyska}, O.~{Alves}, M.~{Rashkovetskyi} et~al., \emph{{HOD-dependent systematics for luminous red galaxies in the DESI 2024 BAO analysis}}, \href{https://doi.org/10.1088/1475-7516/2025/01/133}{\emph{\jcap} {\bfseries 2025} (2025) 133} [\href{https://arxiv.org/abs/2404.03008}{{\ttfamily 2404.03008}}].

\bibitem{KP4s11-Garcia-Quintero}
C.~{Garcia-Quintero}, J.~{Mena-Fern{\'a}ndez}, A.~{Rocher}, S.~{Yuan}, B.~{Hadzhiyska}, O.~{Alves} et~al., \emph{{HOD-dependent systematics in Emission Line Galaxies for the DESI 2024 BAO analysis}}, \href{https://doi.org/10.1088/1475-7516/2025/01/132}{\emph{\jcap} {\bfseries 2025} (2025) 132} [\href{https://arxiv.org/abs/2404.03009}{{\ttfamily 2404.03009}}].

\bibitem{KP5s1-Maus}
M.~{Maus}, Y.~{Lai}, H.E.~{Noriega}, S.~{Ramirez-Solano}, A.~{Aviles}, S.~{Chen} et~al., \emph{{A comparison of effective field theory models of redshift space galaxy power spectra for DESI 2024 and future surveys}}, \href{https://doi.org/10.1088/1475-7516/2025/01/134}{\emph{\jcap} {\bfseries 2025} (2025) 134} [\href{https://arxiv.org/abs/2404.07272}{{\ttfamily 2404.07272}}].

\bibitem{KP5s2-Maus}
M.~{Maus}, S.~{Chen}, M.~{White}, J.~{Aguilar}, S.~{Ahlen}, A.~{Aviles} et~al., \emph{{An analysis of parameter compression and Full-Modeling techniques with Velocileptors for DESI 2024 and beyond}}, \href{https://doi.org/10.1088/1475-7516/2025/01/138}{\emph{\jcap} {\bfseries 2025} (2025) 138} [\href{https://arxiv.org/abs/2404.07312}{{\ttfamily 2404.07312}}].

\bibitem{KP5s3-Noriega}
H.E.~{Noriega}, A.~{Aviles}, H.~{Gil-Mar{\'\i}n}, S.~{Ramirez-Solano}, S.~{Fromenteau}, M.~{Vargas-Maga{\~n}a} et~al., \emph{{Comparing Compressed and Full-Modeling analyses with FOLPS: implications for DESI 2024 and beyond}}, \href{https://doi.org/10.1088/1475-7516/2025/01/136}{\emph{\jcap} {\bfseries 2025} (2025) 136} [\href{https://arxiv.org/abs/2404.07269}{{\ttfamily 2404.07269}}].

\bibitem{KP5s4-Lai}
Y.~{Lai}, C.~{Howlett}, M.~{Maus}, H.~{Gil-Mar{\'\i}n}, H.E.~{Noriega}, S.~{Ram{\'\i}rez-Solano} et~al., \emph{{A comparison between ShapeFit compression and Full-Modelling method with PyBird for DESI 2024 and beyond}}, \href{https://doi.org/10.1088/1475-7516/2025/01/139}{\emph{\jcap} {\bfseries 2025} (2025) 139} [\href{https://arxiv.org/abs/2404.07283}{{\ttfamily 2404.07283}}].

\bibitem{KP5s5-Ramirez}
S.~{Ramirez-Solano}, M.~{Icaza-Lizaola}, H.E.~{Noriega}, M.~{Vargas-Maga{\~n}a}, S.~{Fromenteau}, A.~{Aviles} et~al., \emph{{Full Modeling and parameter compression methods in configuration space for DESI 2024 and beyond}}, \href{https://doi.org/10.1088/1475-7516/2025/01/129}{\emph{\jcap} {\bfseries 2025} (2025) 129} [\href{https://arxiv.org/abs/2404.07268}{{\ttfamily 2404.07268}}].

\bibitem{KP5s6-Zhao}
{R.~Zhao et al.}, \emph{{Impact and mitigation of imaging systematics for DESI 2024 full shape analysis}}, {\emph{in preparation} (2025) }.

\bibitem{DESI2024.III.KP4}
{DESI Collaboration}, A.G.~Adame, J.~Aguilar, S.~Ahlen, S.~Alam, D.M.~Alexander et~al., \emph{{DESI 2024 III: Baryon Acoustic Oscillations from Galaxies and Quasars}}, \href{https://doi.org/10.48550/arXiv.2404.03000}{\emph{arXiv e-prints} (2024) arXiv:2404.03000} [\href{https://arxiv.org/abs/2404.03000}{{\ttfamily 2404.03000}}].

\bibitem{DESI2024.V.KP5}
{DESI Collaboration}, A.G.~{Adame}, J.~{Aguilar}, S.~{Ahlen}, S.~{Alam}, D.M.~{Alexander} et~al., \emph{{DESI 2024 V: Full-Shape Galaxy Clustering from Galaxies and Quasars}}, \href{https://doi.org/10.48550/arXiv.2411.12021}{\emph{arXiv e-prints} (2024) arXiv:2411.12021} [\href{https://arxiv.org/abs/2411.12021}{{\ttfamily 2411.12021}}].

\bibitem{DESI2024.VI.KP7A}
{DESI Collaboration}, A.G.~{Adame}, J.~{Aguilar}, S.~{Ahlen}, S.~{Alam}, D.M.~{Alexander} et~al., \emph{{DESI 2024 VI: cosmological constraints from the measurements of baryon acoustic oscillations}}, \href{https://doi.org/10.1088/1475-7516/2025/02/021}{\emph{\jcap} {\bfseries 2025} (2025) 021} [\href{https://arxiv.org/abs/2404.03002}{{\ttfamily 2404.03002}}].

\bibitem{DESI2024.VII.KP7B}
{DESI Collaboration}, A.G.~{Adame}, J.~{Aguilar}, S.~{Ahlen}, S.~{Alam}, D.M.~{Alexander} et~al., \emph{{DESI 2024 VII: Cosmological Constraints from the Full-Shape Modeling of Clustering Measurements}}, \href{https://doi.org/10.48550/arXiv.2411.12022}{\emph{arXiv e-prints} (2024) arXiv:2411.12022} [\href{https://arxiv.org/abs/2411.12022}{{\ttfamily 2411.12022}}].

\bibitem{Sailer24}
N.~{Sailer}, J.~{Kim}, S.~{Ferraro}, M.S.~{Madhavacheril}, M.~{White}, I.~{Abril-Cabezas} et~al., \emph{{Cosmological constraints from the cross-correlation of DESI Luminous Red Galaxies with CMB lensing from Planck PR4 and ACT DR6}}, \href{https://doi.org/10.48550/arXiv.2407.04607}{\emph{arXiv e-prints} (2024) arXiv:2407.04607} [\href{https://arxiv.org/abs/2407.04607}{{\ttfamily 2407.04607}}].

\bibitem{Kim2024}
J.~{Kim}, N.~{Sailer}, M.S.~{Madhavacheril}, S.~{Ferraro}, I.~{Abril-Cabezas}, J.N.~{Aguilar} et~al., \emph{{The Atacama Cosmology Telescope DR6 and DESI: structure formation over cosmic time with a measurement of the cross-correlation of CMB lensing and luminous red galaxies}}, \href{https://doi.org/10.1088/1475-7516/2024/12/022}{\emph{\jcap} {\bfseries 2024} (2024) 022} [\href{https://arxiv.org/abs/2407.04606}{{\ttfamily 2407.04606}}].

\bibitem{Sailer2025}
N.~{Sailer}, J.~{DeRose}, S.~{Ferraro}, S.-F.~{Chen}, R.~{Zhou}, M.~{White} et~al., \emph{{Evolution of structure growth during dark energy domination: insights from the cross-correlation of DESI galaxies with CMB lensing and galaxy magnification}}, {\emph{arXiv e-prints} (2025) arXiv:2503.24385} [\href{https://arxiv.org/abs/2503.24385}{{\ttfamily 2503.24385}}].

\bibitem{Karim2025}
T.~{Karim}, S.~{Singh}, M.~{Rezaie}, D.~{Eisenstein}, B.~{Hadzhiyska}, J.S.~{Speagle} et~al., \emph{{Measuring {\ensuremath{\sigma}} $_{8}$ using DESI Legacy Imaging Surveys Emission-Line galaxies and Planck CMB lensing, and the impact of dust on parameter inference}}, \href{https://doi.org/10.1088/1475-7516/2025/02/045}{\emph{\jcap} {\bfseries 2025} (2025) 045} [\href{https://arxiv.org/abs/2408.15909}{{\ttfamily 2408.15909}}].

\bibitem{pullenConstrainingGravityLargest2016a}
A.R.~Pullen, S.~Alam, S.~He and S.~Ho, \emph{Constraining {Gravity} at the {Largest} {Scales} through {CMB} {Lensing} and {Galaxy} {Velocities}}, \href{https://doi.org/10.1093/mnras/stw1249}{\emph{Monthly Notices of the Royal Astronomical Society} {\bfseries 460} (2016) 4098}.

\bibitem{singhProbingGravityJoint2019}
S.~Singh, S.~Alam, R.~Mandelbaum, U.~Seljak, S.~Rodriguez-Torres and S.~Ho, \emph{Probing gravity with a joint analysis of galaxy and {CMB} lensing and {SDSS} spectroscopy}, \href{https://doi.org/10.1093/mnras/sty2681}{\emph{Monthly Notices of the Royal Astronomical Society} {\bfseries 482} (2019) 785}.

\bibitem{wenzlAtacamaCosmologyTelescope2024}
L.~Wenzl, R.~An, N.~Battaglia, R.~Bean, E.~Calabrese, S.-F.~Chen et~al., \emph{The {Atacama} {Cosmology} {Telescope}: {DR6} {Gravitational} {Lensing} and {SDSS} {BOSS} cross-correlation measurement and constraints on gravity with the \${E}\_g\$ statistic},  May, 2024.
\newblock 10.48550/arXiv.2405.12795.

\bibitem{wenzlConstrainingGravityNew2024}
L.~Wenzl, R.~Bean, S.-F.~Chen, G.S.~Farren, M.S.~Madhavacheril, G.A.~Marques et~al., \emph{Constraining gravity with a new precision \${E}\_g\$ estimator using {Planck} + {SDSS} {BOSS}}, \href{https://doi.org/10.1103/PhysRevD.109.083540}{\emph{Physical Review D} {\bfseries 109} (2024) 083540}.

\bibitem{Chen24}
S.-F.~{Chen}, M.M.~{Ivanov}, O.H.E.~{Philcox} and L.~{Wenzl}, \emph{{Suppression without Thawing: Constraining Structure Formation and Dark Energy with Galaxy Clustering}}, \href{https://doi.org/10.1103/PhysRevLett.133.231001}{\emph{\prl} {\bfseries 133} (2024) 231001} [\href{https://arxiv.org/abs/2406.13388}{{\ttfamily 2406.13388}}].

\bibitem{Chen22_2}
S.-F.~{Chen}, M.~{White}, J.~{DeRose} and N.~{Kokron}, \emph{{Cosmological analysis of three-dimensional BOSS galaxy clustering and Planck CMB lensing cross correlations via Lagrangian perturbation theory}}, \href{https://doi.org/10.1088/1475-7516/2022/07/041}{\emph{\jcap} {\bfseries 2022} (2022) 041} [\href{https://arxiv.org/abs/2204.10392}{{\ttfamily 2204.10392}}].

\bibitem{deBelsunce2025}
R.~{de Belsunce}, A.~{Krolewski}, S.~{Chiarenza}, E.~{Chaussidon}, S.~{Ferraro}, B.~{Hadzhiyska} et~al., \emph{{Cosmology from Planck CMB Lensing and DESI DR1 Quasar Tomography}}, \href{https://doi.org/10.48550/arXiv.2506.22416}{\emph{arXiv e-prints} (2025) arXiv:2506.22416} [\href{https://arxiv.org/abs/2506.22416}{{\ttfamily 2506.22416}}].

\bibitem{DESI_legacy}
A.~{Dey}, D.J.~{Schlegel}, D.~{Lang}, R.~{Blum}, K.~{Burleigh}, X.~{Fan} et~al., \emph{{Overview of the DESI Legacy Imaging Surveys}}, \href{https://doi.org/10.3847/1538-3881/ab089d}{\emph{\aj} {\bfseries 157} (2019) 168} [\href{https://arxiv.org/abs/1804.08657}{{\ttfamily 1804.08657}}].

\bibitem{DESI2024.I.DR1}
{DESI Collaboration}, M.~{Abdul-Karim}, A.G.~{Adame}, D.~{Aguado}, J.~{Aguilar}, S.~{Ahlen} et~al., \emph{{Data Release 1 of the Dark Energy Spectroscopic Instrument}}, \href{https://doi.org/10.48550/arXiv.2503.14745}{\emph{arXiv e-prints} (2025) arXiv:2503.14745} [\href{https://arxiv.org/abs/2503.14745}{{\ttfamily 2503.14745}}].

\bibitem{LRG.TS.Zhou.2023}
R.~{Zhou}, B.~{Dey}, J.A.~{Newman}, D.J.~{Eisenstein}, K.~{Dawson}, S.~{Bailey} et~al., \emph{{Target Selection and Validation of DESI Luminous Red Galaxies}}, \href{https://doi.org/10.3847/1538-3881/aca5fb}{\emph{\aj} {\bfseries 165} (2023) 58} [\href{https://arxiv.org/abs/2208.08515}{{\ttfamily 2208.08515}}].

\bibitem{DESIDR2_bao}
{DESI Collaboration}, M.A.~{Karim}, J.~{Aguilar}, S.~{Ahlen}, S.~{Alam}, L.~{Allen} et~al., \emph{{DESI DR2 Results II: Measurements of Baryon Acoustic Oscillations and Cosmological Constraints}}, \href{https://doi.org/10.48550/arXiv.2503.14738}{\emph{arXiv e-prints} (2025) arXiv:2503.14738} [\href{https://arxiv.org/abs/2503.14738}{{\ttfamily 2503.14738}}].

\bibitem{Zhou23}
R.~{Zhou}, S.~{Ferraro}, M.~{White}, J.~{DeRose}, N.~{Sailer}, J.~{Aguilar} et~al., \emph{{DESI luminous red galaxy samples for cross-correlations}}, \href{https://doi.org/10.1088/1475-7516/2023/11/097}{\emph{\jcap} {\bfseries 2023} (2023) 097} [\href{https://arxiv.org/abs/2309.06443}{{\ttfamily 2309.06443}}].

\bibitem{ChenDeRose24}
S.~{Chen}, J.~{DeRose}, R.~{Zhou}, M.~{White}, S.~{Ferraro}, C.~{Blake} et~al., \emph{{Not all lensing is low: An analysis of DESI$\times$DES using the Lagrangian Effective Theory of LSS}}, \href{https://doi.org/10.48550/arXiv.2407.04795}{\emph{arXiv e-prints} (2024) arXiv:2407.04795} [\href{https://arxiv.org/abs/2407.04795}{{\ttfamily 2407.04795}}].

\bibitem{Carron_2022}
J.~Carron, M.~Mirmelstein and A.~Lewis, \emph{Cmb lensing from planck pr4 maps}, \href{https://doi.org/10.1088/1475-7516/2022/09/039}{\emph{Journal of Cosmology and Astroparticle Physics} {\bfseries 2022} (2022) 039}.

\bibitem{ACT:2023dou}
F.J.~{Qu}, B.D.~{Sherwin}, M.S.~{Madhavacheril}, D.~{Han}, K.T.~{Crowley}, I.~{Abril-Cabezas} et~al., \emph{{The Atacama Cosmology Telescope: A Measurement of the DR6 CMB Lensing Power Spectrum and Its Implications for Structure Growth}}, \href{https://doi.org/10.3847/1538-4357/acfe06}{\emph{\apj} {\bfseries 962} (2024) 112} [\href{https://arxiv.org/abs/2304.05202}{{\ttfamily 2304.05202}}].

\bibitem{ACT:2023ubw}
N.~{MacCrann}, B.D.~{Sherwin}, F.J.~{Qu}, T.~{Namikawa}, M.S.~{Madhavacheril}, I.~{Abril-Cabezas} et~al., \emph{{The Atacama Cosmology Telescope: Mitigating the Impact of Extragalactic Foregrounds for the DR6 Cosmic Microwave Background Lensing Analysis}}, \href{https://doi.org/10.3847/1538-4357/ad2610}{\emph{\apj} {\bfseries 966} (2024) 138} [\href{https://arxiv.org/abs/2304.05196}{{\ttfamily 2304.05196}}].

\bibitem{ACT:2023kun}
M.S.~{Madhavacheril}, F.J.~{Qu}, B.D.~{Sherwin}, N.~{MacCrann}, Y.~{Li}, I.~{Abril-Cabezas} et~al., \emph{{The Atacama Cosmology Telescope: DR6 Gravitational Lensing Map and Cosmological Parameters}}, \href{https://doi.org/10.3847/1538-4357/acff5f}{\emph{\apj} {\bfseries 962} (2024) 113} [\href{https://arxiv.org/abs/2304.05203}{{\ttfamily 2304.05203}}].

\bibitem{Maniyar:2021msb}
A.S.~Maniyar, Y.~Ali-Ha\"\i{}moud, J.~Carron, A.~Lewis and M.S.~Madhavacheril, \emph{{Quadratic estimators for CMB weak lensing}}, \href{https://doi.org/10.1103/PhysRevD.103.083524}{\emph{Phys. Rev. D} {\bfseries 103} (2021) 083524} [\href{https://arxiv.org/abs/2101.12193}{{\ttfamily 2101.12193}}].

\bibitem{Madhavacheril:2020ido}
M.S.~Madhavacheril, K.M.~Smith, B.D.~Sherwin and S.~Naess, \emph{{CMB lensing power spectrum estimation without instrument noise bias}},  \href{https://arxiv.org/abs/2011.02475}{{\ttfamily 2011.02475}}.

\bibitem{2013MNRAS.431..609N}
T.~{Namikawa}, D.~{Hanson} and R.~{Takahashi}, \emph{{Bias-hardened CMB lensing}}, \href{https://doi.org/10.1093/mnras/stt195}{\emph{\mnras} {\bfseries 431} (2013) 609} [\href{https://arxiv.org/abs/1209.0091}{{\ttfamily 1209.0091}}].

\bibitem{Osborne:2013nna}
S.J.~Osborne, D.~Hanson and O.~Dor\'e, \emph{{Extragalactic Foreground Contamination in Temperature-based CMB Lens Reconstruction}}, \href{https://doi.org/10.1088/1475-7516/2014/03/024}{\emph{JCAP} {\bfseries 03} (2014) 024} [\href{https://arxiv.org/abs/1310.7547}{{\ttfamily 1310.7547}}].

\bibitem{Sailer:2020lal}
N.~Sailer, E.~Schaan and S.~Ferraro, \emph{{Lower bias, lower noise CMB lensing with foreground-hardened estimators}}, \href{https://doi.org/10.1103/PhysRevD.102.063517}{\emph{Phys. Rev. D} {\bfseries 102} (2020) 063517} [\href{https://arxiv.org/abs/2007.04325}{{\ttfamily 2007.04325}}].

\bibitem{Sailer:2022jwt}
N.~Sailer, S.~Ferraro and E.~Schaan, \emph{{Foreground-immune CMB lensing reconstruction with polarization}}, \href{https://doi.org/10.1103/PhysRevD.107.023504}{\emph{Phys. Rev. D} {\bfseries 107} (2023) 023504} [\href{https://arxiv.org/abs/2211.03786}{{\ttfamily 2211.03786}}].

\bibitem{Maksimova21}
N.A.~Maksimova, L.H.~Garrison, D.J.~Eisenstein, B.~Hadzhiyska, S.~Bose and T.P.~Satterthwaite, \emph{{AbacusSummit: a massive set of high-accuracy, high-resolution N-body simulations}}, \href{https://doi.org/10.1093/mnras/stab2484}{\emph{Monthly Notices of the Royal Astronomical Society} {\bfseries 508} (2021) 4017} [\href{https://arxiv.org/abs/https://academic.oup.com/mnras/article-pdf/508/3/4017/40811763/stab2484.pdf}{{\ttfamily https://academic.oup.com/mnras/article-pdf/508/3/4017/40811763/stab2484.pdf}}].

\bibitem{Garrison21}
L.H.~Garrison, D.J.~Eisenstein, D.~Ferrer, N.A.~Maksimova and P.A.~Pinto, \emph{{The abacus cosmological N-body code}}, \href{https://doi.org/10.1093/mnras/stab2482}{\emph{Monthly Notices of the Royal Astronomical Society} {\bfseries 508} (2021) 575} [\href{https://arxiv.org/abs/https://academic.oup.com/mnras/article-pdf/508/1/575/40458823/stab2482.pdf}{{\ttfamily https://academic.oup.com/mnras/article-pdf/508/1/575/40458823/stab2482.pdf}}].

\bibitem{Padmanabhan_recon2009}
N.~{Padmanabhan}, M.~{White} and J.D.~{Cohn}, \emph{{Reconstructing baryon oscillations: A Lagrangian theory perspective}}, \href{https://doi.org/10.1103/PhysRevD.79.063523}{\emph{\prd} {\bfseries 79} (2009) 063523} [\href{https://arxiv.org/abs/0812.2905}{{\ttfamily 0812.2905}}].

\bibitem{Noh_recon2009}
Y.~{Noh}, M.~{White} and N.~{Padmanabhan}, \emph{{Reconstructing baryon oscillations}}, \href{https://doi.org/10.1103/PhysRevD.80.123501}{\emph{\prd} {\bfseries 80} (2009) 123501} [\href{https://arxiv.org/abs/0909.1802}{{\ttfamily 0909.1802}}].

\bibitem{White_recon2015}
M.~{White}, \emph{{Reconstruction within the Zeldovich approximation}}, \href{https://doi.org/10.1093/mnras/stv842}{\emph{\mnras} {\bfseries 450} (2015) 3822} [\href{https://arxiv.org/abs/1504.03677}{{\ttfamily 1504.03677}}].

\bibitem{Chen_recon2019}
S.-F.~{Chen}, Z.~{Vlah} and M.~{White}, \emph{{The reconstructed power spectrum in the Zeldovich approximation}}, \href{https://doi.org/10.1088/1475-7516/2019/09/017}{\emph{\jcap} {\bfseries 2019} (2019) 017} [\href{https://arxiv.org/abs/1907.00043}{{\ttfamily 1907.00043}}].

\bibitem{Chen_BOSSrecon2022}
S.-F.~{Chen}, Z.~{Vlah} and M.~{White}, \emph{{A new analysis of galaxy 2-point functions in the BOSS survey, including full-shape information and post-reconstruction BAO}}, \href{https://doi.org/10.1088/1475-7516/2022/02/008}{\emph{\jcap} {\bfseries 2022} (2022) 008} [\href{https://arxiv.org/abs/2110.05530}{{\ttfamily 2110.05530}}].

\bibitem{Taylor22}
P.L.~{Taylor} and K.~{Markovi{\v{c}}}, \emph{{The Covariance of Photometric and Spectroscopic Two-Point Statistics: Implications for Cosmological Parameter Inference}}, {\emph{arXiv e-prints} (2022) arXiv:2205.14167} [\href{https://arxiv.org/abs/2205.14167}{{\ttfamily 2205.14167}}].

\bibitem{PlanckParams18}
{Planck Collaboration}, N.~{Aghanim}, Y.~{Akrami}, M.~{Ashdown}, J.~{Aumont}, C.~{Baccigalupi} et~al., \emph{{Planck 2018 results. VI. Cosmological parameters}}, \href{https://doi.org/10.1051/0004-6361/201833910}{\emph{\aap} {\bfseries 641} (2020) A6} [\href{https://arxiv.org/abs/1807.06209}{{\ttfamily 1807.06209}}].

\bibitem{FKP1994}
H.A.~{Feldman}, N.~{Kaiser} and J.A.~{Peacock}, \emph{{Power-Spectrum Analysis of Three-dimensional Redshift Surveys}}, \href{https://doi.org/10.1086/174036}{\emph{\apj} {\bfseries 426} (1994) 23} [\href{https://arxiv.org/abs/astro-ph/9304022}{{\ttfamily astro-ph/9304022}}].

\bibitem{Hand2017}
N.~{Hand}, Y.~{Li}, Z.~{Slepian} and U.~{Seljak}, \emph{{An optimal FFT-based anisotropic power spectrum estimator}}, \href{https://doi.org/10.1088/1475-7516/2017/07/002}{\emph{\jcap} {\bfseries 2017} (2017) 002} [\href{https://arxiv.org/abs/1704.02357}{{\ttfamily 1704.02357}}].

\bibitem{Landy1993}
S.D.~{Landy} and A.S.~{Szalay}, \emph{{Bias and Variance of Angular Correlation Functions}}, \href{https://doi.org/10.1086/172900}{\emph{\apj} {\bfseries 412} (1993) 64}.

\bibitem{Padmanabhan_2012}
N.~Padmanabhan, X.~Xu, D.J.~Eisenstein, R.~Scalzo, A.J.~Cuesta, K.T.~Mehta et~al., \emph{A 2 per cent distance toz= 0.35 by reconstructing baryon acoustic oscillations – i. methods and application to the sloan digital sky survey: A 2 per cent distance to z = 0.35}, \href{https://doi.org/10.1111/j.1365-2966.2012.21888.x}{\emph{Monthly Notices of the Royal Astronomical Society} {\bfseries 427} (2012) 2132–2145}.

\bibitem{Chuang_2014}
C.-H.~Chuang, F.-S.~Kitaura, F.~Prada, C.~Zhao and G.~Yepes, \emph{Ezmocks: extending the zel’dovich approximation to generate mock galaxy catalogues with accurate clustering statistics}, \href{https://doi.org/10.1093/mnras/stu2301}{\emph{Monthly Notices of the Royal Astronomical Society} {\bfseries 446} (2014) 2621–2628}.

\bibitem{Hartlap07}
J.~{Hartlap}, P.~{Simon} and P.~{Schneider}, \emph{{Why your model parameter confidences might be too optimistic. Unbiased estimation of the inverse covariance matrix}}, \href{https://doi.org/10.1051/0004-6361:20066170}{\emph{\aap} {\bfseries 464} (2007) 399} [\href{https://arxiv.org/abs/astro-ph/0608064}{{\ttfamily astro-ph/0608064}}].

\bibitem{Percival2014}
W.J.~{Percival}, A.J.~{Ross}, A.G.~{S{\'a}nchez}, L.~{Samushia}, A.~{Burden}, R.~{Crittenden} et~al., \emph{{The clustering of Galaxies in the SDSS-III Baryon Oscillation Spectroscopic Survey: including covariance matrix errors}}, \href{https://doi.org/10.1093/mnras/stu112}{\emph{\mnras} {\bfseries 439} (2014) 2531} [\href{https://arxiv.org/abs/1312.4841}{{\ttfamily 1312.4841}}].

\bibitem{Baleato24}
A.~{Baleato Lizancos} and M.~{White}, \emph{{Harmonic analysis of discrete tracers of large-scale structure}}, \href{https://doi.org/10.1088/1475-7516/2024/05/010}{\emph{\jcap} {\bfseries 2024} (2024) 010} [\href{https://arxiv.org/abs/2312.12285}{{\ttfamily 2312.12285}}].

\bibitem{Alonso18}
D.~{Alonso}, J.~{Sanchez}, A.~{Slosar} and {LSST Dark Energy Science Collaboration}, \emph{{A unified pseudo-C$_{{\ensuremath{\ell}}}$ framework}}, \href{https://doi.org/10.1093/mnras/stz093}{\emph{\mnras} {\bfseries 484} (2019) 4127} [\href{https://arxiv.org/abs/1809.09603}{{\ttfamily 1809.09603}}].

\bibitem{Wolz_2025}
K.~Wolz, D.~Alonso and A.~Nicola, \emph{Catalog-based pseudo-$c_\ell$s}, \href{https://doi.org/10.1088/1475-7516/2025/01/028}{\emph{Journal of Cosmology and Astroparticle Physics} {\bfseries 2025} (2025) 028}.

\bibitem{Kitanidis21}
E.~{Kitanidis} and M.~{White}, \emph{{Cross-correlation of Planck CMB lensing with DESI-like LRGs}}, \href{https://doi.org/10.1093/mnras/staa3927}{\emph{\mnras} {\bfseries 501} (2021) 6181} [\href{https://arxiv.org/abs/2010.04698}{{\ttfamily 2010.04698}}].

\bibitem{White22}
M.~{White}, R.~{Zhou}, J.~{DeRose}, S.~{Ferraro}, S.-F.~{Chen}, N.~{Kokron} et~al., \emph{{Cosmological constraints from the tomographic cross-correlation of DESI Luminous Red Galaxies and Planck CMB lensing}}, {\emph{arXiv e-prints} (2021) arXiv:2111.09898} [\href{https://arxiv.org/abs/2111.09898}{{\ttfamily 2111.09898}}].

\bibitem{Hall_2025}
A.~{Hall} and N.~{Tessore}, \emph{{Pixelization effects in cosmic shear angular power spectra}}, \href{https://doi.org/10.48550/arXiv.2501.08718}{\emph{arXiv e-prints} (2025) arXiv:2501.08718} [\href{https://arxiv.org/abs/2501.08718}{{\ttfamily 2501.08718}}].

\bibitem{Euclid2025_LIX}
{Euclid Collaboration}, N.~{Tessore}, B.~{Joachimi}, A.~{Loureiro}, A.~{Hall}, G.~{Ca{\~n}as-Herrera} et~al., \emph{{Euclid preparation: LIX. Angular power spectra from discrete observations}}, \href{https://doi.org/10.1051/0004-6361/202452018}{\emph{\aap} {\bfseries 694} (2025) A141} [\href{https://arxiv.org/abs/2408.16903}{{\ttfamily 2408.16903}}].

\bibitem{Garc_a_Garc_a_2019}
C.~García-García, D.~Alonso and E.~Bellini, \emph{Disconnected pseudo-$c_\ell$ covariances for projected large-scale structure data}, \href{https://doi.org/10.1088/1475-7516/2019/11/043}{\emph{Journal of Cosmology and Astroparticle Physics} {\bfseries 2019} (2019) 043–043}.

\bibitem{Bianchi_2017}
D.~Bianchi and W.J.~Percival, \emph{Unbiased clustering estimation in the presence of missing observations}, \href{https://doi.org/10.1093/mnras/stx2053}{\emph{Monthly Notices of the Royal Astronomical Society} {\bfseries 472} (2017) 1106–1118}.

\bibitem{Bianchi20}
D.~{Bianchi} and L.~{Verde}, \emph{{Confronting missing observations with probability weights: Fourier space and generalized formalism}}, \href{https://doi.org/10.1093/mnras/staa1267}{\emph{\mnras} {\bfseries 495} (2020) 1511} [\href{https://arxiv.org/abs/1912.08803}{{\ttfamily 1912.08803}}].

\bibitem{KP3s5-Pinon}
M.~{Pinon}, A.~{de Mattia}, P.~{McDonald}, E.~{Burtin}, V.~{Ruhlmann-Kleider}, M.~{White} et~al., \emph{{Mitigation of DESI fiber assignment incompleteness effect on two-point clustering with small angular scale truncated estimators}}, \href{https://doi.org/10.1088/1475-7516/2025/01/131}{\emph{\jcap} {\bfseries 2025} (2025) 131} [\href{https://arxiv.org/abs/2406.04804}{{\ttfamily 2406.04804}}].

\bibitem{deMattia2019}
A.~{de Mattia} and V.~{Ruhlmann-Kleider}, \emph{{Integral constraints in spectroscopic surveys}}, \href{https://doi.org/10.1088/1475-7516/2019/08/036}{\emph{\jcap} {\bfseries 2019} (2019) 036} [\href{https://arxiv.org/abs/1904.08851}{{\ttfamily 1904.08851}}].

\bibitem{Reid16}
B.~{Reid}, S.~{Ho}, N.~{Padmanabhan}, W.J.~{Percival}, J.~{Tinker}, R.~{Tojeiro} et~al., \emph{{SDSS-III Baryon Oscillation Spectroscopic Survey Data Release 12: galaxy target selection and large-scale structure catalogues}}, \href{https://doi.org/10.1093/mnras/stv2382}{\emph{\mnras} {\bfseries 455} (2016) 1553} [\href{https://arxiv.org/abs/1509.06529}{{\ttfamily 1509.06529}}].

\bibitem{Ross_2020}
A.J.~Ross, J.~Bautista, R.~Tojeiro, S.~Alam, S.~Bailey, E.~Burtin et~al., \emph{The completed sdss-iv extended baryon oscillation spectroscopic survey: Large-scale structure catalogues for cosmological analysis}, \href{https://doi.org/10.1093/mnras/staa2416}{\emph{Monthly Notices of the Royal Astronomical Society} {\bfseries 498} (2020) 2354–2371}.

\bibitem{Chaussidon_2021}
E.~Chaussidon, C.~Yèche, N.~Palanque-Delabrouille, A.~de Mattia, A.D.~Myers, M.~Rezaie et~al., \emph{Angular clustering properties of the desi qso target selection using dr9 legacy imaging surveys}, \href{https://doi.org/10.1093/mnras/stab3252}{\emph{Monthly Notices of the Royal Astronomical Society} {\bfseries 509} (2021) 3904–3923}.

\bibitem{Rezaie_2020}
M.~Rezaie, H.-J.~Seo, A.J.~Ross and R.C.~Bunescu, \emph{Improving galaxy clustering measurements with deep learning: analysis of the decals dr7 data}, \href{https://doi.org/10.1093/mnras/staa1231}{\emph{Monthly Notices of the Royal Astronomical Society} {\bfseries 495} (2020) 1613–1640}.

\bibitem{ACT:2023oei}
{\scshape ACT} collaboration, \emph{{The Atacama Cosmology Telescope: Cosmology from Cross-correlations of unWISE Galaxies and ACT DR6 CMB Lensing}}, \href{https://doi.org/10.3847/1538-4357/ad31a5}{\emph{Astrophys. J.} {\bfseries 966} (2024) 157} [\href{https://arxiv.org/abs/2309.05659}{{\ttfamily 2309.05659}}].

\bibitem{Farren:2024rla}
G.S.~Farren et~al., \emph{{Atacama Cosmology Telescope: Multiprobe cosmology with unWISE galaxies and ACT DR6 CMB lensing}}, \href{https://doi.org/10.1103/PhysRevD.111.083516}{\emph{Phys. Rev. D} {\bfseries 111} (2025) 083516} [\href{https://arxiv.org/abs/2409.02109}{{\ttfamily 2409.02109}}].

\bibitem{Chen22}
S.-F.~{Chen}, Z.~{Vlah} and M.~{White}, \emph{{A new analysis of galaxy 2-point functions in the BOSS survey, including full-shape information and post-reconstruction BAO}}, \href{https://doi.org/10.1088/1475-7516/2022/02/008}{\emph{\jcap} {\bfseries 2022} (2022) 008} [\href{https://arxiv.org/abs/2110.05530}{{\ttfamily 2110.05530}}].

\bibitem{Mat08a}
T.~{Matsubara}, \emph{{Resumming cosmological perturbations via the Lagrangian picture: One-loop results in real space and in redshift space}}, \href{https://doi.org/10.1103/PhysRevD.77.063530}{\emph{\prd} {\bfseries 77} (2008) 063530} [\href{https://arxiv.org/abs/0711.2521}{{\ttfamily 0711.2521}}].

\bibitem{Mat08b}
T.~{Matsubara}, \emph{{Nonlinear perturbation theory with halo bias and redshift-space distortions via the Lagrangian picture}}, \href{https://doi.org/10.1103/PhysRevD.78.083519}{\emph{\prd} {\bfseries 78} (2008) 083519} [\href{https://arxiv.org/abs/0807.1733}{{\ttfamily 0807.1733}}].

\bibitem{Cat00}
P.~{Catelan}, C.~{Porciani} and M.~{Kamionkowski}, \emph{{Two ways of biasing galaxy formation}}, \href{https://doi.org/10.1046/j.1365-8711.2000.04023.x}{\emph{\mnras} {\bfseries 318} (2000) L39} [\href{https://arxiv.org/abs/astro-ph/0005544}{{\ttfamily astro-ph/0005544}}].

\bibitem{Ang15}
R.~{Angulo}, M.~{Fasiello}, L.~{Senatore} and Z.~{Vlah}, \emph{{On the statistics of biased tracers in the Effective Field Theory of Large Scale Structures}}, \href{https://doi.org/10.1088/1475-7516/2015/09/029}{\emph{\jcap} {\bfseries 9} (2015) 029} [\href{https://arxiv.org/abs/1503.08826}{{\ttfamily 1503.08826}}].

\bibitem{Des18}
V.~{Desjacques}, D.~{Jeong} and F.~{Schmidt}, \emph{{Large-scale galaxy bias}}, \href{https://doi.org/10.1016/j.physrep.2017.12.002}{\emph{\physrep} {\bfseries 733} (2018) 1} [\href{https://arxiv.org/abs/1611.09787}{{\ttfamily 1611.09787}}].

\bibitem{ChenCasWhi19}
S.-F.~{Chen}, E.~{Castorina} and M.~{White}, \emph{{Biased tracers of two fluids in the Lagrangian picture}}, \href{https://doi.org/10.1088/1475-7516/2019/06/006}{\emph{Journal of Cosmology and Astro-Particle Physics} {\bfseries 2019} (2019) 006} [\href{https://arxiv.org/abs/1903.00437}{{\ttfamily 1903.00437}}].

\bibitem{FujitaVlah20}
T.~Fujita and Z.~Vlah, \emph{{Perturbative description of biased tracers using consistency relations of LSS}}, \href{https://doi.org/10.1088/1475-7516/2020/10/059}{\emph{JCAP} {\bfseries 10} (2020) 059} [\href{https://arxiv.org/abs/2003.10114}{{\ttfamily 2003.10114}}].

\bibitem{Chen20}
S.-F.~{Chen}, Z.~{Vlah} and M.~{White}, \emph{{Consistent modeling of velocity statistics and redshift-space distortions in one-loop perturbation theory}}, \href{https://doi.org/10.1088/1475-7516/2020/07/062}{\emph{\jcap} {\bfseries 2020} (2020) 062} [\href{https://arxiv.org/abs/2005.00523}{{\ttfamily 2005.00523}}].

\bibitem{Alcock79}
C.~{Alcock} and B.~{Paczynski}, \emph{{An evolution free test for non-zero cosmological constant}}, \href{https://doi.org/10.1038/281358a0}{\emph{\nat} {\bfseries 281} (1979) 358}.

\bibitem{Modi20}
C.~Modi, S.-F.~Chen and M.~White, \emph{Simulations and symmetries}, \href{https://doi.org/10.1093/mnras/staa251}{\emph{\mnras} {\bfseries 492} (2020) 5754}.

\bibitem{Hadzhiyska:2021xbv}
B.~Hadzhiyska, C.~Garc\'\i{}a-Garc\'\i{}a, D.~Alonso, A.~Nicola and A.~Slosar, \emph{{Hefty enhancement of cosmological constraints from the DES Y1 data using a hybrid effective field theory approach to galaxy bias}}, \href{https://doi.org/10.1088/1475-7516/2021/09/020}{\emph{JCAP} {\bfseries 09} (2021) 020} [\href{https://arxiv.org/abs/2103.09820}{{\ttfamily 2103.09820}}].

\bibitem{anzu21}
N.~Kokron, J.~DeRose, S.-F.~Chen, M.~White and R.H.~Wechsler, \emph{The cosmology dependence of galaxy clustering and lensing from a hybrid n-body-perturbation theory model}, \href{https://doi.org/10.1093/mnras/stab1358}{\emph{\mnras} {\bfseries 505} (2021) 1422}.

\bibitem{Zennaro2022}
M.~{Zennaro}, R.E.~{Angulo}, S.~{Contreras}, M.~{Pellejero-Ib{\'a}{\~n}ez} and F.~{Maion}, \emph{{Priors on Lagrangian bias parameters from galaxy formation modelling}}, \href{https://doi.org/10.1093/mnras/stac1673}{\emph{\mnras} {\bfseries 514} (2022) 5443} [\href{https://arxiv.org/abs/2110.05408}{{\ttfamily 2110.05408}}].

\bibitem{DeRose_2023}
J.~DeRose, N.~Kokron, A.~Banerjee, S.-F.~Chen, M.~White, R.~Wechsler et~al., \emph{Aemulus $\nu$: precise predictions for matter and biased tracer power spectra in the presence of neutrinos}, \href{https://doi.org/10.1088/1475-7516/2023/07/054}{\emph{Journal of Cosmology and Astroparticle Physics} {\bfseries 2023} (2023) 054}.

\bibitem{Ibanez2023}
M.~{Pellejero Iba{\~n}ez}, R.E.~{Angulo}, M.~{Zennaro}, J.~{St{\"u}cker}, S.~{Contreras}, G.~{Aric{\`o}} et~al., \emph{{The bacco simulation project: bacco hybrid Lagrangian bias expansion model in redshift space}}, \href{https://doi.org/10.1093/mnras/stad368}{\emph{\mnras} {\bfseries 520} (2023) 3725} [\href{https://arxiv.org/abs/2207.06437}{{\ttfamily 2207.06437}}].

\bibitem{Nicola2024}
A.~{Nicola}, B.~{Hadzhiyska}, N.~{Findlay}, C.~{Garc{\'\i}a-Garc{\'\i}a}, D.~{Alonso}, A.~{Slosar} et~al., \emph{{Galaxy bias in the era of LSST: perturbative bias expansions}}, \href{https://doi.org/10.1088/1475-7516/2024/02/015}{\emph{\jcap} {\bfseries 2024} (2024) 015} [\href{https://arxiv.org/abs/2307.03226}{{\ttfamily 2307.03226}}].

\bibitem{Limber1954}
D.N.~{Limber}, \emph{{The Analysis of Counts of the Extragalactic Nebulae in Terms of a Fluctuating Density Field. II.}}, \href{https://doi.org/10.1086/145870}{\emph{\apj} {\bfseries 119} (1954) 655}.

\bibitem{Kaiser1992}
N.~{Kaiser}, \emph{{Weak Gravitational Lensing of Distant Galaxies}}, \href{https://doi.org/10.1086/171151}{\emph{\apj} {\bfseries 388} (1992) 272}.

\bibitem{Schoeneberg2024BBN}
N.~{Sch{\"o}neberg}, \emph{{The 2024 BBN baryon abundance update}}, \href{https://doi.org/10.1088/1475-7516/2024/06/006}{\emph{\jcap} {\bfseries 2024} (2024) 006} [\href{https://arxiv.org/abs/2401.15054}{{\ttfamily 2401.15054}}].

\bibitem{Heydenreich2025}
S.~{Heydenreich}, A.~{Leauthaud}, C.~{Blake}, Z.~{Sun}, J.U.~{Lange}, T.~{Zhang} et~al., \emph{{Lensing Without Borders: Measurements of galaxy-galaxy lensing and projected galaxy clustering in DESI DR1}}, \href{https://doi.org/10.48550/arXiv.2506.21677}{\emph{arXiv e-prints} (2025) arXiv:2506.21677} [\href{https://arxiv.org/abs/2506.21677}{{\ttfamily 2506.21677}}].

\bibitem{Philcox_2022}
O.H.~Philcox, M.M.~Ivanov, G.~Cabass, M.~Simonović, M.~Zaldarriaga and T.~Nishimichi, \emph{Cosmology with the redshift-space galaxy bispectrum monopole at one-loop order}, \href{https://doi.org/10.1103/physrevd.106.043530}{\emph{Physical Review D} {\bfseries 106} (2022) }.

\bibitem{Masot2025}
S.~{Novell Masot}, H.~{Gil-Mar{\'\i}n}, L.~{Verde}, J.~{Aguilar}, S.~{Ahlen}, S.~{Bailey} et~al., \emph{{Full-Shape analysis of the power spectrum and bispectrum of DESI DR1 LRG and QSO samples}}, \href{https://doi.org/10.48550/arXiv.2503.09714}{\emph{arXiv e-prints} (2025) arXiv:2503.09714} [\href{https://arxiv.org/abs/2503.09714}{{\ttfamily 2503.09714}}].

\bibitem{Powell2009}
M.~Powell, \emph{The bobyqa algorithm for bound constrained optimization without derivatives}, {\emph{Technical Report, Department of Applied Mathematics and Theoretical Physics} (2009) }.

\bibitem{Cartis2018}
C.~Cartis, J.~Fiala, B.~Marteau and L.~Roberts, \emph{Improving the flexibility and robustness of model-based derivative-free optimization solvers}, \href{https://doi.org/10.1145/3338517}{\emph{ACM Transactions on Mathematical Software} {\bfseries 45} (2018) }.

\bibitem{Cartis2021}
C.~Cartis, L.~Roberts and O.~Sheridan-Methven, \emph{Escaping local minima with local derivative-free methods: a numerical investigation}, \href{https://doi.org/10.1080/02331934.2021.1883015}{\emph{Optimization} {\bfseries 71} (2021) 1}.

\bibitem{PACT2025}
T.~{Louis}, A.~{La Posta}, Z.~{Atkins}, H.T.~{Jense}, I.~{Abril-Cabezas}, G.E.~{Addison} et~al., \emph{{The Atacama Cosmology Telescope: DR6 Power Spectra, Likelihoods and $\Lambda$CDM Parameters}}, \href{https://doi.org/10.48550/arXiv.2503.14452}{\emph{arXiv e-prints} (2025) arXiv:2503.14452} [\href{https://arxiv.org/abs/2503.14452}{{\ttfamily 2503.14452}}].

\bibitem{Rosenberg22}
E.~{Rosenberg}, S.~{Gratton} and G.~{Efstathiou}, \emph{{CMB power spectra and cosmological parameters from Planck PR4 with CamSpec}}, \href{https://doi.org/10.1093/mnras/stac2744}{\emph{\mnras} {\bfseries 517} (2022) 4620} [\href{https://arxiv.org/abs/2205.10869}{{\ttfamily 2205.10869}}].

\bibitem{deBelsunce21}
R.~{de Belsunce}, S.~{Gratton}, W.~{Coulton} and G.~{Efstathiou}, \emph{{Inference of the optical depth to reionization from low multipole temperature and polarization Planck data}}, \href{https://doi.org/10.1093/mnras/stab2215}{\emph{\mnras} {\bfseries 507} (2021) 1072} [\href{https://arxiv.org/abs/2103.14378}{{\ttfamily 2103.14378}}].

\bibitem{kids1000}
C.~{Heymans}, T.~{Tr{\"o}ster}, M.~{Asgari}, C.~{Blake}, H.~{Hildebrandt}, B.~{Joachimi} et~al., \emph{{KiDS-1000 Cosmology: Multi-probe weak gravitational lensing and spectroscopic galaxy clustering constraints}}, \href{https://doi.org/10.1051/0004-6361/202039063}{\emph{\aap} {\bfseries 646} (2021) A140} [\href{https://arxiv.org/abs/2007.15632}{{\ttfamily 2007.15632}}].

\bibitem{Amon21}
A.~{Amon}, D.~{Gruen}, M.A.~{Troxel}, N.~{MacCrann}, S.~{Dodelson}, A.~{Choi} et~al., \emph{{Dark Energy Survey Year 3 Results: Cosmology from Cosmic Shear and Robustness to Data Calibration}}, {\emph{arXiv e-prints} (2021) arXiv:2105.13543} [\href{https://arxiv.org/abs/2105.13543}{{\ttfamily 2105.13543}}].

\bibitem{Secco21}
L.F.~{Secco}, S.~{Samuroff}, E.~{Krause}, B.~{Jain}, J.~{Blazek}, M.~{Raveri} et~al., \emph{{Dark Energy Survey Year 3 Results: Cosmology from Cosmic Shear and Robustness to Modeling Uncertainty}}, {\emph{arXiv e-prints} (2021) arXiv:2105.13544} [\href{https://arxiv.org/abs/2105.13544}{{\ttfamily 2105.13544}}].

\bibitem{Dalal23}
R.~{Dalal}, X.~{Li}, A.~{Nicola}, J.~{Zuntz}, M.A.~{Strauss}, S.~{Sugiyama} et~al., \emph{{Hyper Suprime-Cam Year 3 results: Cosmology from cosmic shear power spectra}}, \href{https://doi.org/10.1103/PhysRevD.108.123519}{\emph{\prd} {\bfseries 108} (2023) 123519} [\href{https://arxiv.org/abs/2304.00701}{{\ttfamily 2304.00701}}].

\bibitem{Li23}
X.~{Li}, T.~{Zhang}, S.~{Sugiyama}, R.~{Dalal}, R.~{Terasawa}, M.M.~{Rau} et~al., \emph{{Hyper Suprime-Cam Year 3 results: Cosmology from cosmic shear two-point correlation functions}}, \href{https://doi.org/10.1103/PhysRevD.108.123518}{\emph{\prd} {\bfseries 108} (2023) 123518} [\href{https://arxiv.org/abs/2304.00702}{{\ttfamily 2304.00702}}].

\bibitem{DESKids23}
{Dark Energy Survey and Kilo-Degree Survey Collaboration}, T.M.C.~{Abbott}, M.~{Aguena}, A.~{Alarcon}, O.~{Alves}, A.~{Amon} et~al., \emph{{DES Y3 + KiDS-1000: Consistent cosmology combining cosmic shear surveys}}, \href{https://doi.org/10.21105/astro.2305.17173}{\emph{The Open Journal of Astrophysics} {\bfseries 6} (2023) 36} [\href{https://arxiv.org/abs/2305.17173}{{\ttfamily 2305.17173}}].

\bibitem{KiDS-Legacy25}
A.H.~{Wright}, B.~{St{\"o}lzner}, M.~{Asgari}, M.~{Bilicki}, B.~{Giblin}, C.~{Heymans} et~al., \emph{{KiDS-Legacy: Cosmological constraints from cosmic shear with the complete Kilo-Degree Survey}}, \href{https://doi.org/10.48550/arXiv.2503.19441}{\emph{arXiv e-prints} (2025) arXiv:2503.19441} [\href{https://arxiv.org/abs/2503.19441}{{\ttfamily 2503.19441}}].

\bibitem{GarciaGarcia24}
C.~{Garc{\'\i}a-Garc{\'\i}a}, M.~{Zennaro}, G.~{Aric{\`o}}, D.~{Alonso} and R.E.~{Angulo}, \emph{{Cosmic shear with small scales: DES-Y3, KiDS-1000 and HSC-DR1}}, \href{https://doi.org/10.1088/1475-7516/2024/08/024}{\emph{\jcap} {\bfseries 2024} (2024) 024} [\href{https://arxiv.org/abs/2403.13794}{{\ttfamily 2403.13794}}].

\bibitem{Rubin2023}
D.~{Rubin}, G.~{Aldering}, M.~{Betoule}, A.~{Fruchter}, X.~{Huang}, A.G.~{Kim} et~al., \emph{{Union Through UNITY: Cosmology with 2,000 SNe Using a Unified Bayesian Framework}}, \href{https://doi.org/10.48550/arXiv.2311.12098}{\emph{arXiv e-prints} (2023) arXiv:2311.12098} [\href{https://arxiv.org/abs/2311.12098}{{\ttfamily 2311.12098}}].

\bibitem{Baleato2025}
A.~{Baleato Lizancos}, U.~{Seljak}, M.~{Karamanis}, M.~{Bonici} and S.~{Ferraro}, \emph{{Selecting samples of galaxies with fewer Fingers-of-God}}, \href{https://doi.org/10.48550/arXiv.2501.10587}{\emph{arXiv e-prints} (2025) arXiv:2501.10587} [\href{https://arxiv.org/abs/2501.10587}{{\ttfamily 2501.10587}}].

\bibitem{Chevallier2001}
M.~{Chevallier} and D.~{Polarski}, \emph{{Accelerating Universes with Scaling Dark Matter}}, \href{https://doi.org/10.1142/S0218271801000822}{\emph{International Journal of Modern Physics D} {\bfseries 10} (2001) 213} [\href{https://arxiv.org/abs/gr-qc/0009008}{{\ttfamily gr-qc/0009008}}].

\bibitem{Linder2003}
E.V.~{Linder}, \emph{{Exploring the Expansion History of the Universe}}, \href{https://doi.org/10.1103/PhysRevLett.90.091301}{\emph{\prl} {\bfseries 90} (2003) 091301} [\href{https://arxiv.org/abs/astro-ph/0208512}{{\ttfamily astro-ph/0208512}}].

\bibitem{Joyce2016}
A.~{Joyce}, L.~{Lombriser} and F.~{Schmidt}, \emph{{Dark Energy Versus Modified Gravity}}, \href{https://doi.org/10.1146/annurev-nucl-102115-044553}{\emph{Annual Review of Nuclear and Particle Science} {\bfseries 66} (2016) 95} [\href{https://arxiv.org/abs/1601.06133}{{\ttfamily 1601.06133}}].

\bibitem{Grimm2025}
N.~{Grimm}, C.~{Bonvin} and I.~{Tutusaus}, \emph{{Does dark matter fall in the same way as standard model particles? A direct constraint of Euler's equation with cosmological data}}, \href{https://doi.org/10.48550/arXiv.2502.12843}{\emph{arXiv e-prints} (2025) arXiv:2502.12843} [\href{https://arxiv.org/abs/2502.12843}{{\ttfamily 2502.12843}}].

\end{thebibliography}\endgroup
\bibliographystyle{jhep}

\end{document}